\documentclass[onecolumn,amsmath,amssymb,nofootinbib,12pt]{article}
\usepackage{jheppub}
\usepackage{ifpdf}
\pdfoutput=1

\usepackage{graphicx,subcaption}
\usepackage{amsfonts}
\usepackage{amsmath}
\usepackage{amssymb}
\usepackage{epsfig}
\usepackage{pdfpages}
\usepackage{graphicx,epstopdf}
\usepackage[makeroom]{cancel}
\usepackage{array}
\usepackage{ulem}

\newcommand{\RN}[1]{%
  \textup{\uppercase\expandafter{\romannumeral#1}}%
}

\newcommand{\del}{\partial}
\newcommand{\vac}{\text{vac}}
\newcommand{\surf}{\text{surf}}

\newcommand{\bulk}{\text{bulk}}

\newcommand{\corner}{\text{corner}}

\newcommand{\jnt}{\text{jnt}}

\newcommand{\be}{\begin{equation}}
\newcommand{\ee}{\end{equation}}
\newcommand{\bea}{\begin{eqnarray}}
\newcommand{\eea}{\end{eqnarray}}
\newcommand{\beq}{\begin{equation}}
\newcommand{\eeq}{\end{equation}}
\newcommand{\beqa}{\begin{eqnarray}}
\newcommand{\eeqa}{\end{eqnarray}}
\newcommand{\beqar}{\begin{eqnarray*}}
\newcommand{\eeqar}{\end{eqnarray*}}

\newcommand{\labell}[1]{\label{#1}} 

\newcommand{\eg}{{\it e.g.,}\ }
\newcommand{\ie}{{\it i.e.,}\ }

\newcommand{\reef}[1]{(\ref{#1})}
\newcommand{\ssc}{\scriptscriptstyle}
\newcommand{\mt}[1]{\textrm{\tiny #1}}

\newcommand{\CA}{\mathcal{C}_A}

\newcommand{\pa}{\partial}
\newcommand{\dt}{{\delta t}}
\newcommand{\rh}{{r_h}}

\begin{document}

\preprint{arXiv:1709.nnnnn [hep-th]}
\preprint{OU-HET-945}
\title{On the Time Dependence of Holographic Complexity}

\author[a,b]{Dean Carmi,}
\author[a]{Shira Chapman,}
\author[a, c]{Hugo Marrochio,}
\author[a]{Robert C. Myers}
\author[a, d]{and Sotaro Sugishita}
\affiliation[a]{Perimeter Institute for Theoretical Physics, Waterloo, ON N2L 2Y5, Canada}
\affiliation[b]{Raymond and Beverly Sackler Faculty of Exact Sciences\\
School of Physics and Astronomy, Tel-Aviv University, Ramat-Aviv 69978, Israel}
\affiliation[c]{Department of Physics $\&$ Astronomy\\
University of Waterloo, Waterloo, ON N2L 3G1, Canada}
\affiliation[d]{Department of Physics, Osaka University, Toyonaka, Osaka, 560-0043, Japan}

\emailAdd{schapman@perimeterinstitute.ca}
\emailAdd{hmarrochio@perimeterinstitute.ca}
\emailAdd{rmyers@perimeterinstitute.ca}
\emailAdd{sugishita@het.phys.sci.osaka-u.ac.jp}

\date{\today}

\abstract{We evaluate the full time dependence of holographic complexity in various eternal black hole backgrounds using both the complexity=action (CA) and the complexity=volume (CV) conjectures.
We conclude using the CV conjecture that the rate of change of complexity is a monotonically increasing function of time, which saturates from below to a positive constant in the late time limit. Using the CA conjecture for uncharged black holes, the holographic complexity remains constant for an initial period, then briefly decreases but quickly begins to increase. As observed previously, at late times, the rate of growth of the complexity approaches  a constant, which may be associated with Lloyd's bound on the rate of computation. However, we find that this late time limit is approached from above, thus violating the bound.
Adding a charge to the eternal black holes washes out the early time behaviour, \ie complexity immediately begins increasing with sufficient charge, but the late time behaviour is essentially the same as in the neutral case. We also evaluate the complexity of formation for charged black holes and find that it is divergent for extremal black holes, implying that the states at finite chemical potential and zero temperature are infinitely more complex than their finite temperature counterparts.
}

\maketitle

\section{Introduction}\label{sec:Introduction}
In recent years, surprising new connections have been developing between quantum information and quantum gravity. The AdS/CFT correspondence allows us to quantitatively
study these connections in a holographic framework where certain geometric quantities in the bulk spacetime can be related to the entanglement properties of the boundary field theory. The Ryu-Takayanagi construction, which provides a geometrical realization of the entanglement entropy of the boundary CFT, is the prime example of such a relation \cite{Ryu:2006bv, Casini:2011kv, Lewkowycz:2013nqa,Dong:2016hjy}. However, a new concept that has recently entered this discussion is quantum computational complexity. In fact, two holographic proposals have been developed to describe the quantum complexity of states in the boundary theory, namely, the complexity=volume (CV) conjecture \cite{Susskind:2014rva,Stanford:2014jda} and  the complexity=action (CA) conjecture \cite{Brown1,Brown2}.

In the holographic context, quantum complexity quantifies how hard it is to prepare a particular state of interest, by applying a series of (simple) elementary gates to a (simple) reference state, \eg see \cite{johnw,AaronsonRev} for  reviews. However, despite being relatively well understood for spin-chains, only recently complexity models have been developed for quantum field theories \cite{qft1,qft2,koji}. While only considering free scalars \cite{qft1,qft2}, these calculations yield striking similarities to the results produced with holographic complexity. In \cite{koji}, the time dependence of complexity in Abelian gauge theories was studied. Related investigations attempting to better understand complexity from the perspective of the boundary theory have also appeared in \cite{Chemissany:2016qqq,EuclideanComplexity1,EuclideanComplexity2,
EuclideanComplexity3,prep9,Yang:2017nfn}.

A prime arena for discussions of holographic complexity has been the eternal two-sided black hole and this will also be the case in the present paper.  This bulk geometry is dual to the thermofield double state in the boundary theory \cite{MaldacenaEternal},
\begin{equation}
\bigl| \text{TFD}(t_L,t_R) \bigl> = Z^{-1/2} \sum_{\alpha} e^{-E_{\alpha}/(2 T)}\,e^{-iE_\alpha(t_L+t_R)} \,\bigl| E_{\alpha} \bigl>_{L} \bigl| E_{\alpha} \bigl>_{R} \, ,
\label{TFDx}
\end{equation}
where $L$ and $R$ label the quantum states (and times) associated with the left and right boundaries. Hence, we have an entangled state of two copies of the boundary CFT and this entanglement is responsible for the geometric connection in the bulk, \ie the Einstein-Rosen bridge \cite{Hartman:2013qma, Maldacena:2013xja}. A puzzle was to understand the growth of the black hole interior in terms of the boundary degrees of freedom. The conjectured holographic complexity appears to provide an explanation \cite{Susskind:2014rva}, since a characteristic property of quantum complexity is that it continues to grow for very long times after the system has thermalized. In fact, the complexity is conjectured to continue growing until a time scale which is exponential in the number of degrees of freedom in the system \cite{2LawComp}.

Turning to the bulk definitions proposed for holographic complexity, we have the following: The complexity=volume conjecture equates the complexity to the volume\footnote{This extremal volume was also argued to be dual to the quantum information metric, when comparing two vacuum states of boundary theories which differ by a marginal deformation \cite{TheTaka}.} of the extremal/maximal time slice anchored at boundary times $t_{L}$ and $t_{R}$ \cite{Susskind:2014rva,Stanford:2014jda},
\begin{equation}\label{volver}
\mathcal{C}_{\text{V}} =
{\rm max}\!\left[\frac{\mathcal{V(B)}}{G_N \, \ell}\right] \, ,
\end{equation}
where $\ell$ is a certain length scale associated with the geometry
(see figure \ref{VolumeWormhole}). The complexity=action conjecture instead equates the boundary complexity with the gravitational action evaluated on a region of spacetime known as the Wheeler-DeWitt (WDW) patch, \ie the region bounded by the null surfaces anchored at the relevant times on the left and right boundaries   (see, \eg figure \ref{PenroseBHa})
\begin{equation}\label{compAct}
\mathcal{C}_A = \frac{I_{WDW}}{\pi \hbar}\, .
\end{equation}
One might also regard the WDW patch as the domain of dependence of the maximal time slice appearing in the CV conjecture. However, one should keep in mind that there are also certain ambiguities in defining the contributions of the null boundaries to the gravitational action $I_{WDW}$ \cite{RobLuis}.
Various features of these two holographic quantities have been studied --- \eg see \cite{Susskind:2014rva, Stanford:2014jda, Brown1, Brown2, ying1, Formation, diverg,Reynolds:2016rvl}. While eqs.~\reef{volver} and \reef{compAct} do not yield the same results quantitatively for the complexity, they still agree at a qualitative level. Therefore it may be that the differences of these bulk quantities are related to differences in the microscopic definition of complexity in the boundary theory, \eg in the choice of elementary gates \cite{Formation}.

One striking result found with the CA proposal is that the late time growth rate is proportional to $2 M/\pi$, independent of the boundary curvature and the spacetime dimension \cite{Brown1,Brown2}. Further it was suggested that this saturation of the growth rate is related to Lloyd's bound on the rate of computation by a system with energy $M$ \cite{Lloyd}. Using the CV conjecture, the late time growth rate of the complexity also saturates, but this final rate is only proportional to the mass at high temperatures and with a coefficient that depends on the spacetime dimension \cite{Stanford:2014jda,Formation}. Despite extensive discussions of this late time limit for the time dependence of the holographic complexity, the question of its full time evolution and in particular the rate of change at early times has not been thoroughly investigated.\footnote{However, see \cite{TheTaka} and section 8 of \cite{Brown2}}  Therefore, in the present paper, we study the full time evolution of holographic complexity, for both the CV and the CA proposals, in static two-sided eternal black holes. We consider black holes in various dimensions and with spherical, planar and hyperbolical horizon geometries. We also investigate the properties of complexity for  charged black holes (for $d\ge3$, where $d$ is the spacetime dimension of the boundary theory).
The full time profile in all cases except $d=2$ requires some numerical treatment. We are, however, able to identify certain general features.

For the CA proposal (and in $d\geq 3$), we find that the complexity remains unchanged for some critical time, which is of the order of the thermal scale. Immediately after this time, the rate of change of the complexity is negatively divergent and we observe a short transient period during which the complexity is decreasing. At late times, the rate of change in complexity approaches a constant, previously understood to be associated with Lloyd's bound on the rate of computation. However we observe  a violation of this late time bound since the rate approaches the late time limit from above. We also comment on the role of the arbitrary length scale in the boundary theory associated with the holographic normalization of null-normals and its influence on the rate of change of complexity. For the CV proposal, the rate of change of complexity is a monotonically increasing function of time, and it saturates to a constant at late times. While at high temperatures this late time rate is proportional to the mass, the precise value depends on the boundary curvature for spherical and hyperbolic horizons at finite temperatures. For both conjectures  (and in $d\geq 3$), we also examined the rate of change of complexity for charged black holes, as well as their complexity of formation. In either case, we find that the holographic complexity smoothly approaches to that of the neutral black holes in the limit of zero charge. With the CA approach, adding a charge washes out the curious early time behaviour, \ie complexity immediately begins increasing with sufficient charge, but the late time violation is essentially the same as in the neutral case. Further, the complexity of formation for charged extremal black holes is divergent in either case, implying that the holographic states at finite chemical potential and zero temperature are infinitely more complex than their finite temperature counterparts.

The remainder of our paper is organized as follows: In section \ref{sec:EternalAction}, we investigate the full time evolution of complexity for the thermofield double state \reef{TFDx}, dual to an eternal AdS black hole, using the CA conjecture. We consider different boundary geometries and different dimensions, and investigate how the holographic complexity approaches the late time limit. In section \ref{sec:EternalVolume}, we study the time evolution of complexity using the CV conjecture. We consider various geometries and dimensions, and prove that it approaches its late time limit from below. In section \ref{sec:ChargedEternal}, we analyze Reissner-Nordstrom AdS charged black holes, their complexity of formation and how they violate a proposed generalization of Lloyd's bound. Finally, we discuss some implications of our results, as well as possible future directions, in section \ref{sec:Discussion}. We relegate certain details of the calculations to the appendices. In appendix \ref{app:BTZnonsymmetric}, we present additional details for the action calculation for BTZ black holes. Extra examples of the time dependence of complexity for   uncharged black holes in $d=3$ using the CA conjecture and for spherical and hyperbolic geometries in $d=3$ and $d=4$ using the CV conjecture are presented in appendix \ref{app:MoreAction}. We present a late time expansion of the uncharged CV results in appendix \ref{app:MoreVolume}.  In appendix \ref{CformCharged}, we show the details of the calculation of the complexity of formation for charged black holes, both using the CA and CV proposals. In appendix \ref{app:EternalAmb}, we discuss the influence of ambiguities associated with the presence of null boundaries on the CA proposal results.

\section{Complexity=Action}\label{sec:EternalAction}

In this section, we study the time evolution of holographic complexity using the complexity=action (CA) conjecture \cite{Brown1,Brown2} for (neutral) eternal AdS black holes in $d+1$ dimensions. The proposed translation between the boundary and the bulk theories states that the quantum complexity of the boundary state is given by the gravitational action evaluated on a bulk region known as the Wheeler-DeWitt (WDW) patch, as in eq.~\reef{compAct}. Our conventions and notation here follow those established in \cite{Formation}. The (neutral) black hole metric for different horizon geometries reads
\begin{equation}\label{HigherDMetric}
d s^{2} = - f(r)\,
\, d t^{2} + \frac{d r^{2}}{f(r)} + r^{2}\, d \Sigma^{2}_{k,d-1}\,,
\end{equation}
with the blackening factor
\begin{equation}\label{BFactor}
f(r) = \frac{r^2}{L^2}+k -
\frac{\omega^{d-2}}{r^{d-2}}\,.
\end{equation}
Here $L$ denotes the AdS curvature scale while $k$ indicates the curvature of the ($d$--1)-dimensional line element $d \Sigma^{2}_{k,d-1}$.
The parameter $k$ assumes three different values, $\{+1, 0, -1\}$, which correspond to spherical, planar, and hyperbolic horizon geometries, respectively.
In the expressions below, we will use $\Omega_{k,d-1}$ to denote the dimensionless volume of the relevant spatial geometry. For $k=1$, this is just the  volume of a ($d$--1)-dimensional unit sphere, \ie  $\Omega_{1,d-1} = 2 \pi^{d/2}/\Gamma\left({d}/{2}\right)$, while for  hyperbolic and planar geometries, we must introduce an infrared regulator to produce a finite volume (\eg see eq.~(2.3) in \cite{Formation}).

The relation between the position of the horizon $r_h$ and the `mass' parameter $\omega$ is \cite{energy, count}
\begin{equation}
\omega^{d-2}= r_h^{d-2}\left(\frac{r_h^2}{L^2}+k\right)\, ,
\label{horiz}
\end{equation}
which is then related to the mass of the black hole with
\begin{equation}\label{Mass}
M = \frac{(d-1) \, \Omega_{k,d-1}}{16 \pi \, G_N}
\, \omega^{d-2} \,  .
\end{equation}
We will also use the temperature and entropy of the black hole given by
\begin{equation}\label{eq:ST}
S =  \frac{\Omega_{k,d-1} }{4 G_N}\,r_h^{d-1}\,,\qquad
T=\frac{1}{4\pi }\left.\frac{\partial f}{\partial r}\right|_{r=r_h}=\frac{1}{4\pi \,r_h}\left(d\,\frac{r_h^2}{L^2} + (d-2)\,k \right)\,.
\end{equation}

To describe the null sheets bounding the WDW patch, it is convenient to define the tortoise coordinate, and its asymptotic value:
\begin{equation}
r^*(r) = \int \frac{d r}{f(r)}\,, \qquad
r_\infty^* = \lim_{r\rightarrow \infty}r^*(r)\,.
\label{tort2}
\end{equation}
We then define the Eddington-Finkelstein coordinates, $u$ and $v$, describing out- and in-going null rays, respectively,
\begin{equation}
v = t+r^*(r)\,, \qquad u =  t-r^*(r)\,. \label{EFcoord}
\end{equation}

\subsection{Evaluating the Action} \label{eval2}
\begin{figure}
\centering
\includegraphics[scale=0.0569]{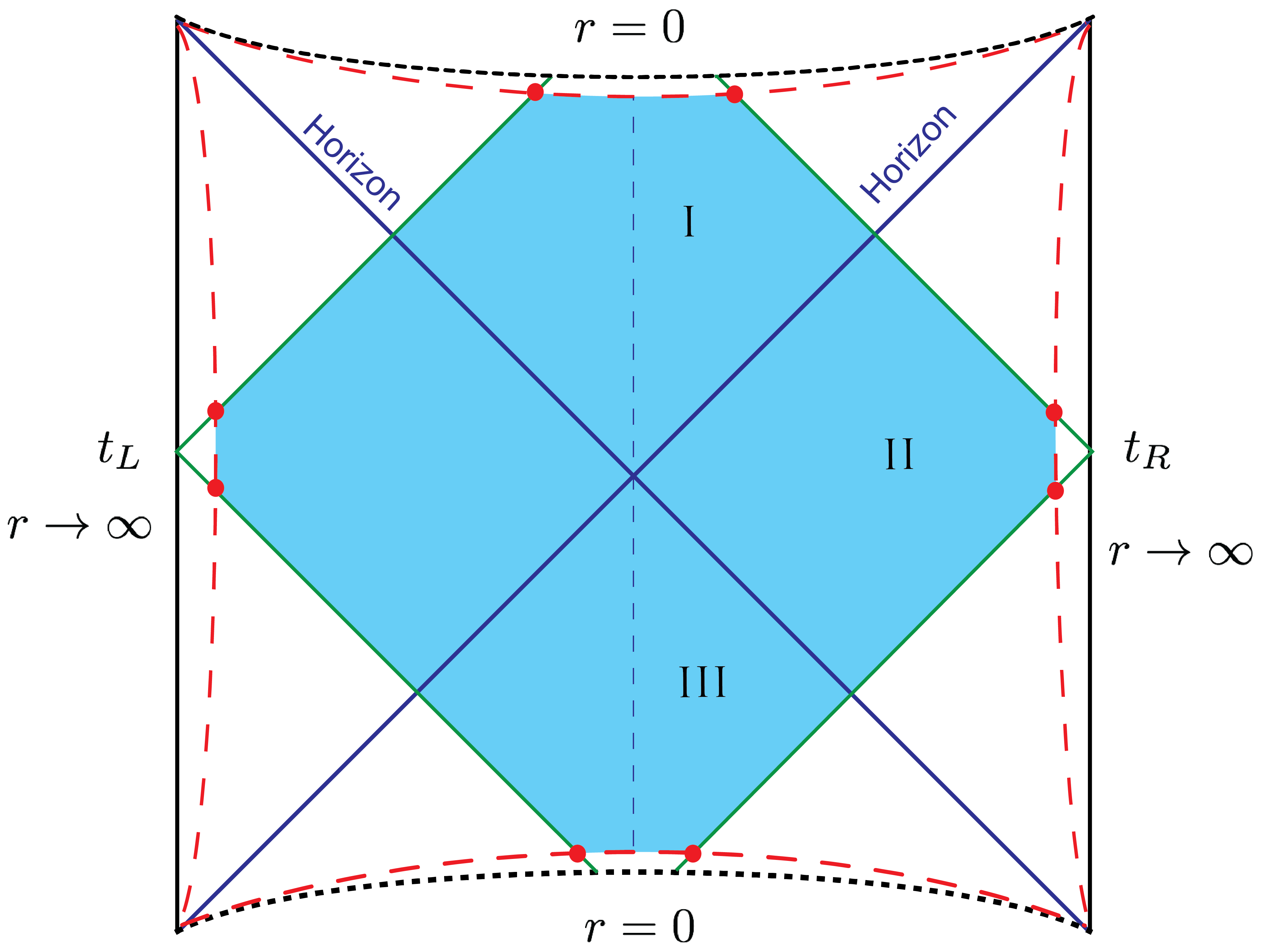}~
\includegraphics[scale=0.057]{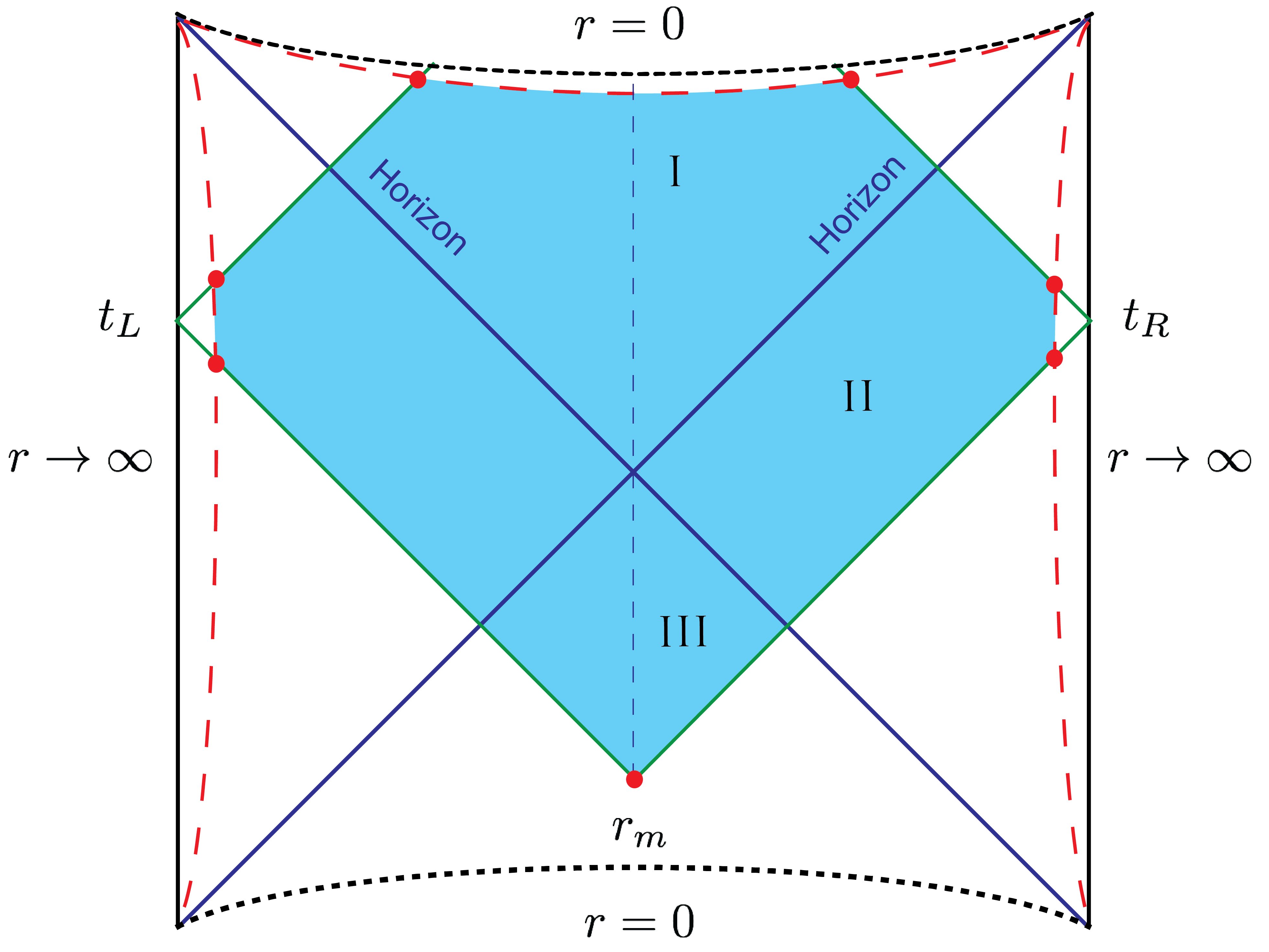}
\caption{Penrose diagram of the WDW patch of an eternal AdS black hole, moving forward in time in a symmetric way ($t_L=t_R$).}
\label{PenroseBHa}
\end{figure}

The causal structure of the black holes described by the metric \eqref{HigherDMetric} is illustrated by the Penrose diagram in figure \ref{PenroseBHa}.\footnote{Small hyperbolic black holes are an exception since their causal structure resembles that of charged black holes. We will comment on this case at the end of appendix \ref{CformCharged}, where we discuss further properties of charged black holes.} We are considering the holographic complexity of the boundary state on the constant time slices, denoted by $t_L$ and $t_R$, on the two asymptotic boundaries. The corresponding WDW patch (also depicted in figure \ref{PenroseBHa}) is then bounded by the light sheets sent from these two asymptotic time slices. We will be interested in the time dependence of the complexity and therefore in the time dependence of the gravitational action evaluated on this patch as the boundary time increases.\footnote{The geometry is symmetric under $t\to-t$ and we only consider the behaviour of the complexity for $t>0$. We briefly comment on the decrease of the complexity found for $t<0$ in section \ref{tary}.}
The result depends only on $t=t_L+t_R$ and not on each of the boundary times separately due to the invariance of the system under boosts in Kruskal coordinates, \ie under shifts $t_L\to t_L+\Delta t$ and $t_R\to t_R-\Delta t$. In terms of the boundary theory, this corresponds to the invariance of the thermofield double state \reef{TFDx} under an evolution with the Hamiltonian $H=H_L-H_R$. In any event, we can therefore deduce the rate of change of the holographic complexity for a general choice of time slices from the result for the symmetric configuration with times $t_L=t_R\equiv t/2$.

For our calculations, there are two different regimes to be considered with respect to the position of the WDW patch. The first, illustrated in the left panel of figure \ref{PenroseBHa}, is when the WDW patch is in contact with the past singularity.  In the second regime, shown in the right panel, the past light sheets from the left and right boundaries intersect before hitting the past singularity.  The critical time $t_c$ separating the two regimes is easily found to be
\begin{equation}\label{eq:tcrit}
t_{c} = 2(r^*_{\infty}-r^*(0))\, ,
\end{equation}
for the symmetric scenario (\ie $t_L=t_R=t/2$).
Generally, we can only find closed form expression for  $t_c$ in specific dimensions. However,
for  planar black holes (\ie $k=0$ in eq.~\reef{BFactor}), the solution can be written in a closed form for any $d$ as:
\begin{equation}\label{tcritplan}
t_{c} = \frac{2\pi }{ d} \frac{L^2}{r_h} \textrm{cot} \bigg(\frac{\pi}{d} \bigg) = \frac{1}{2 T} \, \textrm{cot} \bigg(\frac{\pi}{d} \bigg) \, ,
\end{equation}
where $T= d\, r_h /(4 \pi L^2)$ is the boundary temperature \reef{eq:ST} in this case.

In the following, we evaluate the various contributions to the gravitational action for both the $0<t<t_c$ and $t>t_c$ regimes. We use these results to compute the rate of change of the holographic complexity using eq.~\reef{compAct}. In fact, we will find that the action does not change in the initial time period, $0\le t\le t_c$, while it does change as $t$ changes for $t>t_c$.  The gravitational action can be written as follows \cite{RobLuis}:
\begin{equation}\label{THEEACTION}
\begin{split}
I = & \frac{1}{16 \pi G_N} \int_\mathcal{M} d^{d+1} x \sqrt{-g} \left(\mathcal R + \frac{d(d-1)}{L^2}\right) \\
&\quad+ \frac{1}{8\pi G_N} \int_{\mathcal{B}} d^d x \sqrt{|h|} K + \frac{1}{8\pi G_N} \int_\Sigma d^{d-1}x \sqrt{\sigma} \eta
\\
&\quad -\frac{1}{8\pi G_N} \int_{\mathcal{B}'}
d\lambda\, d^{d-1} \theta \sqrt{\gamma} \kappa
+\frac{1}{8\pi G_N} \int_{\Sigma'} d^{d-1} x \sqrt{\sigma} a\,.
\end{split}
\end{equation}
The first line contains the standard Einstein-Hilbert action including the Ricci scalar $\mathcal{R}$ and the cosmological constant $\Lambda=-d(d-1)/(2L^2)$. The second line begins with the Gibbons-Hawking-York (GHY) surface term \cite{York,GH} for smooth timelike and spacelike segments of the boundary, which is defined in terms of the trace of the extrinsic curvature $K$. In the second contribution there are the Hayward joint terms \cite{Hay1,Hay2}, which appear at the intersection of two such boundary segments and which are defined in terms of the ``boost angle" $\eta$ between the corresponding normal vectors. The last line contains the additional terms which are required when the boundary includes null segments \cite{RobLuis}.\footnote{See also \cite{Parattu:2015gga, Hopfmuller:2016scf,Jubb:2016qzt,Wieland:2017zkf} for other developments on the action with null boundaries and corners.}  First of these is the surface term for the null segments defined in terms of the $\kappa$, which measures the failure of the null generators to be affinely parametrized. Secondly, there are the joint terms at the intersection of these null boundary segments with any other boundary segment, where $a$ relates the normals on the intersecting segments. We follow the conventions of appendix A of \cite{diverg} and the precise definition of all of the boundary terms can be found there.

Let us recall that there are certain ambiguities associated with the null surface and joint contributions \cite{RobLuis}. In the following, we adopt the natural conventions presented in the discussion of \cite{RobLuis}. In particular, we choose the normals to the null boundary segments to be affinely parameterized. This sets $\kappa=0$ and hence we do not have to consider the corresponding boundary terms for the null segments in eq.~\reef{THEEACTION}. We also fix the conventions for $a$ so that the action is additive --- see \cite{RobLuis} or appendix A of \cite{diverg}. Finally, we are left with a freedom to rescale each of the null normals ${\bf k}_i$ by an overall constant. We fix this ambiguity by setting ${\bf k}_i\cdot \hat t = \pm \alpha$ at the asymptotic boundary, where $\hat t=\pa_t$ is the time-like vector describing the time flow in the boundary theory and $\alpha$ is some positive constant. If we assume that $\hat t$ is future directed on all boundaries, then the $+$ and $-$ sign is chosen here for the future and past null boundaries of the WDW patch, respectively.  We will see that different choices of the normalization constant $\alpha$ will modify subleading contributions to $d\CA/dt$ in a late time expansion in the following. More generally,
we also comment on how making different choices to fix these ambiguities might effect our results for the time rate of change of the holographic complexity in appendix \ref{app:EternalAmb}.

Finally, we observe that the action of the WDW patch is divergent because this spacetime region extends all the way to the asymptotic AdS boundary and so
we regulate the calculation of the complexity in the standard way (\eg see \cite{count,sken1,sken2}) by introducing a cutoff surface at $r=r_{\max}$. In general, a potential subtlety is choosing the cutoff surface in a consistent way to allow for the comparison of WDW actions evaluated in different spacetimes. However, as described in \cite{Formation}, one can describe the different geometries in a canonical way using the Fefferman-Graham expansion and then we set the radial cutoff surface at $z=\delta$. As usual, $\delta$ plays the role of a short-distance cutoff in the dual boundary theory. This highlights the fact that the divergence in the action is a UV divergence in the holographic complexity related to establishing correlations in the boundary CFT down to the cutoff scale \cite{Formation,diverg}. Further, we note that in the present calculations, this UV divergence will be independent of time and so it does not influence the time rate of change of the holographic complexity. We will also need to introduce a regulator surface at $r=\epsilon_0$ near the past and future singularities.

\subsubsection{Initial times: $t<t_c$}
For times before $t_c$, the action \reef{THEEACTION} contains three nonvanishing contributions: the bulk contribution; the GHY surface contributions from the regulator surfaces at the past and future singularities, as well as from the UV cutoff surfaces; and the null joint terms where the null boundaries of the WDW patches intersect the regulator surfaces at the past and future singularities, as well as the intersections with the UV cutoff surface. We will evaluate all these contributions in turn and demonstrate that the total action is independent of time in the interval $t_c\ge t\ (\ge -t_c)$.\footnote{See comment about negative times around eq.~\eqref{negatime}.} Due to the symmetry of the configuration that we have chosen, we can evaluate the contributions for the right side of the Penrose diagram (in the left panel of figure \ref{PenroseBHa}) and then simply multiply the result by a factor of two.

\vspace{0.5em}

{\bf Bulk contribution:}
We divide the WDW patch into three regions: I, the region behind the future horizon; II, the region outside both horizons; and III, the regions behind the past horizon --- see figure \ref{PenroseBHa}. The corresponding bulk contributions to the action read:
\begin{equation}
\begin{split}
I_{\bulk}^{\text{I}}=&\, -\frac{d \, \Omega_{k,d-1}}{8\pi G_N L^2} \int_{\epsilon_0}^{r_h} r^{d-1} \left(\frac{t}{2}+r^*_{\infty}-r^*(r)\right) d r
\\
I_{\bulk}^{\text{II}}= &\, -\frac{d \, \Omega_{k,d-1}}{4\pi G_N L^2} \int_{r_h}^{r_\text{max}} r^{d-1}\, \Big( r^*_{\infty}-r^*(r) \Big)\, d r
\\
I_{\bulk}^{\text{III}}= &\, -\frac{d \, \Omega_{k,d-1}}{8\pi G_N L^2} \int_{\epsilon_0}^{r_h} r^{d-1} \left(-\frac{t}{2}+r^*_{\infty}-r^*(r)\right) d r
\end{split}
\end{equation}
where $r_{\text{max}}$ is a UV cutoff. Summing these three contributions, we are left with:
\begin{equation}\label{BULKtSMLRtc}
I_{\bulk}^{0}= -\frac{d \, \Omega_{k,d-1}}{2\pi G_N L^2} \int_{\epsilon_0}^{r_\text{max}} r^{d-1} (r^*_{\infty}-r^*(r)) d r\, ,
\end{equation}
where an extra factor of two was included to account for the two sides of the Penrose diagram in figure \ref{PenroseBHa}. We see that the time dependences in $I_{\bulk}^{\text{I}}$ and $I_{\bulk}^{\text{III}}$ precisely cancel and hence the total bulk contribution is time independent.

\vspace{0.5em}

{\bf GHY surface contributions:}
There are three different GHY surface contributions to be considered: those coming from the regulator surfaces at the future and past singularities, and the surface contribution at the UV cutoff surface.\footnote{Recall that we chose the null normals to be affinely parametrized and hence the null surface contributions vanish, \ie $\kappa=0$.} We use the following (outward-directed unit) normal vectors to evaluate the corresponding extrinsic curvatures
\begin{eqnarray}
r=r_{\max} :&& \qquad {\bf s}=s_\mu dx^\mu = \frac{dr}{\sqrt{f(r_{\max})}}\\
r=\epsilon_0: &&\qquad {\bf t}=t_\mu dx^\mu = - \frac{dr}{\sqrt{-f(\epsilon_0)}}
\nonumber
\end{eqnarray}
where the second normal applies for both regulator surfaces next to the past and future singularities. For a constant $r$ surface in the metric \reef{HigherDMetric}, the trace of the extrinsic curvature is given by
\begin{equation}
K=\frac{n_r}{2} \left(\del_r f(r) + \frac{2(d-1)}{r} f(r)\right)\,,
\end{equation}
and as a result, we obtain
\begin{equation}
\begin{split}
&I_{\text{surf}}^{\text{future}}=-\frac{	r^{d-1} \Omega_{k,d-1}}{8\pi G_N} \left(\partial_r f(r)+\frac{2(d-1)}{r}f(r)\right)\left(\frac{t}{2}+r^*_{\infty}-r^*(r)\right)\biggr{|}_{r=\epsilon_0},
\\
&I_{\text{surf}}^{\text{past}}=-\frac{	r^{d-1} \Omega_{k,d-1}}{8\pi G_N} \left(\partial_r f(r)+\frac{2(d-1)}{r}f(r)\right)\left(-\frac{t}{2}+r^*_{\infty}-r^*(r)\right)\biggr{|}_{r=\epsilon_0},
\\
&I_{\text{surf}}^{\text{cutoff}}=\frac{	r^{d-1} \Omega_{k,d-1}}{8\pi G_N} \left(\partial_r f(r)+\frac{2(d-1)}{r}f(r)\right)
\left(r^*_{\infty}-r^*(r)\right)
\biggr{|}_{r=r_{\max}}.
\end{split}
\label{harvey}
\end{equation}
We see that the surface contribution $I_{\text{surf}}^{\text{cutoff}}$ at the UV cutoff surface is independent of time. Further we note that this contribution is identical in the regime $t>t_c$. Therefore, the UV surface terms do not contribute to the time dependence of holographic complexity and we will ignore them both here and in the next section. For $t<t_c$, we see that the time dependence of the GHY surface contributions from the past and future singularities precisely cancels leaving:
\begin{equation}\label{SURFtSMLRtc}
I_{\text{surf,~sing}}^{0}=-\frac{	r^{d-1} \Omega_{k,d-1}}{4\pi G_N} \left(\partial_r f(r)+\frac{2(d-1)}{r}f(r)\right)\left(r^*_{\infty}-r^*(r)\right)\biggr{|}_{r=\epsilon_0}.
\end{equation}
We note again that this contribution is independent of time for all $t<t_c$.

\vspace{0.5em}

{\bf Null joint contributions:}
There are a number of null joint contributions to be considered.  In particular, we have the joint contributions at the intersections of the null boundaries of the WDW patch with the regulator surfaces at the past and future singularities and those at their intersections  with the UV cutoff surface. These contributions were carefully evaluated in \cite{Formation} --- see eqs.~(2.34) and (2.35) of the reference --- and they are not modified in the present case. However, two key observations are that the null joint contributions at the singularities vanish, while those at the UV cutoff surface have no time dependence. Hence neither of these terms contribute to the time rate of change of holographic complexity.

\vspace{0.5em}

{\bf Total Action:} Hence as our calculations above demonstrate, the total gravitational action of the WDW patch is independent of time for the initial time period $t<t_c$. If we denote its value by $I_0$,\footnote{Note that $I_0=I_\mt{WDW}(t_L=t_R=0)$ and so this result is identical to the action evaluated in \cite{Formation}. In particular,  the complexity of formation of the thermofield double state in the boundary is given by $I_0$ minus  twice the corresponding action of the WDW patch in vacuum AdS.} then in this early time interval, we have
\beq
0\le t\le t_c\,:\qquad \frac{d\mathcal{C}_A}{dt} = \frac1\pi\,\frac{dI_0}{dt}=0\,.
\label{nochange}
\eeq

\subsubsection{Later times: $t>t_c$}

For times $t>t_c$, the same three sets of terms make nonvanishing contributions to the action of the WDW patch, \ie the bulk term, the GHY surface terms and the null joint terms, and so we again evaluate each of these contributions in turn. We again use the symmetry of the configuration to only explicitly evaluate the contributions for the right side of the Penrose diagram (in the right panel of figure \ref{PenroseBHa}) and then simply multiply the result by a factor of two.

\vspace{0.5em}

{\bf Bulk contribution:}
As before, we split the WDW patch into three regions which we denote as I, II and III --- see figure \ref{PenroseBHa}. The corresponding bulk contributions to the gravitational action become:
\begin{equation}
\begin{split}
I_{\bulk}^{\text{I}}=&\, -\frac{d \, \Omega_{k,d-1}}{8\pi G_N L^2} \int_{0}^{r_h} r^{d-1} \left(\frac{t}{2}+r^*_{\infty}-r^*(r)\right) d r
\\
I_{\bulk}^{\text{II}}= &\, -\frac{d \, \Omega_{k,d-1}}{8\pi G_N L^2} \int_{r_h}^{r_\text{max}} r^{d-1}\, 2\, \Big( r^*_{\infty}-r^*(r) \Big)\, d r
\\
I_{\bulk}^{\text{III}}= &\, -\frac{d \, \Omega_{k,d-1}}{8\pi G_N L^2} \int_{r_m}^{r_h} r^{d-1} \left(-\frac{t}{2}+r^*_{\infty}-r^*(r)\right) d r
\end{split}
\end{equation}
where $r_m$ is the radius behind the past horizon where the null boundary sheets from the left and right boundaries intersect.  This position is determined by the following equation:
\begin{equation}\label{meeting134}
\frac{t}{2} - r^*_{\infty} + r^*(r_m)=0\, .
\end{equation}
Generally, this is a transcendental equation and we can only determine $r_m$ numerically. Combining the above results,
we obtain the total bulk contribution
\begin{equation}\label{BULKtBGRtc}
I_{\bulk} = I_\bulk^0 - \frac{d \, \Omega_{k,d-1}}{4\pi G_N L^2} \int_{0}^{r_m} r^{d-1} \left( \frac{t}{2}-r^*_{\infty}+r^*(r)\right) d r \, ,
\end{equation}
where we have again included a factor of two to account for the equal contributions coming from the two sides of the WDW patch shown in figure \ref{PenroseBHa}. We have also introduced $I_\bulk^0$, which was defined in eq.~\eqref{BULKtSMLRtc} and which is time independent.

\vspace{0.5em}

{\bf GHY surface contributions:} For $t>t_c$, the WDW patch does not reach the past singularity and so only the regulator surface at the future singularity contributes here. The expression takes the same form as in eq.~\reef{harvey} and as a result we obtain
\begin{equation}
\begin{split}
&I_{\text{surf}}^{\text{future}}=-\frac{f(r)	r^{d-1} \Omega_{k,d-1}}{8\pi G_N} \left(\frac{\partial_r f(r)}{f(r)}+\frac{2(d-1)}{r}\right)\left(\frac{t}{2}+r^*_{\infty}-r^*(r)\right)\biggr{|}_{r=\epsilon_0}\,.
\end{split}
\end{equation}
We also have the GHY contribution from the UV cutoff surface as in eq.~\reef{harvey}. However, this contribution is time independent and so we ignore it here.

Using eq.~\eqref{SURFtSMLRtc}, the above expression can be rewritten as follows
\begin{equation}\label{SURFtBGRtc}
I_{\surf} = I_{\surf,\, \text{sing}}^0-\frac{r^{d-1} \Omega_{k,d-1}}{8\pi G_N} \left(\partial_r f(r)+\frac{2(d-1)}{r} f(r)\right)\left(\frac{t}{2}-r^*_{\infty}+r^*(r)\right)\biggr{|}_{r=\epsilon_0}\,.
\end{equation}
The difference $I_{\surf}-I_{\surf,\, \text{sing}}^0$ encodes the change in the GHY contribution to the holographic complexity after $t=t_c$.

\vspace{0.5em}

{\bf Null Joint Contribution:} There are null joint contributions from the intersection of the null boundaries with the regulator surface at the future singularity and with the UV cutoff surface. However, as in the previous section, the former vanish while the latter are independent of time. Therefore neither of these contribute to $d{\mathcal C}_A/dt$. The last joint contribution to consider when $t>t_c$ is that from the intersection of the two past null boundaries  at $r=r_m$.
To evaluate this term, we use the following outward-directed null normal vectors:
\beq
{\rm Right}\,:\quad{\bf k}_\mt{R} = -\alpha dt +\alpha \frac{dr}{f(r)}\,;\qquad
{\rm Left}\,:\quad{\bf k}_\mt{L} = \alpha dt + \alpha\frac{dr}{f(r)}\,.
\label{rapido}
\eeq
Here we have assumed that the Killing vector $\pa_t$ describes a flow from right to left  for the region behind the past horizon in figure \ref{PenroseBHa}. The joint term can then be evaluated as
\begin{equation}\label{CORNERtBGRtc}
I_{\jnt} = -\frac{\Omega_{d-1} r_m^{d-1}}{8 \pi G_N} \log \frac{|f(r_m)|}{\alpha^2}\, ,
\end{equation}
This term depends on $t$ through the implicit time dependence of $r_m$, as determined by eq.~\reef{meeting134}.
We would like to stress that this contribution is sensitive to the ambiguities discussed in \cite{RobLuis}, \ie through its dependence on the normalization constant $\alpha$. We discuss this issue further in appendix \ref{app:EternalAmb}.

\vspace{0.5em}

{\bf Total Action:}
The total action for $t>t_c$ is given by the sum of eqs.~\eqref{BULKtBGRtc}, \eqref{SURFtBGRtc} and \eqref{CORNERtBGRtc} plus some time independent contributions from the UV cutoff surfaces and the null junctions.
It is sometimes convenient to express our various contributions in terms of $\delta t = t-t_c$.
As a consequence, the equation for the position $r_m$ of the past null junction becomes
\begin{equation}\label{meeting1}
\frac{\delta t}{2}+ r^*(r_m) - r^*(0) = 0\, .
\end{equation}
The total gravitational action can then be expressed as
\begin{equation}
I = I_0 + \delta I \qquad{\rm with}\quad
\delta I = \delta I_{\text{bulk}} + \delta I_{\text{surf}}
+ I_{\jnt}.
\end{equation}
where
\begin{align}\label{Bulk1}
\delta I_{\text{bulk}}
\equiv &
 I_{\text{bulk}}-I^0_{\text{bulk}}
=  - \frac{d \, \Omega_{k,d-1}}{4\pi G_N L^2} \int_{0}^{r_m} dr r^{d-1} \left(\frac{\delta t}{2}+r^*(r)-r^*(0)\right)\,,
\\
\label{surf1}
\delta I_{\text{surf}} \equiv &
 I_{\text{surf}}-I^0_{\text{surf}}  = -\frac{	r^{d-1} \Omega_{k,d-1}}{8\pi G_N} \left(\partial_r f(r)+\frac{2(d-1)}{r}f(r)\right)\frac{\delta t}{2}\biggr|_{r=\epsilon_0}\,,
\\
\label{corners1}
I_{\jnt} = &-\frac{\Omega_{d-1} r_{m}^{d-1}}{8 \pi G_N} \log \frac{|f(r_{m})|}{\alpha^2}\,.
\end{align}
We note that $\delta I$ is finite, \ie independent of the UV cutoff $\delta$. Further it vanishes in the limit $\delta t \rightarrow 0$, which can be seen by explicitly substituting the blackening factor \reef{BFactor} into eqs.~\eqref{Bulk1}-\eqref{corners1}. However, we will show below that the rate of change of the holographic complexity is discontinuous at $t=t_c$.

\subsection{Time Dependence of Complexity}\label{timeder}

Here we examine the time dependence of the holographic complexity. As we already noted above in eq.~\reef{nochange}, initially, we have
\beq
0\le t\le t_c\,:\qquad \frac{d\mathcal{C}_A}{dt} = \frac1\pi\,\frac{dI_0}{dt}=0\,,
\label{nochange2}
\eeq
where $t_c$ was defined in eq.  \eqref{eq:tcrit}.

For later times $t>t_c$, we obtain the time derivative of complexity by differentiating eqs.~\eqref{meeting1}-\eqref{corners1} with respect to time. From eq.~\eqref{meeting1}, we  find the time dependence of the meeting point $r_m$ to be
\begin{equation}\label{forchain1}
\frac{d r_m}{d t} = -\frac{f(r_m)}{2}\,.
\end{equation}
Differentiating eq.~\eqref{Bulk1} yields
\begin{equation}
\frac{d I_{\bulk}}{d t} = \frac{d \,\delta I_{\bulk}}{d t} =-\frac{\Omega_{k,d-1}}{8 \pi G_N L^2} r_m^d\,,
\end{equation}
where in obtaining this result, we  used eq.~\eqref{meeting1} to demonstrate that  the contribution coming from differentiating the upper limit of integration vanishes.
Evaluating the GHY surface term \reef{surf1} at $r=\epsilon_0$ and then taking the $\epsilon_0\rightarrow 0$ limit yields
\begin{equation}
\frac{d I_{\surf}}{d t} = \frac{d\, \delta I_{\surf}}{d t} =\frac{\omega^{d-2} d \, \Omega_{k,d-1}}{16\pi G_N} \, .
\end{equation}
Finally, differentiating the null joint term \reef{corners1} gives
\begin{equation}
\frac{d I_{\jnt}}{d t} = \frac{\Omega_{k,d-1} r_{m}^{d-2}}{16 \pi G_N} \left[(d-1) f(r_m)  \log \frac{|f(r_{m})|}{\alpha^2}+r_{m} \del_r f(r_m)\right]\,.
\end{equation}
where we have used eq.~\eqref{forchain1}. Using the explicit form of the blackening factor \reef{BFactor} and summing the three terms above,  eq.~\eqref{compAct} yields the rate of growth of holographic complexity as
\begin{equation}\label{tder1}
t>t_c\,:\qquad
\frac{d \mathcal{C}_A}{d t}=\frac{1}{\pi} \left( 2M +
\frac{\Omega_{k,d-1} (d-1) r_{m}^{d-2}}{16 \pi G_N}  f(r_m)  \log \frac{|f(r_{m})|}{\alpha^2}\right)\,.
\end{equation}
Of course, this result reproduces the expected rate of growth at late times \cite{Brown1,Brown2}, \ie $d\mathcal{C}_A/dt=2M/\pi$, since in this limit $r_m$  approaches $r_h$ and so the second term on the right vanishes with $f(r_m\to r_h) \to 0^-$.   We provide further comments on the properties of our result \reef{tder1} below.

\subsubsection{Comments} \label{tary}

As already noted above, this result \reef{tder1} reproduces the expected rate of growth at late times since in this limit $r_m$  approaches $r_h$ and so $f(r_m\to r_h) \to 0^-$.  We also note that at late times with $r_m$ approaching $r_h$ from below, $f(r_m)$ is small and negative and therefore the correction to $d\mathcal{C}_A/dt=2M/\pi$ in eq.~\eqref{tder1} is positive! That is, $d\mathcal{C}_A/dt$ approaches the late time limit from above. Recall that \cite{Brown1,Brown2} suggested that the late time limit of $d\mathcal{C}_A/dt$ may be related to Lloyd's bound $2M/\pi$  for the rate of computation for a system of energy $M$ \cite{Lloyd}. Therefore we see here a (small) violation of Lloyd's bound in the eternal black hole.

\vspace{0.5em}

\noindent{\bf Late time expansion:}
To get a better understanding of the late time behaviour, it is possible to solve the equation for $r_m$ in a late time expansion. We do this by defining the regular part of the blackening factor $F(r)$:
\begin{equation}
f(r) \equiv F(r) (r-r_h)
\end{equation}
and decomposing  the inverse blackening factor as
\beq
\frac{1}{f(r)} = \frac{1}{F(r_h) \, (r-r_h)} + \frac{F(r_h) -F(r) }{F(r_h)F(r) \, (r-r_h)}\, .
\eeq
This leads to the following form of the tortoise coordinate
\begin{equation}
r^*(r) = \frac{1}{F(r_h)} \log \frac{|r-r_h|}{\tilde \ell} + \int^r \frac{F(r_h) -F(r) }{F(r_h)F(r) (r-r_h)}\, d r
\end{equation}
where $\tilde \ell$ is an unspecified integration constant. Using eqs.~\reef{eq:ST} and \eqref{meeting1}, we can solve for $r_m$ at late times as
\begin{equation}
r_m = r_h \,\left(1- c_1 e^{-2\pi T(t-t_c)}\right) + \cdots
\label{george}
\end{equation}
with
\begin{equation}
c_1 =  \exp\left[- \int^{r_h}_0\!\!dr\, \frac{F(r_h) -F(r) }{F(r)  (r-r_h)}\right] >0,
\end{equation}
and where  the ellipsis stands for corrections which are higher order in $(r_h-r_m)$, \ie which would decay at least as fast as $e^{-4\pi T t}$).
Substituting this expression \reef{george} into eq.~\eqref{tder1}, we obtain at the first corrections to the rate of change in complexity in the $t\rightarrow\infty$ limit
\begin{equation}
\frac{d \mathcal{C}_A}{d t}=\frac{2M}{\pi}  +
2 (d-1)\,   c_1 \, S \, T^2\, e^{-2\pi T (t-t_c)}\left(t-t_c-\frac{1}{2 \pi T}\log\left[\frac{4\pi c_1 Tr_h}{\alpha^2} \right]\right) +\cdots\,.
\label{latetimeSub}
\end{equation}
We see that the final factor will always become positive for sufficiently late times and hence the bound conjectured by \cite{Brown1,Brown2} will be violated.

\vspace{0.5em}

\noindent{\bf Early times:} It is also interesting to look at an early  time expansion of the expression \eqref{tder1}. At very early times after $t_c$, $r_m$ is very close to the past singularity, \ie as $\delta t=t-t_c\to0$, $r_m\to 0$. As a consequence, $f(r_m)\sim-\omega^{d-2}/r_m^{d-2}$ and the second term in eq. \eqref{tder1}  diverges to minus infinity (as long as $d\ge3$). More explicitly, one can show that this leading divergence as $\delta t \rightarrow 0$ is logarithmic with
\beq
\frac{d \CA}{d t}\bigg|_{\dt\to0} \longrightarrow\  - \frac{(d-2) M}{(d-1) \pi} \log \left( \frac{2 \omega}{\alpha^{2(d-1)/(d-2)} (d-1) \delta t} \right)\qquad
{\rm for}\ \ d\ge3\,.
\label{pop}
\eeq
 Despite this divergence, we note again that the complexity itself remains finite as $\delta t\to 0$ and it is only its derivative which is divergent.
We would also like to stress again, that these results are influenced by the ambiguities in the corner term mentioned in \cite{RobLuis}. We explore this issue further in appendix \ref{app:EternalAmb}. We also examine the case $d=2$, \ie BTZ black holes, in detail in the following section.

\vspace{0.5em}

\noindent{\bf Averaging:} The discussion above indicates that the action changes very rapidly in the vicinity of $\dt=0$ --- see also the examples in section \ref{samples}.  However, one might argue that the holographic complexity does not have a good definition on time scales smaller than $\beta=1/T$ in the context of the eternal black hole.\footnote{We thank Lenny Susskind, Dan Roberts and Brian Swingle for correspondence on this point.} Hence we might average the rate of change in complexity over time scales which are longer than the thermal time scale. We can define a simple averaged rate of change in complexity as follows:
\begin{equation}\label{averageC}
\left[\frac{d\mathcal{C}_A}{dt} \right]_{\gamma;\mt{avg}} = \frac1{\gamma\,\beta} \int_{t-\gamma\,\beta/2}^{t+\gamma\,\beta/2} \frac{d\mathcal{C}_A}{dt'} \, dt' = \frac{\CA(t+\gamma\,\beta/2)-\CA(t-\gamma\,\beta/2)}{\gamma\,\beta}\,,
\end{equation}
where $\gamma$ is some numerical factor of order one. In the second expression,
we see that we have essentially constructed a discrete time derivative on a time step $\Delta t=\gamma/T$.

Let us comment on the properties of this averaged rate: First, we note that $\left[\frac{d\mathcal{C}_A}{dt} \right]_{\gamma;\mt{avg}}$ remains continuous at all times. However, its time derivative will be discontinuous at $|t\pm \frac{\gamma \beta}{2}|=t_c$ because of the discontinuity in $d\CA/dt$ noted above.
When $\gamma \beta/2 < t_c$ there will generically be a short period of time right after $t=t_c-\gamma\beta/2$ for which this averaged rate will be negative. After this period, the rate will rise quickly to positive values. Note that this averaging does not remove the (small) violation of Lloyd's bound, discussed above. We will return to discuss this time averaging in more detail in section \ref{sec:Discussion}.

\vspace{0.5em}

\noindent{\bf Negative Times:}
In our setup, the complexity is a symmetric function of time $\mathcal{C}_A(t) = \mathcal{C}_A(-t)$. Of course, this implies that the time derivative is anti-symmetric
\begin{equation}\label{negatime}
\frac{d\mathcal{C}_A}{dt}(t) =- \frac{d\mathcal{C}_A}{dt}(-t).
\end{equation}
Our system therefore admits a regime of decreasing complexity, at least for large negative times. This situation is unstable --- an arbitrary small perturbation would cause the complexity to start increasing again. A discussion of this issue can be found in subsection [2.1] of \cite{Brown:2016wib}.

\vspace{0.5em}

\noindent{\bf Dependence on the boundary curvature:} Given the black hole metric in eqs.~\reef{HigherDMetric} and \reef{BFactor}, it is clear that $L$ is the AdS curvature scale. However, implicitly, $L$ also plays the role of the curvature of the boundary metric in the cases $k=\pm1$. Hence when we express our results in terms of quantities of the boundary theory, it is perfectly consistent for the final answer to depend on $L$. However, if we introduce a separate curvature scale $R$ for the boundary metric, it becomes a consistency test to demonstrate that we can eliminate the AdS scale from our expressions.

Hence let us consider the AdS black hole metric
\beq
d s^2 = -f(r)\, \frac{L^2}{R^2 }\,d \tau^2 + \frac{dr^2}{f(r)}+ r^2 d \Sigma_{k,d-1}^2\,,
\label{metric2}
\eeq
where $f(r)$ is still given by eq.~\reef{BFactor}. Now scaling the metric in the asymptotic region $r\to\infty$ by $R^2/r^2$ yields the boundary metric
\beq
ds_{bdy}^2 = -d \tau^2 + R^2\, d \Sigma_{k,d-1}^2\,,
\label{body2}
\eeq
where the curvature of the spatial geometry is now set by
$R$.\footnote{Notice that for the planar geometry, \ie $k=0$, there is no curvature scale and hence $R$ becomes some arbitrary length scale in the boundary theory. Further, for $k=0$ in eq.~\eqref{HigherDMetric}, we implicitly had chosen the boundary metric $d \Sigma_{0, d-1}^2=\sum^{d-1}_{i=1}dx_i^2/L^2$, following \cite{Formation}. Normalizing with the AdS curvature scale $L$ was required to ensure that the line element was dimensionless. Here, it is more natural to set $d \Sigma_{0, d-1}^2=\sum^{d-1}_{i=1}d\tilde x_i^2/R^2$, so that the boundary metric \eqref{body2} is independent of $R$ (and $L$). Of course, this is equivalent to rescaling the (spatial) boundary coordinates as $\tilde x_i= (R/L)\, x_i$. \label{foot88}}
Of course, the only real change between eqs.~\reef{HigherDMetric} and \reef{metric2} is that we have rescaled the time variable, \ie  $\tau= (R/L)\, t$. So essentially all of our computations follow identically for the `new' geometry to those that were performed above. However, the scaling of the time coordinate appears in various places, such as the definition of the null coordinates in eq.~\reef{EFcoord} or of the null normals in eq.~\reef{rapido}. Another important difference is in the definition of various quantities which characterize the boundary state in terms of the geometric parameters appearing in the bulk. In particular, eqs.~\reef{Mass} and \reef{eq:ST} are replaced with the following
\beqa
&&M = \frac{(d-1) \, \Omega_{k,d-1}}{16 \pi \, G_N}\,\frac{L}R \,\omega^{d-2} \,,\qquad\qquad\qquad
S = \frac{\Omega_{k,d-1}}{4G_N}\,r_h^{d-1}\,,  \labell{there}\\
&& \qquad T
=\frac{L}{4\pi R}\left.\frac{\partial f}{\partial r}\right|_{r=r_h}=\frac{L}{4\pi R\,r_h}\left(d\,\frac{r_h^2}{L^2} + (d-2)\,k \right)\,,
\nonumber
\eeqa
and the spatial volume of boundary becomes $V=\Omega_{k,d-1} R^{d-1}$.
Given these changes, the critical time is given by
\begin{equation}\label{eq:criticalTaume}
\tau_c = \frac{2 R}{L} \left( r^{*}_\infty - r^{*}(0) \right)
\end{equation}
and our result \reef{tder1} for the rate of change of the complexity becomes
\begin{equation}\label{tder44}
\tau>\tau_c\,:\qquad
\frac{d \mathcal{C}_A}{d \tau}=\frac{1}{\pi} \left( 2M +
\frac{\Omega_{k,d-1} (d-1) r_{m}^{d-2}}{16 \pi G_N}\,\frac{L}R\,  f(r_m)  \log \frac{L^2\,|f(r_{m})|}{R^2\,\alpha^2}\right)\,,
\end{equation}
where the equation for the meeting point can be written as
\begin{equation}
\delta \tau= -\frac{2R}L\left(r^*(r_m) - r^*(0)\right)\, .
\label{rhombus}
\end{equation}

Now we would like to recast this result \reef{tder44} in terms of boundary quantities. We do so by first defining a dimensionless radial coordinate $x=r/r_h$. Next we note that from eq.~\reef{there}, we see that the dimensionless ratio
of geometric scales $\rh/L$ in the bulk is determined by the dimensionless product of boundary quantities $RT$. In particular, we find
\beq
\frac{r_h}L=\frac{2\pi\,RT}d\left(1+ \sqrt{1- \frac{d(d-2)\,k}{(2\pi\,RT)^2}}\right) \equiv{2\pi RT}\ \tilde g(RT)\,.
\label{there3}
\eeq
Now examining the blackening factor, we can write:
\beqa
f(r) &=& \frac{r^2}{L^2} +k+
\frac{r_h^{d-2}}{r^{d-2}}\left(\frac{r_h^2}{L^2}+k\right)
\labell{there2}
\\
&=&\frac{r_h^2}{L^2}\left(x^2+\frac{k\, L^2}{r_h^2}-\frac{1}{x^{d-2}}\left(1+\frac{k\, L^2}{r_h^2}\right)\right)  \equiv \frac{r_h^2}{L^2} \,\tilde f(x,RT) \, .
\nonumber
\eeqa
Further, combining the above expressions in eq.~\reef{rhombus},  we have
\beq\label{xmDimless}
{\pi}\,\tilde{g}(RT)\, T\delta\tau=- \int_0^{x_m}\frac{dx}{\tilde f(x,RT)}\,,
\eeq
which demonstrates that $x_m$ is implicitly a function of the (dimensionless) boundary quantities, $T\delta \tau$ and $RT$.
Further, these results allow us to translate the rate of change in complexity \reef{tder44} for $\tau>\tau_c$ to the form\footnote{Let us note that for planar horizons, \ie for $k=0$, eq.~\reef{there3} yields $\tilde g = 2/d$ while eq.~\reef{there2} simply gives $\tilde f(x_{m}, RT)=(x^d-1)/x^{d-2}$. Hence $d \mathcal{C}_A/d \tau$ does not actually depend on $R T$ for $k=0$.}
\begin{equation}\label{tder45}
\frac{d \mathcal{C}_A}{d \tau}=\frac{1}{\pi} \left( 2M +
ST (d-1)\,\tilde g(RT)\, x_m^{d-2} \tilde f(x_m,RT) \, \log\!\left[ \frac{2\pi LT}{\alpha}\,\tilde{g}(RT)\,|\tilde f(x_{m},RT)|^{1/2}\right]\right)\,.
\end{equation}
Here we see that the right-hand side is expressed in terms of boundary quantities, except for a single factor of $L$ appearing in the argument of the logarithm. Of course, this argument also contains a factor of the (dimensionless) normalization constant $\alpha$, which is arbitrary.  Precisely, the same situation arose in \cite{diverg} in investigating the structure of the UV divergences in holographic complexity. Following \cite{diverg}, it is natural to choose $\alpha = L/\ell$ which eliminates the errant factor of $L$ but introduces some new scale $\ell$ in the boundary theory. Hence this choice raises the question of what the most appropriate choice for $\ell$ would be. For simplicity in the following, we will set $\ell=R$, the curvature scale in the $k = \pm 1$ boundary geometries \reef{body2}. As noted in  the planar case (see footnote \ref{foot88}), $R$ remains an arbitrary length scale in the boundary theory. We return to discuss this point in section \ref{sec:Discussion}.

\subsection{Examples} \label{samples}

In this subsection, we present two specific examples in which  we solve explicitly for the meeting point and evaluate the rate of change in complexity for all times $t>t_c$. First, we will consider BTZ black holes ($d=2$) for which analytic results can be obtained. Further details of the results for this special case are given in appendix \ref{app:BTZnonsymmetric}. Next, we consider numerical solutions for $d=4$ with various horizon geometries. As a further example we consider the case $d=3$ in appendix \ref{app:MoreAction}.

\subsubsection{BTZ Black Holes} \label{btzztb}
For BTZ black holes, most of the expressions can be evaluated analytically. The evaluation of the action given in section \ref{eval2} strictly applies only to $d>2$ and so we must derive the results separately here for the BTZ case. While we review the salient calculations below, further details are also given for this special case in appendix \ref{app:BTZnonsymmetric}. Following eq.~\reef{metric2}, we write the BTZ metric as
\begin{align}
d s^2 = -f(r)\, \frac{L^2}{R^2 }\,d \tau^2  + \frac{dr^2}{f(r)}+ r^2 d \phi^2 \,,
\label{btz2}
\end{align}
where the blackening factor, mass, temperature and entropy are then given by
\begin{equation}\label{BTZblackfactor}
f(r) = \frac{r^2-r_h^2}{L^2}\,,
\qquad
M=\frac{r_h^2}{8G_N L R}\,,
\qquad
T=\frac{r_h}{2\pi L R}\,,
\qquad
S=\frac{\pi r_h}{2 G_N}\,.
\end{equation}
As described in section \ref{tary}, with the coordinates in eq.~\reef{btz2}, the boundary geometry is  fixed  by a new independent scale $R$. In particular,  the boundary metric is given by
\begin{align}
ds^2 = -d \tau^2 + R^2\, d \phi^2\,,
\label{bdyBTZ}
\end{align}
and hence a constant $\tau$ slice is a circle with the circumference $2\pi R$.\footnote{Note that $\beta=1/T$ should satisfy $\beta<2\pi R$ so that the BTZ black hole solution is the dominant saddle point in the gravitational path integral. Further note that, $R$ is associated with the spatial size of the boundary here, rather than a curvature scale as in eq.~\reef{body2}. \label{footy23}}

We can evaluate the tortoise coordinate \reef{tort2} analytically as
\begin{equation}
r^*(r) = \frac{L^2}{2r_h} \log \frac{|r-r_h|}{r+r_h}\,, \quad\implies\ \
r^*_{\infty}=r^*_{0}=\tau_{c}=0\,.
\label{tort3}
\end{equation}
The latter, \ie $\tau_{c}=0$, means that the action of the BTZ black hole starts changing right away for $\tau>0$. This is due to the fact that for the boundary time slice at $\tau=0$, \ie $\tau_R=\tau_L=0$, the null rays coming from the left and right boundaries to define the past and future boundaries of the WDW patch meet at the singularity at $r=0$. Given eq.~\reef{tort3}, the meeting point relation in eq.~\eqref{meeting1} can be solved analytically for general times,
\begin{equation}
r_m = r_h\, \tanh \left(\frac{r_h \tau}{2 LR}\right)\,.
\label{meat2}
\end{equation}

Now in evaluating the action, eqs.~\eqref{Bulk1}--\eqref{corners1} are not modified up to some factors of $L/R$ coming from rescaling the time coordinate --- see the details in appendix \ref{app:BTZnonsymmetric} --- and their sum still reflects the change in complexity from what it was at $\tau=0$. The growth rate \reef{tder44} is then not modified for $d=2$ and substituting in the BTZ blackening factor \eqref{BTZblackfactor} and the meeting point \reef{meat2} then yields
\begin{align}\label{eq:BTZActionEternalRate}
\frac{d\mathcal{C}_A}{d\tau} =\frac{r_h^2}{4 \pi G_N LR} \left(1+\text{sech}^2\left(\frac{r_h \tau}{2 LR}\right)\, \log\! \left[\frac{R \alpha}{r_h}\cosh \left(\frac{r_h \tau}{2 LR}\right)\right]\right)\,,
\end{align}
where we have also used $\Omega_{+1,1}=2\pi$ above. Further using the expressions for the mass and temperature in eq.~\eqref{BTZblackfactor},
this result can be expressed in terms of boundary quantities as
\beq
\frac{d\mathcal{C}_A}{d\tau} =\frac{2 M}{\pi} \left(1+\text{sech}^2\left(\pi T \tau\right)\, \log\! \left[\frac{\alpha}{2\pi L T}\,\cosh \left(\pi T \tau\right)\right]\right) \, .
\label{btznew2}
\eeq
Of course, the above expression is evaluated for $\tau>0$. One simple consistency check on our result is that in the limit $\tau \rightarrow \infty$, we recover the expected late time result of \cite{Brown1,Brown2}, \ie $d\mathcal{C}_A /dt=2M/\pi$.
As in eq.~\reef{tder45}, we see the appearance of both $L$ and $\alpha$ in the argument of the logarithm. Hence there is some ambiguity about the interpretation of this result in the boundary theory.

Now we can also rewrite eq.~\reef{btznew2} in the following form
\beq
\frac{d\mathcal{C}_A}{d\tau} =\frac{2 M}{\pi} \left(\tanh^2\left({\pi\tau}/{\beta}\right) +\frac{\log\cosh \left({\pi\tau}/{\beta}\right)}{\cosh^2 \left({\pi\tau}/{\beta}\right)}
+ \frac{\log\! \left[\frac{\beta\,e}{2\pi L}\,\alpha\right]}{\cosh^2 \left({\pi\tau}/{\beta}\right)}\right)
\label{btznew3}
\eeq
where we have introduced $\beta=1/T$ and $e$ is simply Euler's number, \ie $\log(e)=1$. This form facilitates a comparison to the analogous result in \cite{Brown2} evaluated with a regulator based on timelike radial geodesics in the bulk, which is
\beq
\frac{d\mathcal{C}_A}{d\tau} =\frac{2 M}{\pi} \left(\tanh^2\left({\pi\tau}/{\beta}\right) +\frac{\log\cosh \left({\pi\tau}/{\beta}\right)}{\cosh^2 \left({\pi\tau}/{\beta}\right)}
- \frac{\log \epsilon}{\cosh^2 \left({\pi\tau}/{\beta}\right)}\right)+{\cal O}(\epsilon)\,.
\label{btznew4}
\eeq
where $\epsilon$ is a dimensionless UV regulator, \ie $\epsilon \sim \delta/\beta$ and $\delta$ is the short-distance cut-off in the boundary theory.\footnote{We thank Ying Zhao for explaining this point.} Interestingly, we see that eqs.~\reef{btznew3} and \reef{btznew4} will be in complete agreement if we choose $\alpha\sim L/\delta$. We return to a discussion of this point in section \ref{sec:Discussion}.

\begin{figure}
\centering
\includegraphics[scale=0.448]{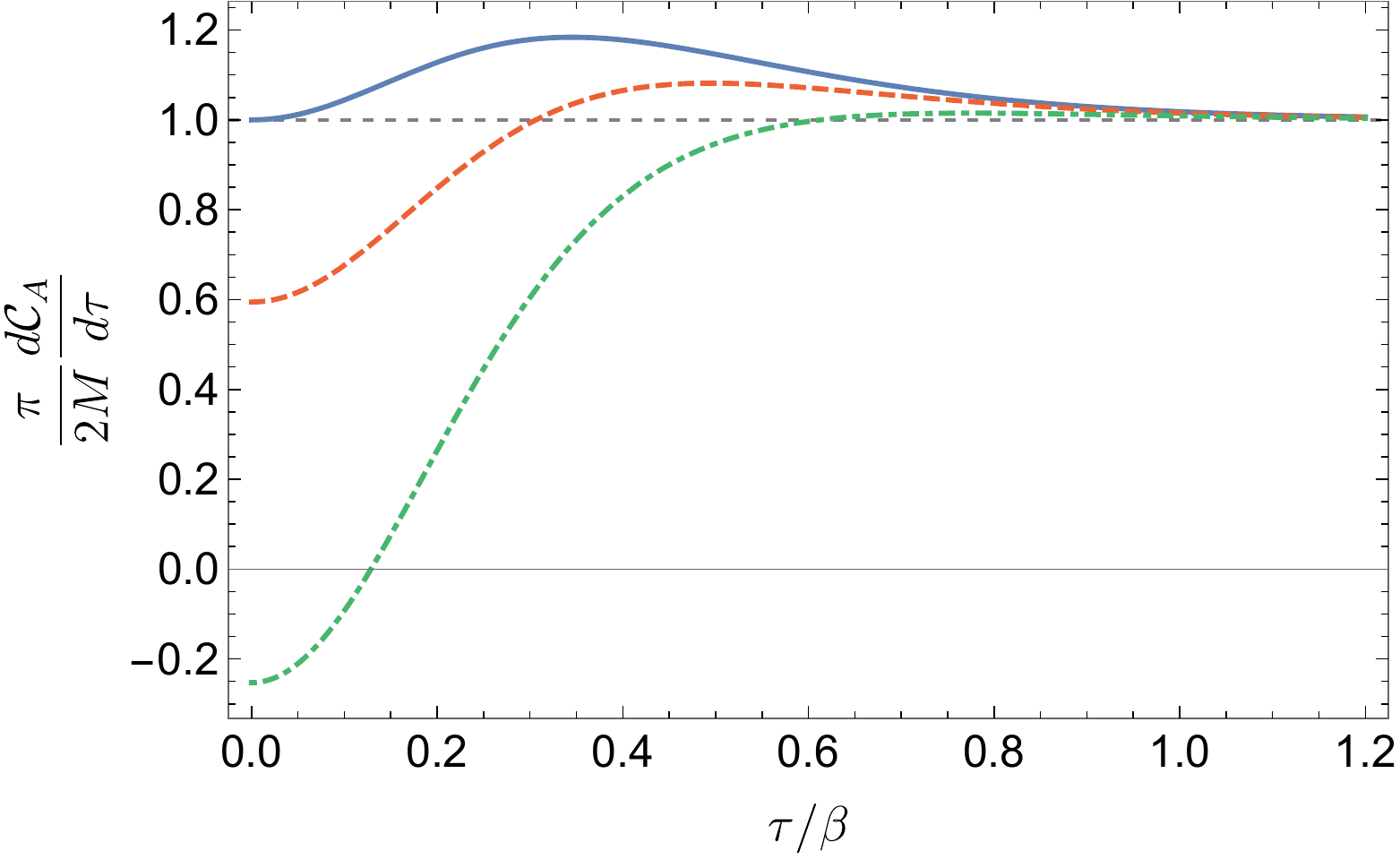}~~
\includegraphics[scale=0.45]{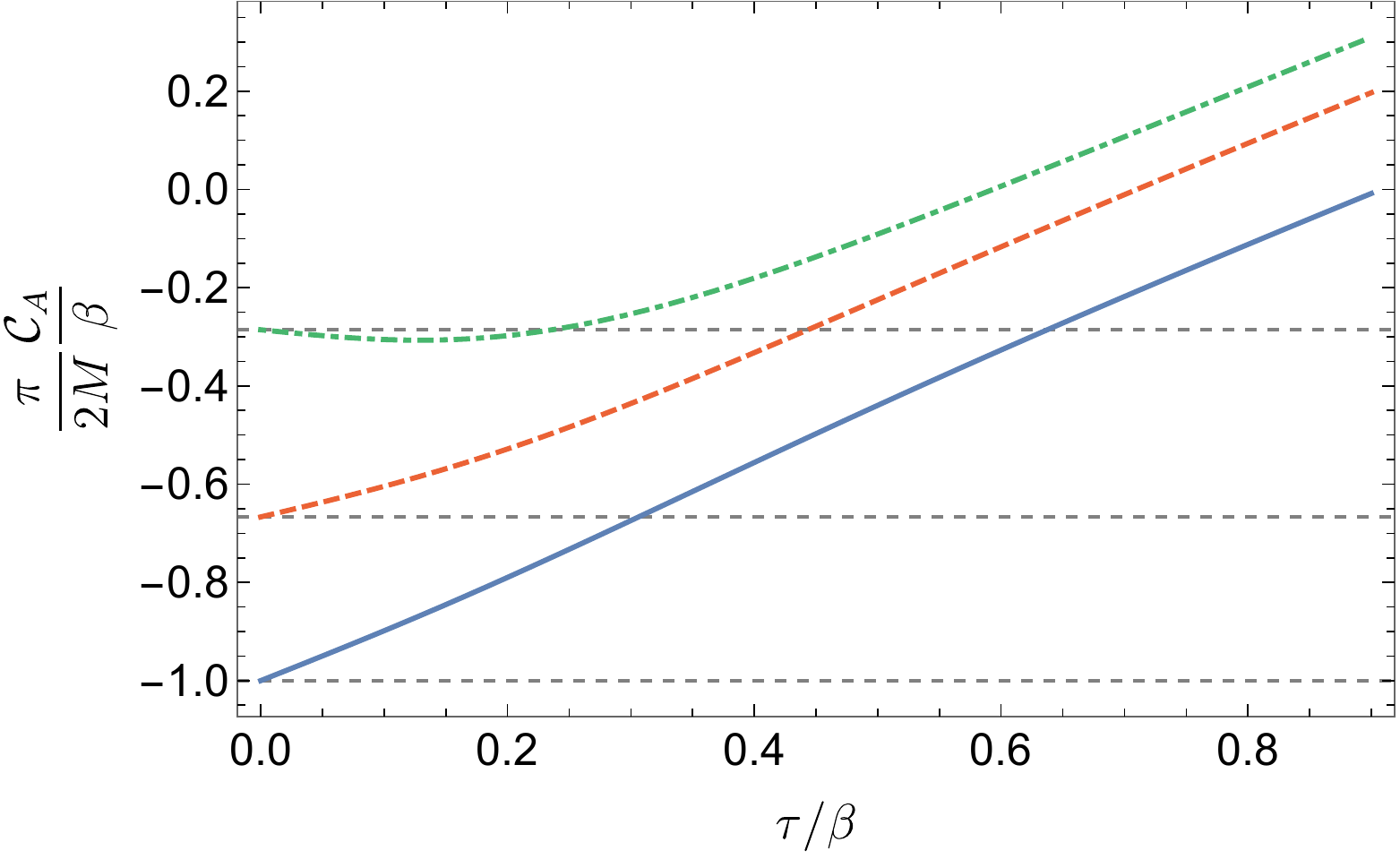}
\caption{Left panel: time derivative of the complexity for the BTZ black hole ($d=2$) from eq.~\reef{btznew2} with $\alpha=L/R$. Right panel: `total' complexity found by integrating $d\mathcal{C}_A/d\tau$. Results are shown for several values of the horizon radius --- $r_h/L=1$ (blue), $r_h/L=1.5$ (dashed red) and $r_h/L=3.5$ (dot-dashed green).
}  \label{fig:BTZ}
\end{figure}

To close this section, we plot both the rate of change of the complexity  \eqref{btznew2} and the total complexity in figure \ref{fig:BTZ} for several values of $r_h/L$. In the figure, we have chosen $\alpha=L/R$ and then in the argument of the logarithmic factor, we have ${2\pi R T}={r_h}/{L}$ using eq.~\reef{BTZblackfactor}. Note that all of the curves for $d\CA/d\tau$ in the left panel exceed the Lloyd bound and further the violation increases for smaller black holes, \ie smaller $r_h/L$, or equivalently smaller temperatures.  The right panel shows the complexity itself, found by integrating $d\mathcal{C}_A/d\tau$. The integration constant is chosen there so that the result of $\mathcal{C}_A(\tau=0)$ corresponds to the complexity of formation \cite{Formation}. In particular, we choose $\mathcal{C}_A(\tau=0)=\mathcal{C}_{\text{form}} = -\frac{  L}{2 G_N}$ --- see eq.~(4.8) in ref.~\cite{Formation}.\footnote{This corresponds to comparing the complexity of the thermofield double state to that of (two copies of) the Neveu-Schwarz vacuum in the boundary theory \cite{couscous}. Comparing to the Ramond vacuum would instead yield $\mathcal{C}_{\text{form}}=0$ \cite{Formation}.} After dividing by $\beta M$, all of these become functions of $r_h/L$.
We provide further details of the calculations and a more extensive discussion of the special case of BTZ black holes in appendix \ref{app:BTZnonsymmetric}.

\subsubsection{$d=4$}\label{UnchargedActd4}
\begin{figure}[h!]
\begin{center}
\includegraphics[scale=0.5]{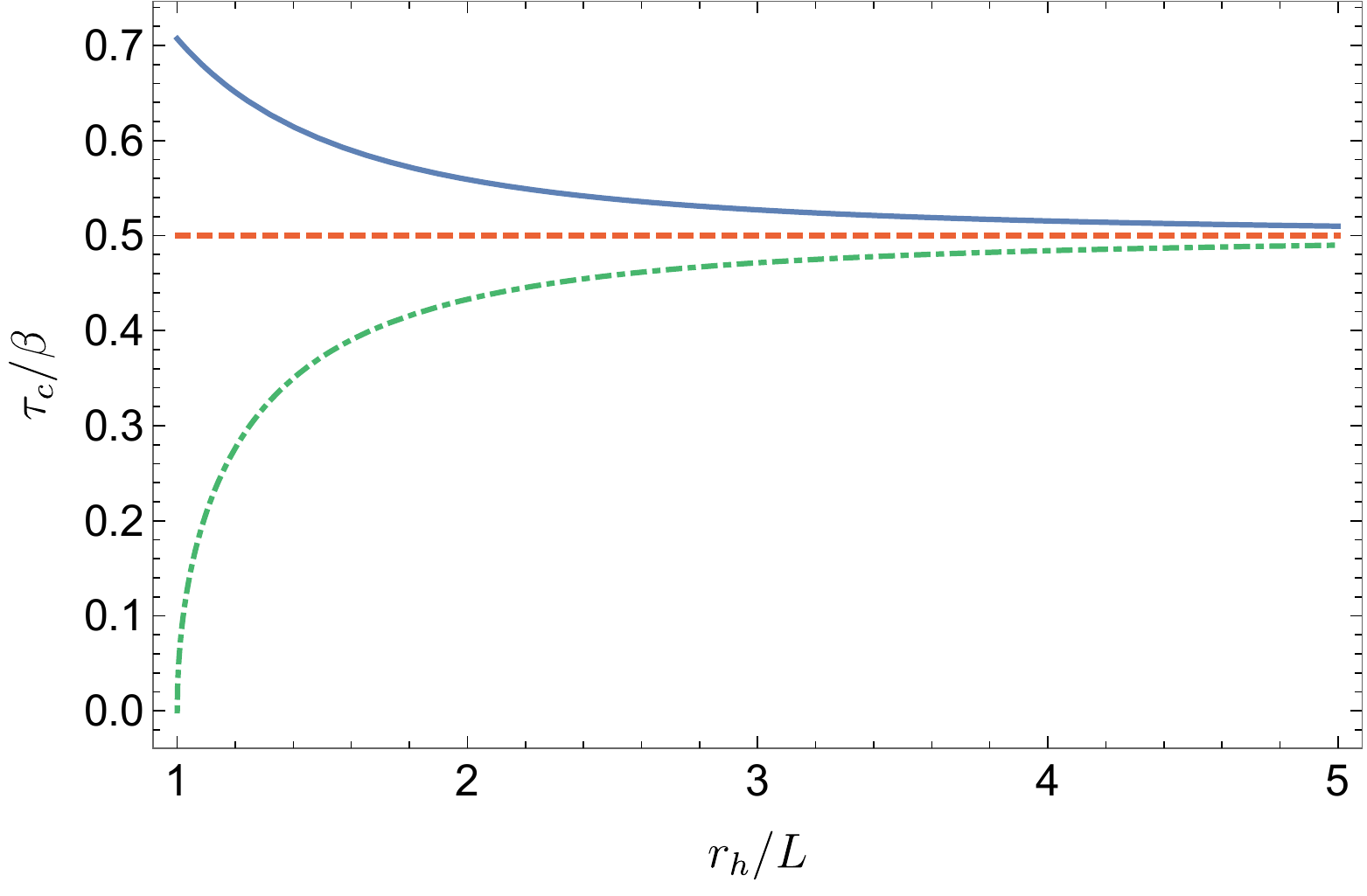}
\caption{Critical time $t_c$ as a function of the horizon radius for $d=4$ for the various horizon geometries, \ie spherical $k=1$ (blue), planar $k=0$ (dashed-red) and large hyperbolic $k=-1$ (dot-dashed green).
Note that we only consider $r_h>L$.
}\label{fig:d4ActionEternaltcOrh}
\end{center}
\end{figure}

To study the case where the boundary theory lives in $d=4$, in principle, we simply substitute this value into  eqs.~\reef{tder44} or \reef{tder45} for $d\mathcal{C}_A/d\tau$, with the blackening factor given by eq.~\reef{BFactor}. Of course, we must evaluate the meeting point $r_m$, or alternatively the dimensionless $x_m$, numerically. For the latter, we introduce the dimensionless radius $x=r/r_h$, as well as $\tilde f(x,RT)=L^2/r_h^2\,f(r)$ from eq.~\reef{there2}. Then following eq.~\reef{xmDimless}, we can then define a dimensionless tortoise coordinate
\begin{align}
&x^{*} (x, R T) \equiv \int \frac{d x}{\tilde f(x, R T)}
\ \ = \frac{r_h}{L^2}\, r^{*}(r)
\nonumber \\
& \quad = \frac{r_h^2}{2r_h^2 + k L^2} \left( \frac{1}{2}\, \log \!\frac{|1-x|}{1+x}  + \frac{\sqrt{r_h^2+ k L^2}}{r_h}\, \tan^{-1}\! \left[ \frac{r_h\,x}{\sqrt{r_h^2 + k L^2}} \right] \right) \, ,
\label{wopwop}
\end{align}
which yields
\begin{equation}
x^*_{\infty} \equiv x^*(\infty, R T)= \frac{\pi}{2} \,r_h\,\frac{\sqrt{r_h^2+k L^2}}{2r_h^2+k L^2}
\qquad{\rm and}\qquad
x^*(0, R T)=0\,.
\end{equation}
It is clear from eq.~\reef{wopwop} that $x^*$ is a function of the ratio $r_h/L$, however, as our notation indicates the latter is implicitly fixed in eq.~\reef{there3} by $RT$ in the boundary theory.
Combining these results with eq.~\reef{eq:criticalTaume} yields the critical time, at which the complexity begins to change,
\beq
\tau_c = \frac{2LR}{r_h}\,\left(x^*_\infty-x^*(0)\right)=\pi  L R\,\frac{ \sqrt{r_h^2+k\,L^2}}{ 2 r_h^2+k\,L^2} = \frac{1}{2 T} \left( 1 + k \left( \frac{L}{r_h} \right)^2  \right) \, .
\label{wop3}
\eeq
Note that for $k=0$, we have $\tau_{c}=1/(2T)$, \ie the critical time does not depend on $R$ for the planar geometry.
Figure \ref{fig:d4ActionEternaltcOrh} shows a plot of $\tau_c$ as a function of $r_h/L$ for the various horizon geometries.

\begin{figure}[h!]
\begin{center}
\includegraphics[scale=0.38]{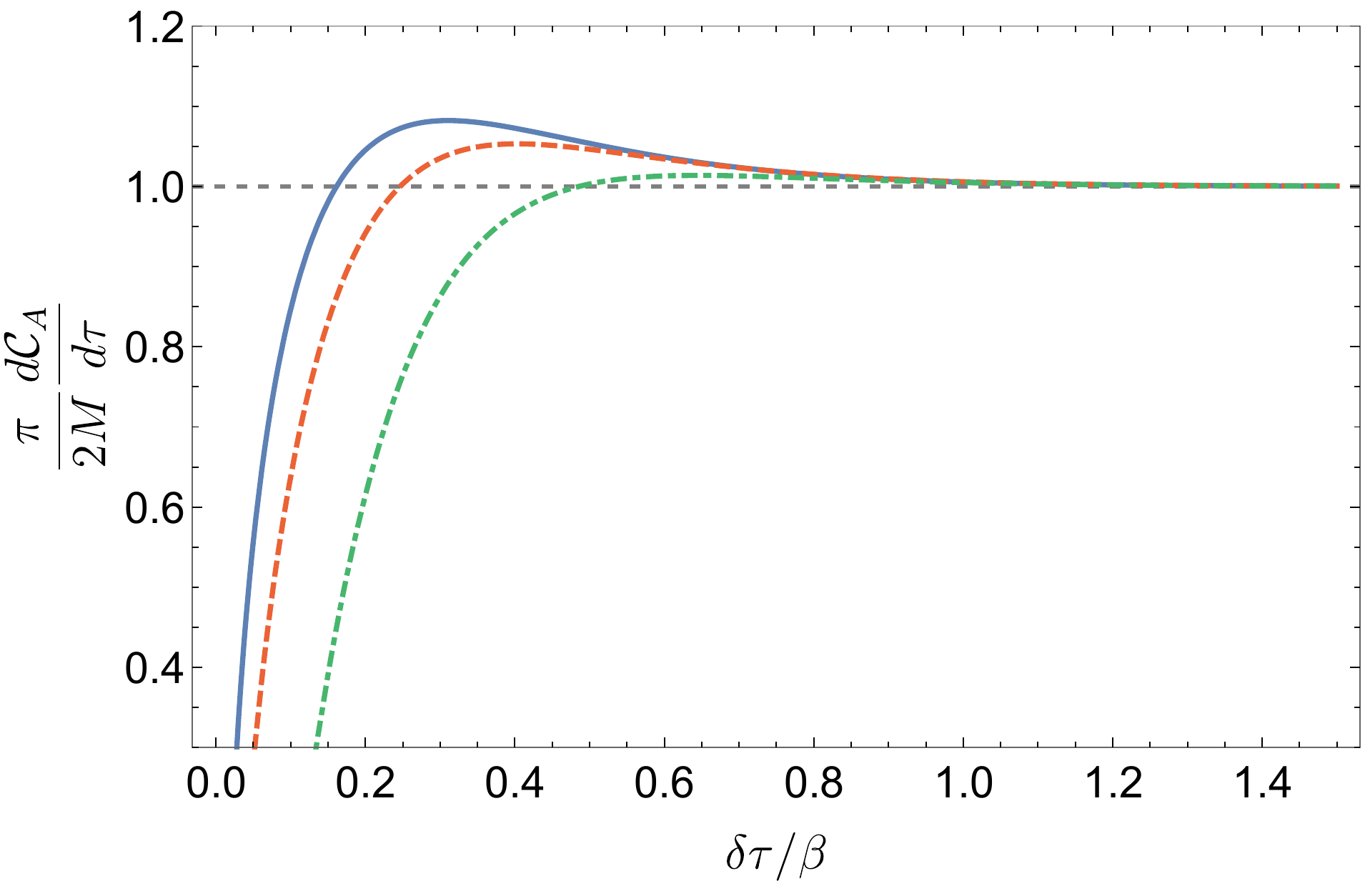}
\includegraphics[scale=0.38]{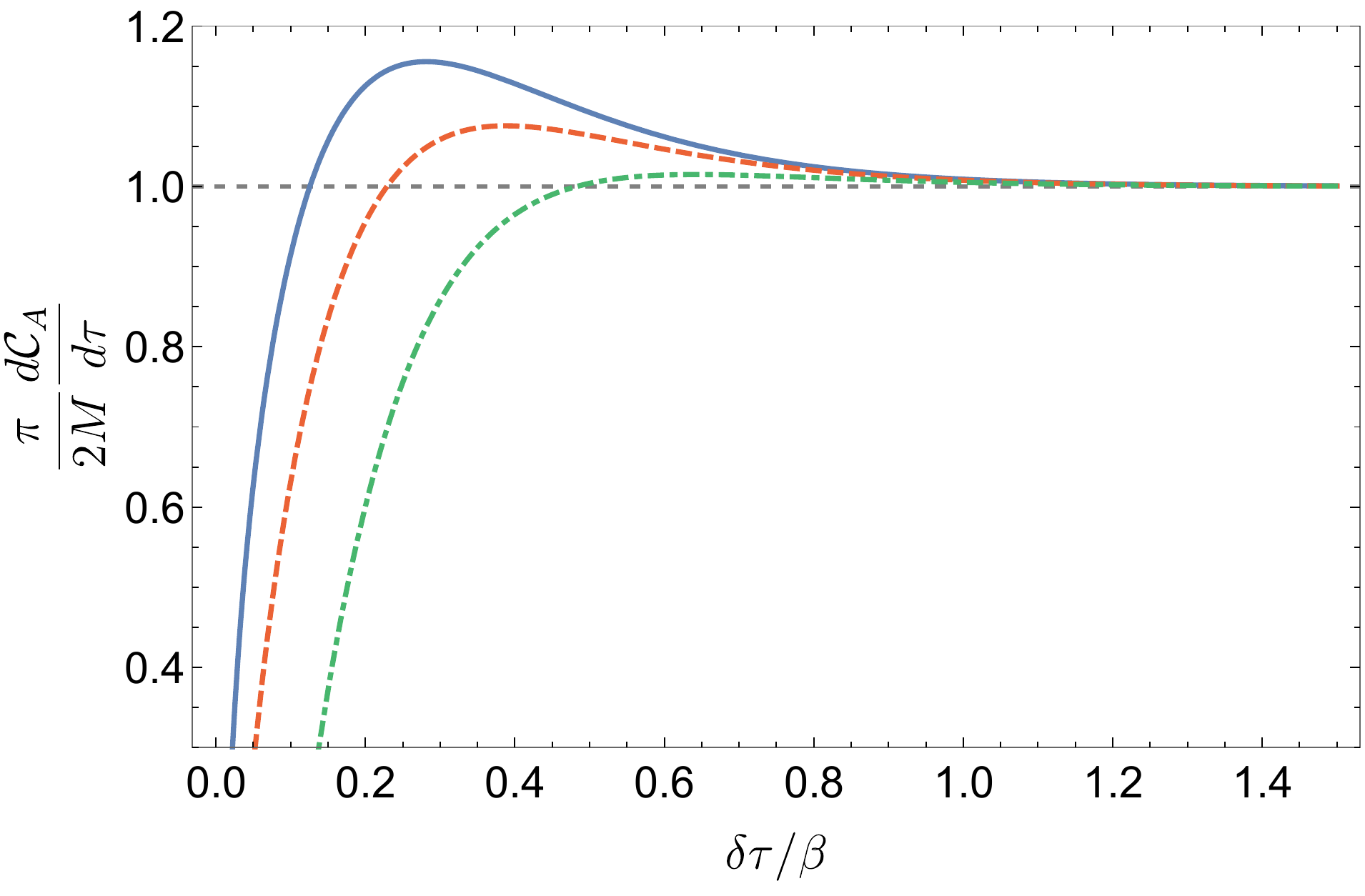}\\[1ex]
\caption{Time derivative of complexity as a function of time for spherical ($k=+1$, left) and planar ($k=0$, right) horizons with $d=4$ boundary dimensions for various values of the horizon radius, \ie $r_h/L=1$ (blue),  $r_h/L=1.5$ (dashed red) and  $r_h/L=3.5$ (dot-dashed green).
We present the plots as a function of $\delta \tau= \tau-\tau_c$ to allow for a meaningful comparison between the different cases. We stress again that  each of the curves has a different value of $\tau_c$ --- see figure \ref{fig:d4ActionEternaltcOrh}. }
\label{d4Rate}
\end{center}
\end{figure}

Now solving numerically for the meeting point $x_m$ using eq.~\reef{xmDimless}, we can evaluate $d\mathcal{C}_A/d\tau$ in eq.~\reef{tder45}, as shown in figure \ref{d4Rate} for spherical ($k=1$) and planar ($k=0$) horizons. As commented above,  we have set $\alpha=L/R$ for simplicity in these plots.
Note that for a fixed $r_h/L$, the planar geometries seem to violate the $2M/\pi$ bound more strongly. We also note that the violation of the bound is stronger for smaller black holes, \ie smaller values of $r_h/L$. A more careful examination shows that generally $d\mathcal{C}_A/d\tau$ is larger for $k=0$ than for $k=+1$ and that this difference between the rate of growth for these two cases grows as the size of the black hole shrinks. Similar results apply for hyperbolic horizon geometries and for other boundary dimensions. We describe our results for the case of $d=3$ for all three horizon geometries in appendix \ref{app:MoreAction}.

\section{Complexity=Volume}\label{sec:EternalVolume}
\begin{figure}[htbp]
\centering
\includegraphics[keepaspectratio, scale=0.3]{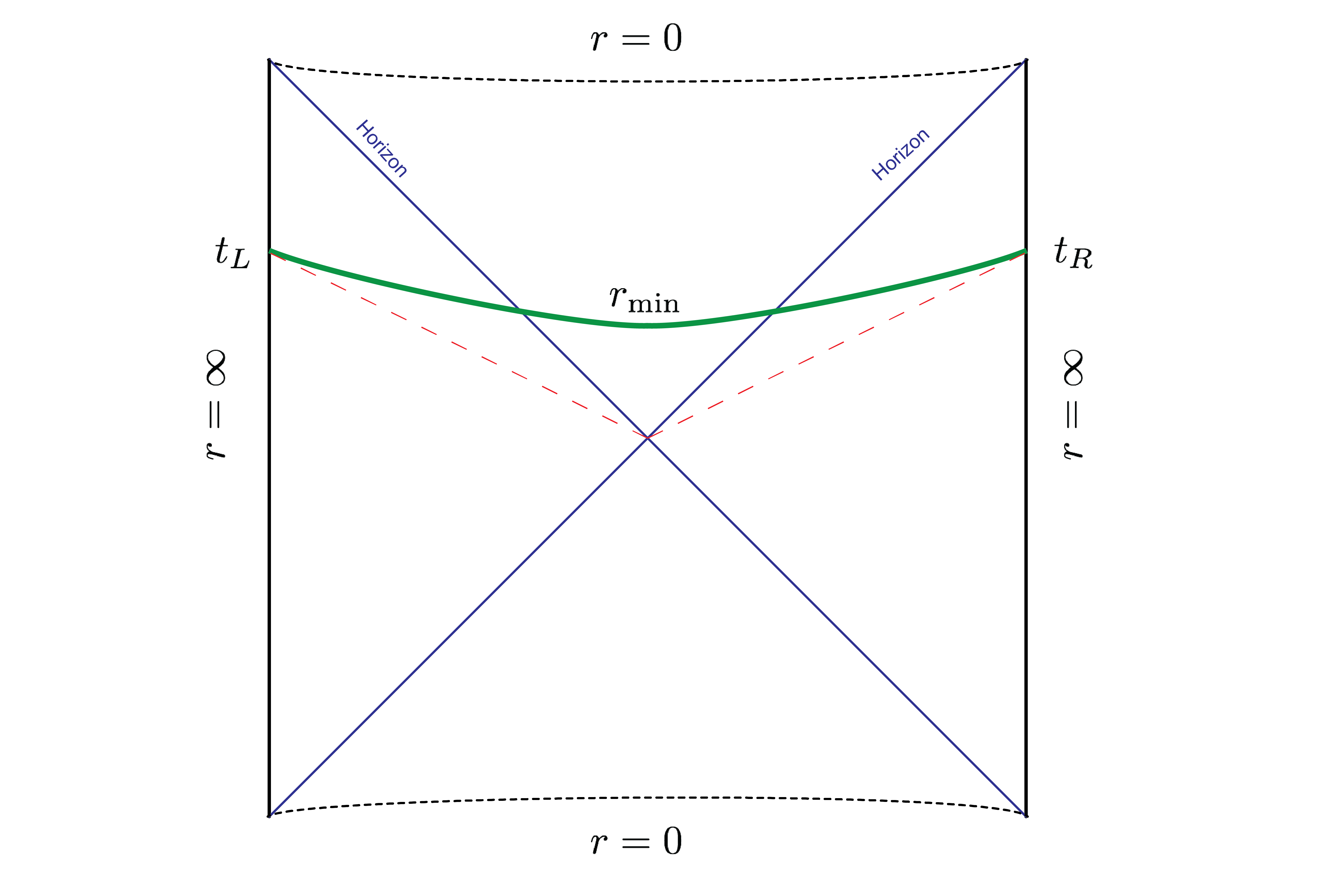}
\caption{A representation of the maximal wormhole connecting the two boundaries anchored at times $t_{L}$ and $t_{R}$ (depicted at symmetric times in the figure). The bridge reaches the minimum distance inside the future horizon at $r_{min}$, and approaches each boundary tangent to constant time slices.}
	\label{VolumeWormhole}
\end{figure}

In this section, we study the time dependence of the complexity for eternal AdS black holes using the complexity=volume conjecture \cite{Susskind:2014rva,Stanford:2014jda}. Applying eq.~\reef{volver}, we must evaluate the volume of the extremal codimension-one bulk surface, whose boundaries correspond to the desired time slices in the two asymptotic boundaries, as shown in figure~\ref{VolumeWormhole}.\footnote{For a proposed generalization for the complexity of subsystems in terms of the co-dimension one volume enclosed by the Ryu-Takayanagi surface, see \cite{diverg, Alishahiha:2015rta, DeanSub}.} As in the previous section, the symmetry of our setup implies that the volume depends only on the total boundary time $t=t_L + t_R$. Thus, it is enough to consider the symmetric case $t_L=t_R$, as we assume from now on.
Further,  in eq. \eqref{volver},  we will simply set $\ell =L$, the AdS radius, to eliminate the ambiguity associated with the choice of the scale $\ell$.

First, we review the computation of the maximal volume following \cite{Stanford:2014jda} and then evaluate its time derivative.  We will see that the time derivative of the extremal volume is determined by a conserved quantity $E$.
With the infalling Eddington-Finkelstein coordinates \reef{EFcoord}, the metric \reef{HigherDMetric} becomes
\begin{equation}
v=t+r^*(r)\,; \qquad
ds^2 = -f(r) \, dv^2 +2\, dv\, dr+r^2 d \Sigma_{k,d-1}^2\,.
\end{equation}
Now, assuming that the extremal surface is `spherically' symmetric,\footnote{That is, the extremal surface has the same symmetry as the spatial slices described by $d \Sigma_{k,d-1}^2$, \eg it is spherically symmetric for $k=+1$.} its profile will be determined by an embedding $r(\lambda)$ and $v(\lambda)$, where $\lambda$ is some radial coordinate intrinsic to the surface. The maximal volume is then obtained by extremizing
\begin{align}
\mathcal{V} = \Omega_{k,d-1} \int d \lambda \, r^{d-1} \sqrt{-f(r) \dot{v}^2 + 2 \dot{v}\dot{r}}
\equiv \Omega_{k,d-1} \int d \lambda \, \mathcal{L}(\dot{v}, r, \dot{r})\, ,
\label{volume_int}
\end{align}
where  the dots indicate derivatives with respect to $\lambda$.
Since the integrand $\mathcal{L}$ does not depend explicitly on $v$, we have a conserved quantity $E$ defined as
\begin{align}
E=-\frac{\partial \mathcal{L}}{\partial \dot{v}} =\frac{r^{d-1} (f \dot{v}-\dot{r})}{\sqrt{-f \dot{v}^2 + 2\dot{v}\dot{r}}}\, .
\end{align}
We will refer to this quantity as the energy. Since the expression in eq.~\reef{volume_int} is reparametrization invariant, we are free to choose $\lambda$ to keep the radial volume element fixed, \ie
\begin{align}
r^{d-1} \sqrt{-f\, \dot{v}^2 + 2\dot{v}\dot{r}} =1 .
\end{align}
The equations determining $r(\lambda)$ and $v(\lambda)$ then simplify to
\begin{align}
E&=r^{2(d-1)} \left(f(r) \dot{v}-\dot r\right), \label{tdot} \\
r^{2(d-1)} \dot{r}^2 &= f(r) + r^{-2(d-1)} E^2 \label{rdot},
\end{align}
and further, the maximal volume can be written as
\begin{align}
\mathcal{V}= 2\Omega_{k,d-1}\int^{r_{max}}_{r_{min}} \frac{d r}{\dot{r}}
=2\Omega_{k,d-1} \int^{r_{max}}_{r_{min}} d r \, \frac{r^{2(d-1)}}{\sqrt{f(r) r^{2(d-1)} +E^2}} \,.
\label{volume_int2}
\end{align}
Here, we are assuming a symmetric configuration where $t_L=t_R$, as described above, and so the integral only runs from a minimum radius $r_{min}$ to the cutoff surface at $r=r_{max}$. The minimal radius is determined by setting $\dot r=0$ in eq.~\reef{rdot}, \ie
\begin{align}
f(r_{min})\, r_{min}^{2(d-1)} +E^2 = 0 \,.
\label{eq_rturn}
\end{align}
Further we note that this turning point is inside the horizon (see figure \ref{VolumeWormhole}) and hence we have $f(r_{min})<0$, $\dot r|_{r=r_{min}}=0$ and $\dot v|_{r=r_{min}}>0$. Therefore we may conclude that $E<0$ by evaluating eq.~\reef{tdot} at this point. Now using eqs.~\eqref{tdot} and \eqref{rdot}, we have
\begin{align}\label{t_r_E}
t_R + r^{*}_{\infty}- r^*(r_{min}) =\int_{v_{min}}^{v_{\infty}} dv= \int^{r=\infty}_{r_{min}} dr \left[\frac{E}{f(r)\sqrt{f(r) r^{2(d-1)} +E^2}}+\frac{1}{f(r)}\right] ,
\end{align}
where the symmetry of our configuration determines $t=0$ at the turning point, \ie $v_{min}=r^*(r_{min})$. One may verify that the integrand in the final expression is well-behaved at the horizon, using the fact that the energy is negative. The integrand also decays as $L^2/r^2$ with $r\to\infty$ and so in the following, we will replace the upper limit of the integral by $r=r_{max}$ because the difference produced by this replacement vanishes as the short-distance cutoff is taken to zero. We will make use of this several times in the derivation below.

Using eq.~\reef{t_r_E}, we can rewrite eq.~\eqref{volume_int2} as follows:
\begin{align}
\frac{\mathcal{V}}{2\Omega_{k,d-1}}
=\int^{r_{max}}_{r_{min}} d r \,
\left[\frac{\sqrt{f(r) r^{2(d-1)} +E^2}}{f(r)}+\frac{E}{f(r)}\right]  - E \left(t_R+r^*_{\infty}-r^{*}(r_{min})\right) .
\end{align}
Next, we would like to take the time derivative of this equation, however, we would like to use the time coordinate introduced in eq.~\reef{metric2}, \ie $\tau=R\,t/L$.
We use eq.~\eqref{eq_rturn} to simplify the contribution from the derivative acting on $r_{min}$ in the lower limit of the integral
to obtain
\begin{align}
\begin{split}
\frac{1}{2\Omega_{k,d-1}} \frac{d\mathcal{V}}{d \tau_R}
=& \frac{d E}{d \tau_R}\, \int^{r_{max}}_{r_{min}} d r \, \left[
\frac{E }{f(r)\sqrt{f(r) r^{2(d-1)} +E^2}}+\frac{1}{f(r)}\right]
\\
&~~~~~~~~
-\frac{d E}{d \tau_R} \left(\frac{L}R\,\tau_R+r^*_\infty - r^*(r_{min})\right) - \frac{L}{R} E\,.
\end{split}
\end{align}
Note that $d E/d \tau_R$ is a constant that characterizes the entire surface and so it was brought outside of the integral in the first term. However, the remaining integral is identical to that appearing in eq.~\eqref{t_r_E} and so we may further simplify the result to
\begin{align}
\frac{d\mathcal{V}}{d \tau_R} = - 2\Omega_{k,d-1} \frac{L}{R} E\,.
\end{align}
Since we set $\tau_R=\tau_L$,
the derivative with respect to $\tau=\tau_R+\tau_L$ is given by simply multiplying the result by a factor of  $1/2$. Hence our final result for the rate of growth of the complexity becomes
\begin{align}
\frac{d\mathcal{C}_V}{d \tau} = \frac{1}{G_N L} \frac{d\mathcal{V}}{d \tau}
= - \frac{\Omega_{k,d-1}}{G_N R} E
= \frac{\Omega_{k,d-1}}{G_N R} \sqrt{-f(r_{min})}\,  r_{min}^{d-1} \, .
\label{dvdt_general}
\end{align}
Therefore, the time derivative of complexity is completely determined by computing either $E$ or $r_{min}$, with eq.~\eqref{eq_rturn}.

However, as in eq.~\reef{tder45}, we would like to show that eq.~\reef{dvdt_general} can be expressed entirely in terms of boundary quantities. After some work, the final result takes the form
\beq
\frac{d\mathcal{C}_V}{d \tau} = \frac{8 \pi M}{(d-1)} \frac{8   \pi^2 R^2 T^2\, \tilde g^2 (RT)}{ 4 \pi^2 R^2 T^2\, \tilde g^2 (RT)+k} \,
\sqrt{-\tilde f(x_{min},RT)}\,   x_{min}^{d-1}\, ,
\label{generalX}
\eeq
where the functions  $\tilde g (RT)$ and $\tilde f(x,RT)$ were defined in eqs.~\reef{there3} and \reef{there2}, respectively. Further, as above, we have introduced the dimensionless radial coordinate $x=r/r_h$. Then defining the corresponding tortoise  coordinate $x^*(x) \equiv \int dx/\tilde f(x,RT)$ and also $x_{E}\equiv E/r_h^{d-1}$, $x_{min}$ is determined by the boundary versions of eqs.~\reef{eq_rturn} and \reef{t_r_E}:
\beqa
&&\qquad\qquad\qquad 0=4\pi^2 R^2 T^2 \,\tilde g^2(RT)\, \tilde f(x_{min},RT)\, x_{min}^{2(d-1)}+x_E^2 \,,
\labell{outsource}\\
&&\frac{\tau_R}{\beta}   +  \frac{x^{*}_{\infty}-x^*(x_{min})}{2\pi } = \int^{x=\infty}_{x_{min}}
\frac{dx\left[x_E+\sqrt{4\pi^2 R^2 T^2 \tilde g^2(RT) \tilde f(x,RT) x^{2(d-1)} +x_E^2}\right]}
{2 \pi  \tilde f(x,RT) \sqrt{4\pi^2 R^2 T^2 \tilde g^2(RT) \tilde f(x,RT) x^{2(d-1)} +x_E^2}}\, .
\nonumber
\eeqa

\subsection{Late Time Behaviour} \label{LateX}

Before examining the full time-dependence of $d\mathcal{C}_V/d \tau$, we would like to study its late time behaviour.
At late times, the maximal surface is (almost) tangent to a special slice of constant $r=\tilde r_{min}$ inside the black hole \cite{Stanford:2014jda}.\footnote{Similar behaviour appears in computing the time dependence of holographic entanglement entropy for regions with components in both asymptotic boundaries \cite{Hartman:2013qma}. However, the  special (codimension-two) surface appearing there extremizes the area rather than the volume.}  To evaluate $\tilde r_{min}$, we first define the function $W(r)$ as appeared in eq.~\eqref{dvdt_general},
\begin{align}
W(r) \equiv  \sqrt{-f(r)}\,  r^{d-1}\,,
\label{w(r)}
\end{align}
and observe that eq.~\eqref{eq_rturn} can be rewritten as $-W(r_{min})^2 +E^2=0$. The latter generally has two positive  roots, with the larger root corresponding to $r_{min}$. However, in the late time limit, $|E|$ increases until the two roots meet at the extremum of $-W(r)^2$, which also corresponds to the extremum of $W(r)$.  Hence $\tilde r_{min}$ is both a root of eq.~\eqref{eq_rturn} and the extremum of $W(r)$. Then $\tilde r_{min}$ can be computed as
\begin{align}
0=W'(\tilde r_{min})
= (d-1) \tilde r_{min}^{d-2} \sqrt{-f(\tilde r_{min})} - \frac{\tilde r_{min}^{d-1} f'(\tilde r_{min})}{2\sqrt{-f(\tilde r_{min})}} \, .
\label{r_m_eq}
\end{align}
Since $d\mathcal{C}_V/d\tau$ in eq.~\reef{dvdt_general} only depends on the time $\tau$ through $r_{min}$,
at late times, we have
\begin{align}
 \frac{d\mathcal{C}_V}{d \tau} = \frac{\Omega_{k,d-1}}{G_N R} \Bigl[W(\tilde r_{min})
 + \frac12 W''(\tilde r_{min}) (r_{min}-\tilde r_{min})^2 + \mathcal{O}((r_{min}-\tilde r_{min})^3)
 \Bigr]\,.
 \label{latetime_dcv/dt}
\end{align}
Hence asymptotically, $d\mathcal{C}_V/d\tau$ approaches the constant value
\begin{align}
\lim_{\tau\rightarrow\infty} \frac{d\mathcal{C}_V}{d \tau}= \frac{\Omega_{k,d-1}}{G_N R} W(\tilde r_{min}) = \frac{\Omega_{k,d-1}}{G_N R} \sqrt{-f(\tilde r_{min})}\, \tilde r_{min}^{d-1}\,.
\label{const_dv/dt}
\end{align}

Further, we observe that $d\mathcal{C}_V/d\tau$ approaches this limit from below because $W''(\tilde r_{min})$ is negative. The latter conclusion is easily produced by noting from eq.~\reef{w(r)}, that $W(r)$ vanishes at both $r=r_h$ and 0 and that $W(r)>0$ inside the horizon. Hence the extremum \reef{r_m_eq} must be a maximum, \ie  $W''(\tilde r_{min})<0$.\footnote{The case of small hyperbolic black holes, \ie $k=-1$ and $r_h<L$, is slightly more complicated since there is an inner horizon --- see appendix \ref{hyperbh}. However, implicitly $r_{min}$ lies in between the two horizons and so one reaches the same conclusion.}
In appendix \ref{app:MoreVolume}, we examine the leading correction to the late time limit \reef{const_dv/dt}  and show that $d\mathcal{C}_V/d\tau$ approaches this asymptotic value with an exponential decay in $\tau$. Next we turn to computing the asymptotic value \eqref{const_dv/dt}.

\paragraph{Planar horizons:}
With $k=0$, eq.~\reef{r_m_eq} can be solved analytically for $\tilde r_{min}$ and we find
\begin{align}
\tilde r_{min} = \Bigl(\frac{\omega^{d-2}L^2}{2}\Bigr)^{\frac{1}{d}}= \frac{r_h}{2^{\frac{1}{d}}} \ ,\label{rtild}
\end{align}
which then leads to
\begin{align}
\sqrt{-f(\tilde r_{min})} \, \tilde r_{min}^{d-1} = \frac{\omega^{d-2}L}{2} \ .
\end{align}
Thus, using eq.~\reef{there}, the asymptotic value \eqref{const_dv/dt} becomes
\begin{align}
\lim_{\tau\to \infty} \frac{d\mathcal{C}_V}{d \tau} = \frac{8 \pi M}{d-1}\, ,
\label{planar_asympt_volume_rate}
\end{align}
for any planar black hole. Of course, this reproduces the result first found in \cite{Stanford:2014jda}.

\begin{figure}[htbp]
	\vspace{2ex}
	\begin{minipage}[b]{0.5\hsize}
		\centering
		\includegraphics[keepaspectratio, scale=0.44]{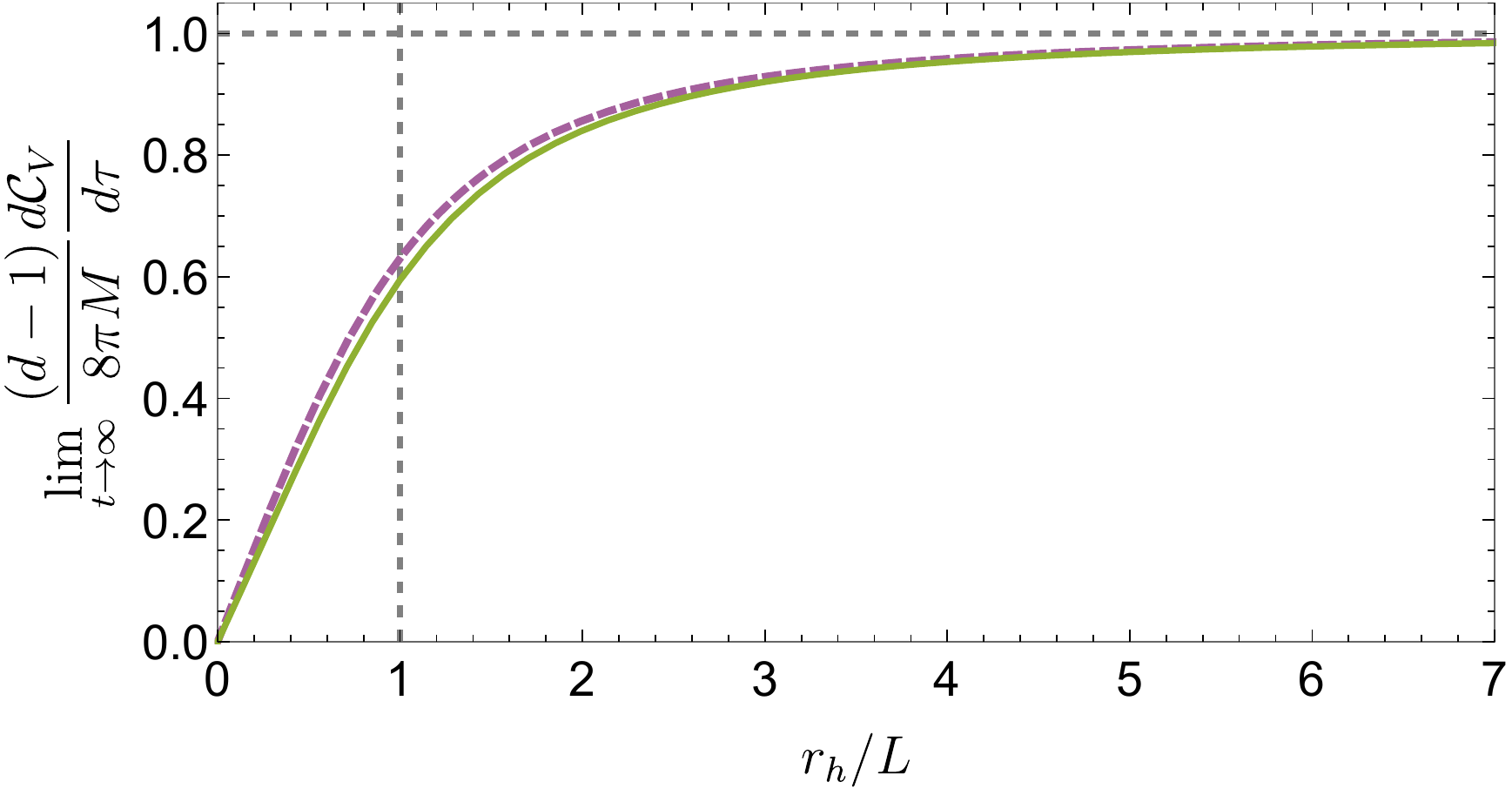}
		\subcaption{}
		\label{asympto_volume_spherical}
	\end{minipage}
	\begin{minipage}[b]{0.5\hsize}
		\centering
		\includegraphics[keepaspectratio, scale=0.44]{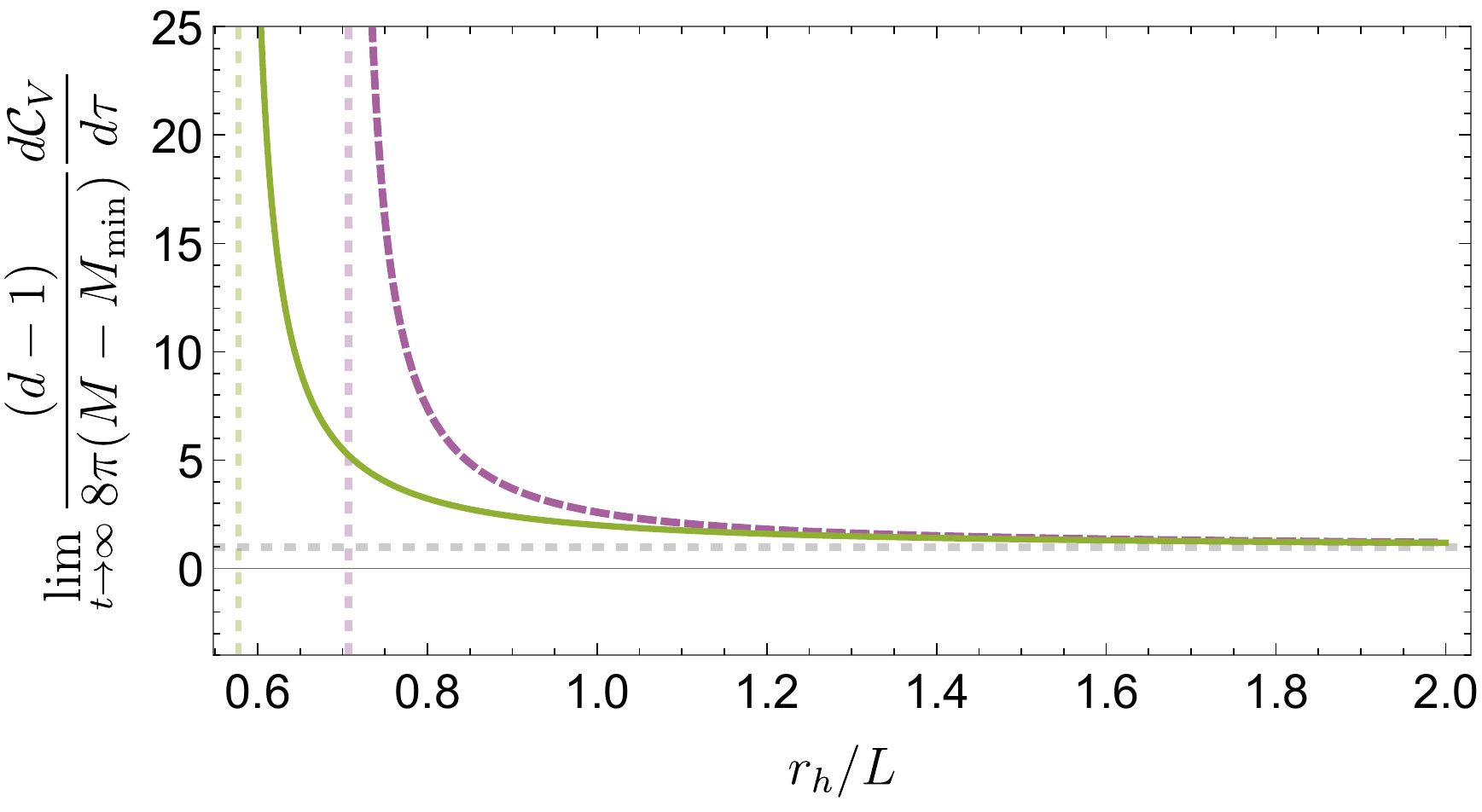}
		\subcaption{}
		\label{asympto_volume_hyper}
	\end{minipage}
	\caption{(a) Late time rate of change in complexity $\frac{d-1}{8\pi M} \lim_{\tau\to \infty} d\mathcal{C}_V/d\tau$ as a function of $r_h/L$ for spherical black holes ($k=1$) in $d=3$ (green) and $d=4$ (dashed purple) dimensions. The vertical dashed line at $r_h/L=1$ indicates the Hawking-Page phase transition below which the dominant saddle point in the bulk partition function is vacuum AdS rather than a (small) spherical black hole.
(b) Plots of $\frac{(d-1)}{8\pi (M-M_{min})} \lim_{\tau\to \infty} d\mathcal{C}_V/d\tau$ as a function of $r_h/L$ for hyperbolic black holes ($k=-1$) in $d=3$ (green) and $d=4$ (dashed purple) dimensions. The vertical lines indicate the minimal values of $r_h/L$ corresponding to extremal small hyperbolic black holes. The gray dashed horizontal line indicates 1, which is approached in the large black hole limit ($r_h\gg L$). }
	\label{asympto_volume}
\end{figure}

\paragraph{Curved horizons:}
Figure \ref{asympto_volume_spherical} shows a plot of the late time limit \eqref{const_dv/dt}
for spherical black holes (with $k=1$) for $d=3$ and 4.
We can see that $d\mathcal{C}_V/d \tau$ approaches the value $8 \pi M/(d-1)$ in the limit $r_h \gg L$, \ie $RT\gg1$.
%


Since the mass of hyperbolic black holes (\ie $k=-1$) can take negative values,
$\frac{d-1}{8\pi M} \lim_{t \to \infty} d\mathcal{C}_V/dt$
would diverge at $M=0$ before reaching the minimal mass.
Hence, we instead present numerical plots of
\begin{align}
\frac{d-1}{8\pi (M-M_{min})} \lim_{\tau \to \infty} \frac{d\mathcal{C}_V}{d \tau},
\label{limit_hyper}
\end{align}
where $M_{min}$ is the minimal value of mass
\begin{align}\label{Mminhyper}
M_{min}=-\frac{(d-1)\Omega_{-1,d-1}}{8 \pi G_N d} \Bigl(\frac{d-2}{d}\Bigr)^{\frac{d-2}{2}} \frac{L^{d-1}}{R}\, .
\end{align}
This corresponds to the mass of the extremal small hyperbolic black holes --- see appendix \ref{hyperbh}.
Figure~\ref{asympto_volume_hyper} presents the late time limit results for $d=3$ and $d=4$ as a function of $r_h/L$. Hence we can see that  eq.~\eqref{limit_hyper} approaches to 1 from above, in the limit  $r_h/L \gg 1$. The divergence in these curves where $r_h/L$ approaches its minimal value, \ie $M\to M_{min}$, is interesting because $d\mathcal{C}_V/d \tau$ actually vanishes in the extremal limit. The horizon radius of the extremal black hole can be written as $r_h^{ext}=\frac{\sqrt{d-2} L}{\sqrt{d}}$. Then we would readily find in the extremal limit that $d\mathcal{C}_V/d \tau\sim (r-r_h^{ext})$ while $M-M_{min}\sim (r-r_h^{ext})^2$. As a consequence, while both the numerator and denominator vanish in this limit, we still obtain a divergent result.
%

Now we proceed to examine the late time behaviour analytically in the limit of large temperatures, \ie for large black holes. First, we expand eq.~\eqref{r_m_eq} in  the limit $r_h \gg L$ to find the leading corrections to $\tilde r_{min}$ compared to its planar value \reef{rtild},
\begin{align}
\begin{split}
\tilde r_{min} =\frac{r_h}{2^\frac{1}{d}}
&\left[1-\frac{ \left(2^{2/d} (d-1)-d\right)  }{d^2 } \frac{L^2}{r_h^2}k \right.
\\
&\left.
+\frac{ (d-1) \left(-d^2+2^{\frac{2}{d}+1} d+2^{4/d} (d-3) (d-1)\right) }{2d^4 }\frac{L^4}{r_h^4}k^2+\mathcal{O}\left(\frac{L^6}{r_h^6}\right)\right]\,.
\end{split}
\end{align}
Using this expression,
the asymptotic value of ${d\mathcal{C}_V}/{d\tau}$ can be written in terms of the following expansion\footnote{This expansion can also be expressed in terms of central charge over the entropy --- see \cite{Formation}.}
\beqa
\lim_{t \to \infty}  \frac{(d-1)}{8\pi M} \frac{d\mathcal{C}_V}{d \tau} &=& \left(1-\frac{M_{min}}M\,\delta_{k,-1}
\right)\left(1-2^{\frac{2}{d}-1} k \frac{L^2}{r_h^2}+\frac{2^{\frac{2}{d}} \left(\gamma+d\right) k^2 }{d^2}\frac{L^4}{r_h^4} +\cdots\right)
\nonumber\\
&=&\left(1+\frac{2  d (d (d-2))^{\frac{d-2}{2}} }{(4\pi)^d (RT)^d}\,\delta_{k,-1}+\cdots
\right)\labell{ok77}\\
&&\qquad\times\left(1-\frac{2^{\frac{2}{d}-1} d^2 k}{(4\pi) ^2 (RT)^2}+\frac{2^{\frac{2}{d}} \left(\gamma- d(d-3)\right) d^2 k^2}{(4\pi) ^4
(RT)^4} +\cdots\right)
\nonumber
\eeqa
where to reduce the clutter in the above expressions, we have defined the coefficient:
\beq
\gamma=2^{\frac2d-3}(3 d-2)(d-2)\,.
\eeq

Let us first focus our attention on the second factor on the right-hand side of eq.~\reef{ok77}. Here the corrections involve (integer) powers of $k/R^2$ and hence we expect that these terms can be expressed as simple powers of the boundary curvature. Of course, these curvature corrections become important when the temperature is comparable to the curvature scale, \ie $RT\sim 1$. However, for high temperatures where the characteristic thermal wavelength is much shorter than the curvature scale, these terms become vanishingly small and the asymptotic growth rate approaches the flat space limit $8\pi M/(d-1)$, as in eq.~\eqref{planar_asympt_volume_rate}.

The above discussion overlooks the first factor on the right-hand side of eq.~\reef{ok77}. This factor only appears for the case of the hyperbolic horizons (\ie $k=-1$) and is related to the fact that the minimal mass is actually negative (rather than zero) for these black holes. Further, we observe that when the boundary dimension $d$ is odd, the first correction in this factor involves an odd power of $1/R$. Therefore while the corrections in this factor are appearing because of the negative curvature in the boundary metric \reef{body2}, they will not generally be expressed in terms of geometric factors involving powers of the curvature tensor.

We also note that the expression in eq.~\reef{ok77} only holds for $d\ge3$ and so the leading correction for $RT\gg1$ always comes from the second factor, \ie the term proportional to $k/(RT)^2$.
Therefore we can conclude that for spherical black holes, the asymptotic value \reef{const_dv/dt} approaches the planar value
\eqref{planar_asympt_volume_rate} from below as $RT\to 0$. Of course, this is in agreement with the results shown in figure \ref{asympto_volume_spherical}, where we see that for all values of $RT$,
\begin{align}
\lim_{\tau\to \infty} \frac{d\mathcal{C}_V}{d \tau} \leq \frac{8\pi M}{d-1}\qquad{\rm for}\ \  k=+1\ .
\label{volume_bound_spherical}
\end{align}
Similarly for hyperbolic black holes, the asymptotic value \reef{const_dv/dt} approaches the planar value
\eqref{planar_asympt_volume_rate} from above in the limit $RT\to 0$. Again, this agrees with the results shown in figure \ref{asympto_volume_hyper}, where we see that for all values of $RT$,
\begin{align}
\lim_{t \to \infty}  \frac{d\mathcal{C}_V}{d \tau} \geq \frac{8\pi }{d-1}\, (M-M_{min})\qquad{\rm for}\ \  k=-1\ .
\label{volume_bound_hyper}
\end{align}

\subsection{General Time Dependence}\label{sub_plot_CV}
To close this section,  we present plots of $d\mathcal{C}_V/d \tau$  for planar black holes in various dimensions for general values of the time. We explore further examples with spherical and hyperbolic horizon geometries in appendix \ref{app:MoreAction}.

\begin{figure}[htbp]
	\vspace{2ex}
	\centering
	\includegraphics[scale=0.7]{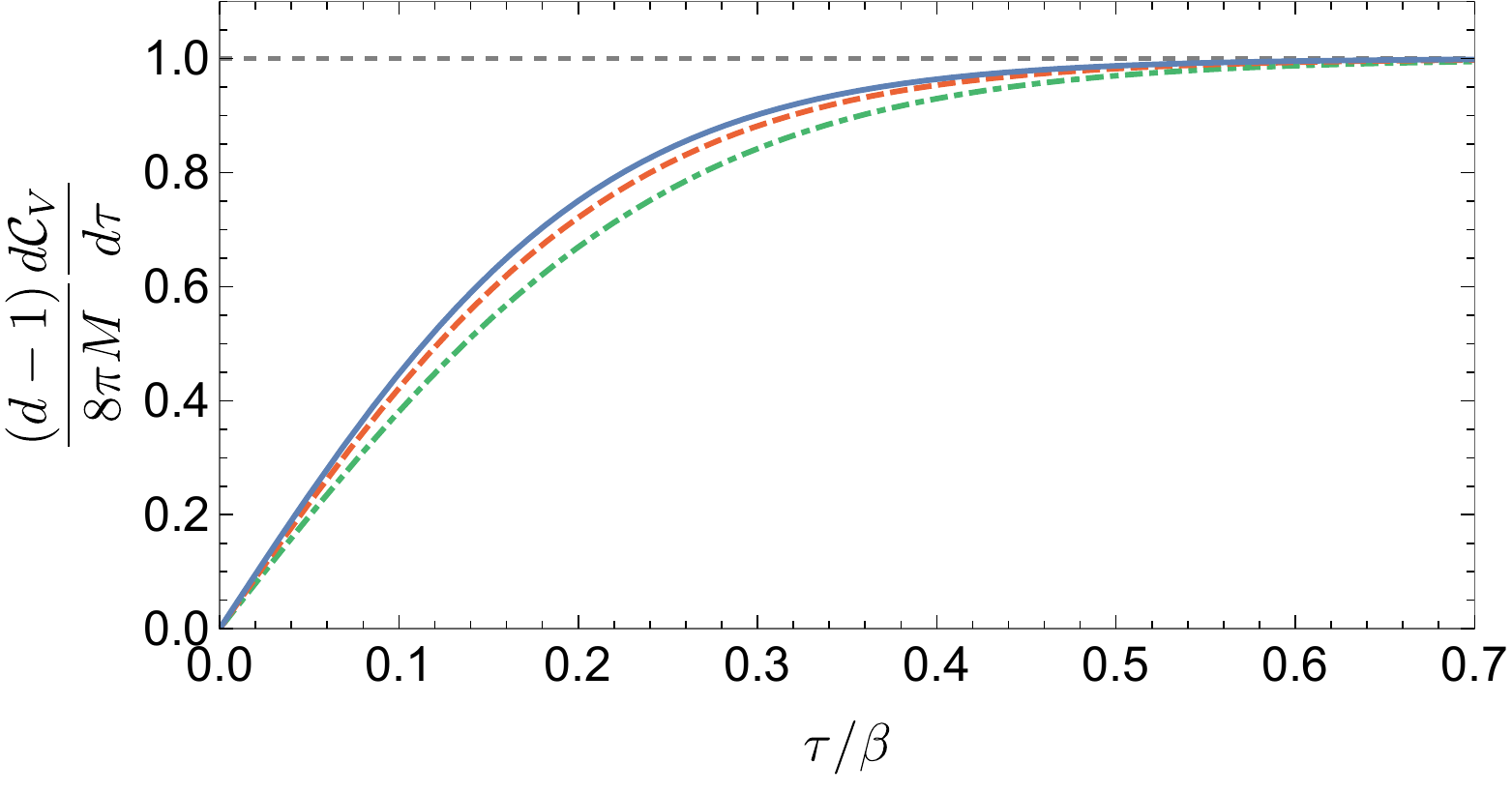}
	\caption{
		Plot of $\frac{d-1}{8\pi M }  \frac{d\mathcal{C}_V}{d\tau}$ for planar $d=4$ (blue), planar $d=3$ (dashed-red) and $d=2$ (dot-dashed green) black holes. All three curves are independent of $r_h/L$ and approach to one at late times.}
	\label{vol_rates}
\end{figure}

In the case that $k=0$ (and $d\ge3$),
if we define $a\equiv \frac{d-1}{8\pi M}  d\mathcal{C}_V/d \tau$,
eq. \eqref{dvdt_general} can be recast in the form
\begin{align}
a= 2 s_{min}^{d/2} \sqrt{1-s_{min}^{d}},
\quad (s_{min}\equiv r_{min}/r_h).
\end{align}
Inverting this equation, we can represent $s_{min}$ as a function of $a$,
\begin{align}
s_{min} = \Bigl(\frac{1+\sqrt{1-a^2}}{2}\Bigr)^{\frac{1}{d}} .
\label{sturn}
\end{align}
Then rewriting eq.~\eqref{t_r_E} in terms of dimensionless quantities,
one can find the relation between $a= \frac{d-1}{8\pi M}  d\mathcal{C}_V/d \tau$ and $\tau/\beta$
\begin{align}
\tau/\beta &= \frac{d\, a}{4 \pi} \int^{\infty}_{s_{min}} \!\! d s
\frac{s^{d-2}}{(1-s^d)\sqrt{s_{min}^{d}(1-s_{min}^{d})-s^{d}(1-s^{d})}}.
\end{align}
Since this relation and eq.~\eqref{sturn} do not depend on $r_h/L$, the plot of $a$ as a function of $\tau/\beta$ has the same form for all values $r_h/L$. Figure \ref{vol_rates} shows the plot and we see that at late times, it approaches to one from below, as discussed above in section \ref{LateX}.

Figure~\ref{vol_rates} shows $\frac{d-1}{8\pi M}  d\mathcal{C}_V/dt$ for the case of $d=2$, \ie BTZ black holes.
A similar derivation to the one presented for planar black holes holds in this case. Again, the result does not depend on the value of $r_h/L$ and approaches to one at late times.

\section{Charged Black Holes}\label{sec:ChargedEternal}

In this section, we study the growth rate of the complexity for charged black holes with $d\ge3$ using both the CA and CV conjectures.
Charged black holes are solutions to Einstein gravity coupled to a Maxwell field with the following action:
\begin{equation}
I = I_{\text{grav}} - \frac{1}{4 g^2} \int d^{d+1}x \sqrt{-g} \, F_{ab}\, F^{ab} \label{maxwell}
\end{equation}
where $I_{\text{grav}}$ is the gravitational action given in eq.~\eqref{THEEACTION}. Note that the gauge coupling $g$ has dimensions of length$^{\frac{d-3}{2}}$.

\begin{figure}
\centering
\includegraphics[scale=0.35]{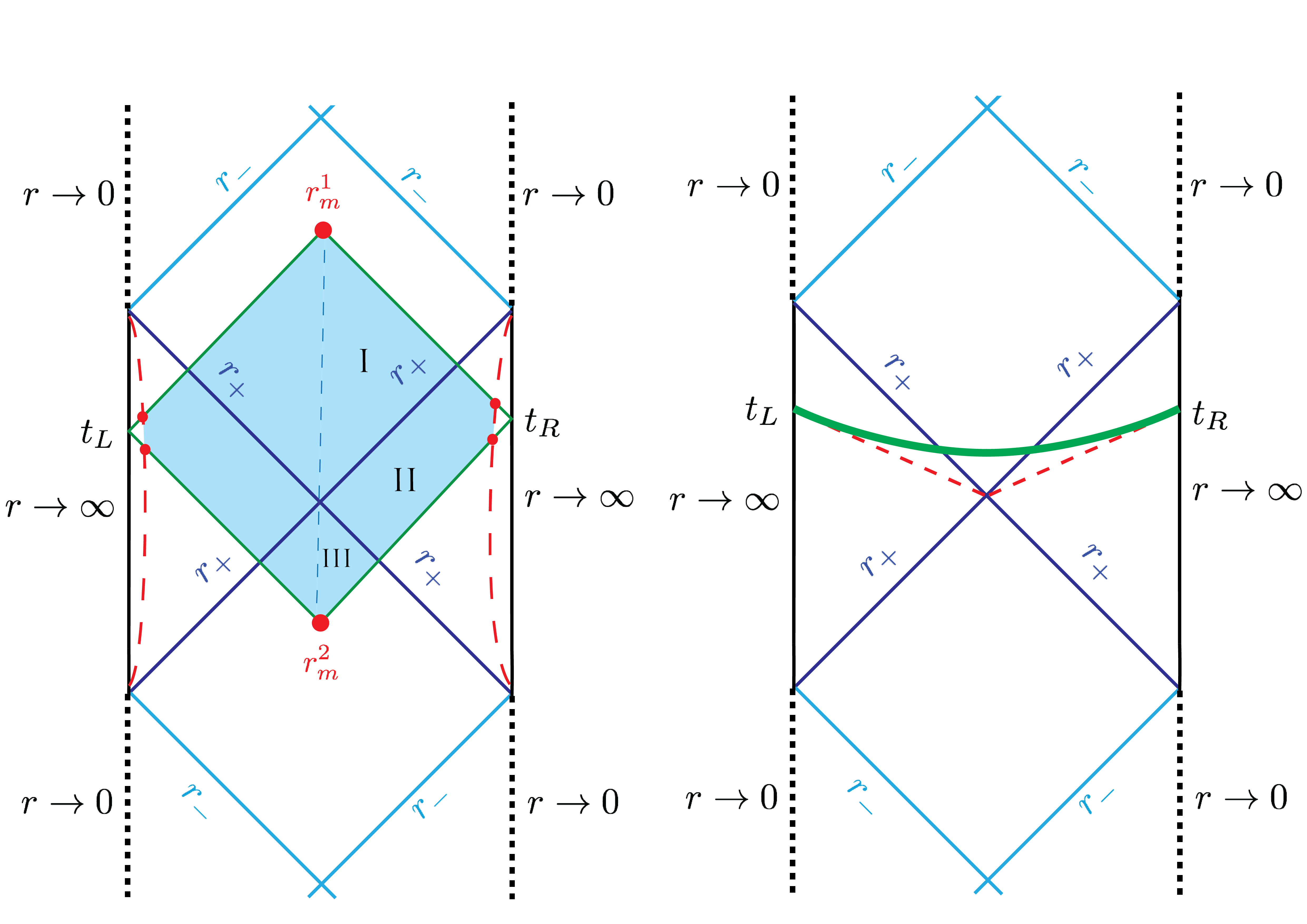}
\caption{Penrose diagrams for a charged black hole. On the left figure we breakdown the action calculation for the Wheeler-DeWitt patch. The future (past) corner approaches the inner (outer) horizon in the late time limit.
On the right, we identify the maximal volume that is evaluated in the CV proposal. As in section \ref{sec:EternalAction} we have for the case of a general boundary size $t= \frac{L}{R} \tau$. }
\label{fig:charged}
\end{figure}

The black hole metric takes the form \reef{metric2} with blackening factor given by, \eg \cite{Chamblin:1999tk,HartLectures}:\footnote{We work with the rescaled time $\tau=R\,t/L$ throughout the following.}
\begin{equation}\label{ChargedMetric}
f(r) = \frac{r^2}{L^2}+k-
\frac{\omega^{d-2}}{r^{d-2}} + \frac{q^2}{r^{2(d-2)}}\,,
\end{equation}
and the Maxwell potential can be written as:\footnote{Our conventions compare to those of \cite{Chamblin:1999tk} (denoted with tildes) as follows: $A_t = \tilde A_t \frac{g}{2 \sqrt{\pi G}}$,  $ Q = \tilde Q \frac{2 \sqrt{\pi G}}{g}$, $\mu = \tilde \mu \frac{g}{2 \sqrt{\pi G}}$; and to those of \cite{HartLectures} by the identification $1/g^2 = \ell^2/G_N$ where $\ell$ is an extra length scale introduced there to distinguish the coupling of the Maxwell field.}
\begin{equation}
A_\tau = \frac{g}{2\sqrt {2\pi G_N}}\,\frac{L}{R}\,\sqrt{\frac{d-1}{d-2}}\left(\frac{q}{r_+^{d-2}}-\frac{q}{r^{d-2}}\right)\,.
\label{potent}
\end{equation}
The new blackening factor \reef{ChargedMetric} has two real roots, $r_+$ and $r_-$ (where $r_+\ge r_-$) corresponding to the outer and inner horizons, respectively. Figure \ref{fig:charged} shows the Penrose diagrams for these charged black holes. We note that the integration constant in $A_\tau$ was chosen such that it vanishes at the outer horizon, which ensures that it is a well behaved differential form at the corresponding bifurcation surface \cite{Chamblin:1999tk}. It will typically be convenient to write our results in terms of $r_+$ and $r_-$ by expressing $\omega^{d-2}$ and $q^2$ in terms of $r_+$ and $r_-$ using the equations $f(r_+)=f(r_-)=0$ --- see below.

Of course, the Maxwell field in the bulk is dual to a conserved current corresponding to a global $U(1)$ symmetry in the boundary theory \eg \cite{HartLectures}. Hence the charged black hole geometry extends the thermofield double state \reef{TFDx} to the entangled state where, as well as a temperature $T$, we have a chemical potential $\mu$ which distinguishes the boundary states by their $U(1)$ charges. We will refer to this as the charged thermofield double state,
\begin{equation}
\bigl| \text{cTFD} (t_L,t_R)\bigl> = Z^{-1/2} \sum_{\alpha,\sigma} e^{-(E_{\alpha}-\mu\, Q_\sigma)/(2 T)}\,e^{-iE_\alpha(t_L+t_R)}  \,\bigl| E_{\alpha}, -Q_\sigma \bigl>_{L} \bigl| E_{\alpha}, Q_\sigma \bigl>_{R} \, ,
\label{TFDq}
\end{equation}
where $L$ and $R$ label the quantum states (and times) at the left and right boundaries. Notice that tracing out the states in either boundary produces the density matrix corresponding to the grand canonical ensemble characterized by $T$ and $\mu$ --- see further discussion below.

The thermodynamic quantities describing the black hole are the same as those given in eq.~\reef{there} with the replacement $r_h\to r_+$\, \ie
\beq
M = \frac{(d-1) \, \Omega_{k,d-1}}{16 \pi \, G_N}\,\frac{L}R \,\omega^{d-2} \,,\quad
S =  \frac{\Omega_{k,d-1} }{4 G_N}\,r_+^{d-1}\,,
\quad
T=\frac{L}{R}\frac{1}{4\pi }\left.\frac{\partial f}{\partial r}\right|_{r=r_+}\,.
\label{charge22}
\eeq
The charge is naturally defined in terms of Gauss' law, \ie
\beq\label{Charge11}
Q=\oint *F =\frac{q \,  	\Omega_{k,d-1}\sqrt{(d-1)(d-2)}}{2 g \sqrt{2\pi G_N}}
\eeq
where the ($d$--1)-form $*F$ is the Hodge dual of the field strength $F_{ab}=\partial_a A_b-\partial_b A_a$. Of course, the Maxwell field in the bulk is dual to a global
symmetry current in the boundary theory.\footnote{The current can be defined by varying the boundary action with respect to the gauge field, \eg \cite{HartLectures}.}  In this holographic context, the charge \reef{Charge11} also corresponds to the integral of the zeroth component of the boundary current over a constant $\tau$ slice. The chemical potential can be determined using the thermodynamic relation $dM = TdS + \mu\, dQ$,
\beq
\mu = \frac{g}{2\sqrt {2\pi G_N}}\,\frac{L}{R}\,\sqrt{\frac{d-1}{d-2}}\,\frac{q}{r_+^{d-2}}\,.
\eeq
Comparing to eq.~\reef{potent}, this also corresponds to the `non-normalizable' mode of the gauge potential, \ie
$\mu=\lim_{r\to\infty}A_\tau$.

We note that the action \reef{maxwell} provides a well defined variational principle where we keep the gauge potential  fixed at the boundary. Hence if we were examining the thermodynamics of these black holes, \eg with the corresponding Euclidean action, then we would be working with the grand canonical ensemble where the chemical potential $\mu$ is fixed. That is, implicitly, our control parameters are the temperature $T$ and the chemical potential $\mu$ \cite{Chamblin:1999tk,HartLectures}. Hence the full geometry of the eternal charged black hole is dual to the charged thermofield double state, given in eq.~\reef{TFDq}.
Alternatively, we could consider a fixed charge ensemble, but this would require adding a boundary term of the form $1/g^2 \int_{\del \mathcal{M}} d^d x \sqrt{\gamma} n^a F_{ab} A^b$ to the action. It would be interesting to pursue this possibility in the context of the complexity=action proposal, where it seems that we would need to include this boundary term on all of the boundaries of the WDW patch.

In order to express our results for the complexity in terms of boundary quantities, it will be useful to also have holographic expressions for the central charges associated with the two-point functions of the boundary stress tensor (\eg \cite{erd,pet,RobMisha}) and currents (\eg \cite{FreedmanMathur, Barnes:2005bw}). That is, for a $d$-dimensional CFT, the leading singularities in the vacuum correlators take the form:
\beq
\langle T_{\mu \nu}(x) T_{\rho\sigma}(0) \rangle = \frac{C_T}{ x^{2d}}\,{\cal I}_{ab,cd}\,,\qquad
\langle J_{\mu}(x) J_{\nu}(0) \rangle  = \, \frac{C_{J}}{x^{2(d-1)}} \,I_{\mu\nu}(x)
\label{correl}
\eeq
where
\beq
{\cal I}_{ab,cd}\equiv
\frac{1}{2}
\left( I_{\mu\nu}(x) I_{\rho\sigma}(x)+ I_{\mu\sigma}(x) I_{\nu\rho(x)} \right)
-\frac{1}{d} \eta_{\mu\nu} \eta_{\rho\sigma}
\,,\qquad
I_{\mu\nu } \equiv \eta_{\mu\nu}-2 \frac{x_\mu x_\nu}{x^2}\,.
\label{correl2}
\eeq
For our holographic framework, the two central charges can then be expressed in terms of bulk parameters as
\beq\label{cjct22}
 C_T=\frac{d+1}{d-1}\, \frac{  \Gamma (d+1)}{8    \pi ^{(d+2)/2}\, \Gamma \left({d}/{2}\right)}\,\frac{L^{d-1}}{G_N}\,,\qquad
C_J=\frac{ (d-2) \Gamma (d)}{2 \pi ^{d/2} \Gamma \left({d}/{2}\right)}\,\frac{L^{d-3}}{g^2} \,.
\eeq

It will be convenient to work in terms of the following dimensionless quantities:
\begin{equation}\label{eq:xyz}
x\equiv\frac{r}{r_+}\,, \qquad
y\equiv\frac{r_-}{r_+}\,, \qquad
z\equiv\frac{L}{r_+}\,.
\end{equation}
Here, $x$ is a dimensionless radial coordinate, while $y$ and $z$ can be expressed in terms of dimensionless boundary quantities. In particular, combining the expressions above yields
\begin{equation}\label{eq:RTnu}
\nu \equiv \sqrt{\frac{C_J}{C_T}}\frac{\mu}{T} = h(y,z)\,, \qquad
RT = \tilde h(y,z)\,.
\end{equation}
Of course, these equations can be inverted and so one can think directly of $y$ and $z$ as boundary quantities.
As we will see, all our result can be expressed as functions of $\nu$ and $RT$, or alternatively of $y$ and $z$. Explicit expressions for $h(y,z)$ and $\tilde h(y,z)$ for the different dimensions and geometries read
 \small
\begin{align}
&h(y,z) = \frac{2 \sqrt{2} \pi  (d-1) \left(y^{\frac{d}{2}-1} \sqrt{1-y^{d-2}} \sqrt{\left(k z^2+1\right)-y^{d-2} \left(k z^2+y^2\right)}\right)}{\sqrt{d (d+1)} \left( (d-2) k z^2+d -2 y^{d-2} \left((d-2) k z^2+d-1\right)+(d-2) y^{2 (d-2)} \left(k z^2+y^2\right) \right)} \, , \nonumber \\
& \nonumber \\
&\tilde h(y,z) = \frac{ (d-2) k z^2+d  -2 y^{d-2} \left((d-2) k z^2+d-1\right)+(d-2) y^{2 (d-2)} \left(k z^2+y^2\right) }{4 \pi  z \left(1-y^{d-2}\right)} \, .
\label{eq:RTnuGenerald}
\end{align}
\normalsize
It is instructive to expand these functions in the small charge limit (\ie small $y$) where one obtains
\small
\begin{align}
\begin{split}
h(y,z)  = &\frac{2 \sqrt{2} \pi  (d-1) \sqrt{1+k z^2} }{\sqrt{d (d+1)} \left(d+(d-2) k z^2\right)} y^{\frac{d}{2}-1} \times
\\
& \times\left[ 1 + \left(1+ \frac{1}{2} \frac{1}{(1+ k z^2)} - \frac{2}{d + (d-2) k z^2} \right) y^{d-2} + \mathcal{O}\left(y^d\right) \right] \\
\\
\tilde h(y,z)  = &\frac{d + (d-2) k z^2}{4 \pi z} - \frac{(1 + k z^2)}{4 \pi z} (d-2) y^{d-2} + \mathcal{O}\left(y^{2 (d-2)}\right)  \, .
\end{split} \label{eq:RTnuSmally}
\end{align}
\normalsize
As expected, the dimensionless quantity $\nu$ goes to zero and $T R$ to the uncharged limit as in eq.~\eqref{there}. From the expansions in eq.~\eqref{eq:RTnuSmally}, we can also conclude that the chemical potential, $\sqrt{\frac{C_J}{C_T}} \mu R = \tilde h(y,z) h(y,z)$ scales as $\propto y^{\frac{d-2}{2}}$ for small charges. Similarly, the blackening factor can be expressed as $f(x,y,z)$ where $x$ was defined in eq.~\eqref{eq:xyz}.

\paragraph{Complexity of Formation:}
The complexity of formation for uncharged black holes was examined in detail in \cite{Formation}. Hence
for completeness, we also examine the `complexity of formation' of charged black holes here and the corresponding calculations are described in detail in appendix \ref{CformCharged}. The  question of interest is what is the additional complexity involved in preparing the two copies of the boundary CFT in the charged entangled thermofield double state \reef{TFDq} compared to preparing each of the CFTs separately in their vacuum state. Using the CA proposal,\footnote{Of course, an analogous calculation can also be performed using the CV proposal, see appendix \ref{CformCharged}.} the bulk calculation consists of evaluating the gravitational action for the WDW patch (anchored at $t_L=t_R=0$) in the charged AdS black hole background and subtracting twice the action for the WDW patch in empty AdS space (\ie $\omega=q=0$). A key feature of this subtraction is that all of the UV (large $r$) divergences cancel leaving a UV-finite result.

We discuss here the charged complexity of formation using the CA conjecture for the planar case, \ie $k =0$, for $d=4$. For small chemical potential, the charged complexity of formation can be written as a series expansion for small $y$,
\begin{equation}
\label{eq:SeriesCoFPlan}
\Delta \mathcal{C}_A = \frac{S}{2 \pi}\left(1 + \left( \frac{20}{3 \pi} +\frac{4}{\pi} \log\!\left[ \frac{y z}{2} \frac{\alpha R}{L}\right] \right) y^3
 + \cdots\right) \, ,
 \end{equation}
where $S$ is the thermal entropy. Of course, we recover the $d=4$ planar result found in \cite{Formation} in the limit of vanishing chemical potential, \ie $y \rightarrow 0$.
We can rewrite the above expression without the explicit $z R$ dependence, using the $k=0$ and $d=4$ instances of eq.~\eqref{eq:RTnuGenerald}, which reads
\begin{equation}\label{eq:RTnuPlanAdS5}
\nu = \frac{3 \pi}{\sqrt{10}} \frac{y \sqrt{1+y^2}}{(2- y^2 -y^4)}  \,, \qquad
T R = \frac{(1-y^2) (2+y^2)}{2 \pi z} \, .
\end{equation}
The expansion of the complexity of formation then becomes
\begin{equation}\label{eq:SeriesCoFPlanWithT}
\Delta \mathcal{C}_A = \frac{S}{2 \pi}\left(1 + \frac{10^{3/2}}{(3\pi)^4}\left( 20 +12 \log \left[ \frac{10^{1/2} }{3 \pi^2} \frac{\alpha\,\nu}{ L\, T } \right] \right) \nu^3 + \cdots \right) \, .
\end{equation}
As in section \ref{sec:EternalAction}, we might simplify the above expression by choosing the normalization of the null normals at infinity to be $\alpha = L/R$, where $R$ is to be interpreted not as the curvature scale, but instead as an arbitrary reference length scale in the boundary theory (for $k=0$).

\begin{figure}
\centering
\includegraphics[scale=0.8]{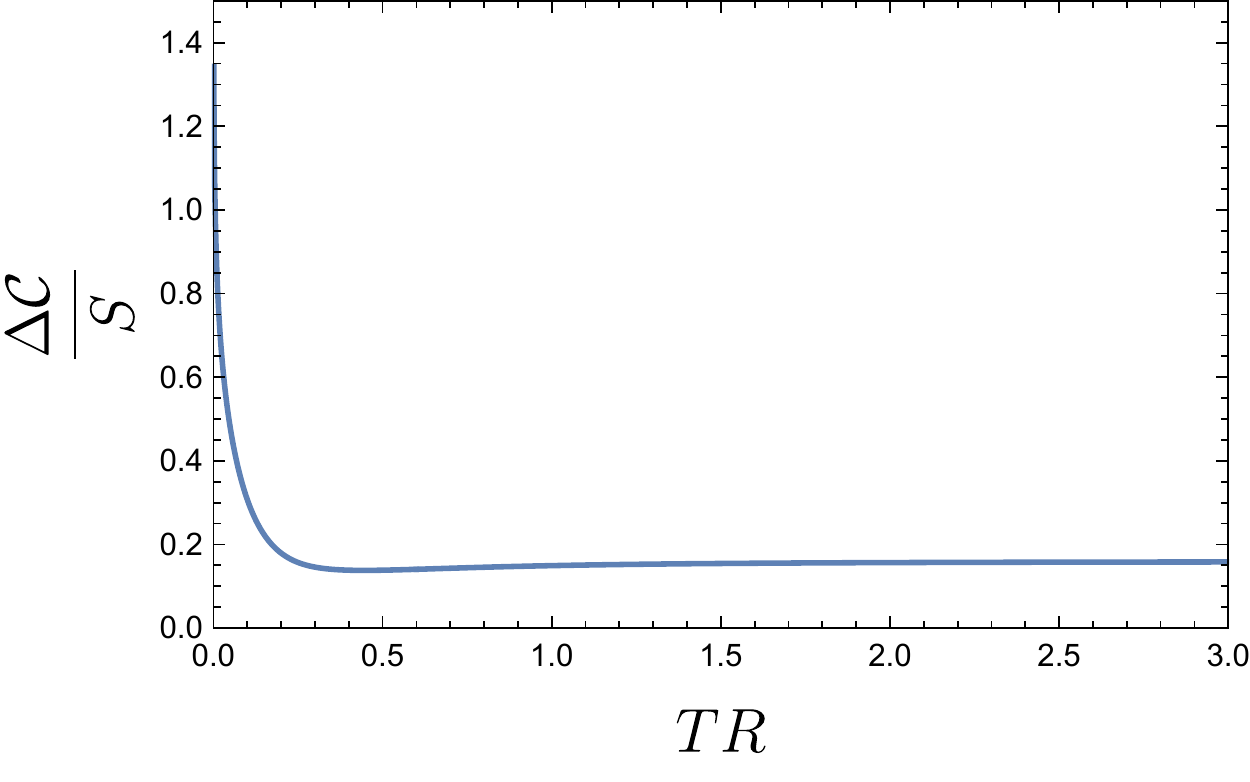}
\caption{Complexity of formation divided by the entropy for the planar charged black hole in $d=4$. Here we are subtracting the complexity of two copies of the vacuum spacetime (\ie the zero mass and zero charge limit of the planar black hole). In this plot, we keep the chemical potential fixed as $\sqrt{\frac{C_{J}}{C_{T}}} \mu R = \frac{1}{2}$. For a fixed chemical potential in the limit of zero temperature (dual to extremal black hole) the complexity of formation is divergent.}
\label{fig:CFixedMu}
\end{figure}

We also use the boundary quantities from eq.~\eqref{eq:RTnuPlanAdS5} to evaluate numerically the complexity of formation fixing the chemical potential and varying the temperature in figure \ref{fig:CFixedMu}. There is an unexpected behaviour when the temperature is very small, as the complexity of formation grows unbounded. The fact that the complexity of formation for extremal black holes of finite chemical potential is divergent suggests that the proposed ground state for large charged black holes in \cite{Brown2} should be revisited. It is also interesting to notice that in this limit of zero temperature with a fixed chemical potential, $d\mathcal{C}_A/d\tau$ goes to zero \cite{Brown2}, as we will show in the following subsection.
We will explore further some features of the charged complexity of formation in appendix \ref{CformCharged}.

\subsection{Complexity=Action}\label{subsec:ChargedEternal}

Next, we examine the time evolution of holographic complexity using the CA proposal for the eternal charged AdS black holes.
The integrand of the bulk action is given by\footnote{To simplify this expression, we have used the trace of Einstein equations, which yields $\mathcal{R} = -\frac{d(d+1)}{L^2} + \frac{d-3}{d-1} \frac{4\pi G_N}{g^2} F_{ab} F^{ab}$.}
\begin{equation}\label{eq:gofr}
I(r)\equiv \frac{1}{16\pi G_N }\left(\mathcal{R}-2 \Lambda\right)- \frac{1}{4  g^2} F_{ab} F^{ab}=
\frac{1}{16\pi  G_N}\left(-\frac{2d}{L^2} +\frac{2  (d-2)q^2}{r^{2(d-1)}}\right)\,.
\end{equation}
We then write the bulk action as
\begin{equation}
I_{\bulk} = \frac{L}R\,\Omega_{k,d-1}\,\int dr\, r^{d-1}\, I(r) \ \int  d\tau
\end{equation}
where we still have to specify the limits of integration.
In particular, we need to find the {future ($r_m^1$)}  and { past ($r_m^2$)}  meeting points of the null sheets bounding the WDW patch --- see figure \ref{fig:charged}.
These satisfy the following relations
\begin{equation}\label{eq:meeting345}
\frac{L}{R}\,\frac{\tau}{2}+ r^*_{\infty} - r^*(r_m^{1})=0,
\qquad
\frac{L}{R}\,\frac{\tau}{2}-r^*_{\infty}+r^*(r_m^{2})=0.
\end{equation}
Note that taking the time derivative of these relations yields:
\begin{equation}\label{eq:meeting3456}
\frac{R}{L}\,\frac{d r_m^{1}}{d \tau} = \frac{f(r_m^{1})}{2},
\qquad
\frac{R}{L}\,\frac{d r_m^{2}}{d \tau} = - \frac{f(r_m^{2})}{2}.
\end{equation}
We again divide the bulk contribution into three separate regions
\begin{equation}
\begin{split}
I_{\bulk}^{\text{I}}= &\,  \, 2 \Omega_{k,d-1}
\int_{r_m^{1}}^{r_+} I(r)r^{d-1} \left(\frac{\tau }{2}+\frac{R}{L}\left(r^*_{\infty}-r^*(r)\right)\right) d r
\\
I_{\bulk}^{\text{II}}= &\,
4\Omega_{k,d-1}
 \int_{r_+}^{r_\text{max}} I(r)  r^{d-1}\, \frac{R}{L}(r^*_{\infty}-r^*(r)) dr
\\
I_{\bulk}^{\text{III}}= &\,
2 \Omega_{k,d-1} \int_{r_m^{2}}^{r_+} I(r)  r^{d-1}
\left(-\frac{\tau }{2}+\frac{R}{L}\left(r^*_{\infty}-r^*(r)\right)\right) dr \,.
\end{split}
\end{equation}
Differentiating with respect to $\tau$ we see once again (as in the neutral case) that the contributions due to differentiating the limits of integration vanish using eq.~\eqref{eq:meeting345}. The contribution outside the black hole (region II) is independent of time.\footnote{This results from the boost invariance of the exterior geometry, as noted in \cite{Brown1,Brown2}.} Hence the only nonvanishing contribution comes from differentiating inside the integrals and we obtain
\begin{equation}
\frac{d I_{\bulk}}{d \tau} =  \frac{L}R\,\Omega_{k,d-1}  \int_{r_m^{1}}^{r_m^{2}} r^{d-1} I(r) dr =
\frac{L}{R} \frac{\, \Omega_{k,d-1}}{8\pi G_N}
\left[\frac{r^d}{L^2}+\frac{q^2}{r^{d-2}}\right]
 \Bigg|^{r_m^{1}}_{r_m^{2}}.
\end{equation}
There are no contributions to $d \mathcal{C}_A/d \tau$ from the surface terms or from the asymptotic boundaries here, but we do expect the two joints (at $r= r_m^1$ and $r_m^2$) to contribute:
\begin{equation}
I_{\corner} = -\frac{\Omega_{k,d-1}} {8 \pi G_N}
\left[
(r_{m}^{1})^{d-1}  \log \left[\frac{L^2|f(r_{m}^{1})|}{R^2 \alpha^2}\right] +
(r_{m}^{2})^{d-1} \log \left[\frac{L^2|f(r_{m}^{2})|}{R^2 \alpha^2}\right] \right].
\end{equation}
Differentiating the corner contribution with respect to $\tau$ then gives
\begin{equation}
\frac{d I_{\corner}}{d \tau} =
-\frac{L}{R}\frac{\Omega_{k,d-1}} {16 \pi G_N}
\left[
(d-1)r ^{d-2} f(r) \log \frac{L^2|f(r)| }{R^2\alpha^2}
+r^{d-1} \del_r f(r)
\right]\Bigg|^{r_m^{1}}_{r_m^{2}}\,,
\end{equation}
where we used eq.~\eqref{eq:meeting345}.  Combining the nonvanishing contributions together leads to
\begin{equation}\label{chargedEqAmazing}
\frac{d \mathcal{C}_A}{d \tau}=
\frac{L}{R}\frac{\Omega_{k,d-1} (d-1) } {8 \pi^2 G_N }
\frac{q^2}{r^{d-2}}\Bigg|^{r_m^{1}}_{r_m^{2}}
-\frac{L}{R}\frac{\Omega_{k,d-1}(d-1)} {16 \pi^2 G_N}
r ^{d-2} f(r) \log \frac{L^2|f(r)| }{R^2\alpha^2}
\Bigg|^{r_m^{1}}_{r_m^{2}}.
\end{equation}
As a consistency check, we note that in the late time limit, we  recover eq.~(3.39) of \cite{RobLuis}:
\begin{equation}\label{ChargedActionLateTime}
\lim_{\tau\rightarrow\infty} \frac{d \mathcal{C}_A}{d \tau}=\frac{\Omega_{k,d-1} (d-1) q^2} {8 \pi^2 G_N }\frac{L}{R}
\frac{1}{r^{d-2}}\Bigg|^{r_-}_{r_+},
\end{equation}
where we have used that $r^1_m\to r_-$ and $r^2_m\to r_+$ in this limit. It is also possible to express this late time rate of change using the black hole mass and the dimensionless quantities from eq.~\eqref{eq:xyz} as
\begin{equation}\label{ChargedActionLateTimeDiff}
\lim_{\tau\rightarrow\infty} \frac{d \mathcal{C}_A}{d \tau}= \frac{2 M}{\pi} \left( \frac{(1- y^{d-2})( (1 - y^d) + k z^2 (1 - y^{d-2}) ) }{(1 - y^{2(d-1)}) + k z^2 (1 - y^{2(d-2)}) } \right)  \, .
\end{equation}
In these variables, the late time limit of the uncharged case is easily obtained with $y \rightarrow 0$.

\begin{figure}
\centering
\includegraphics[scale=0.5]{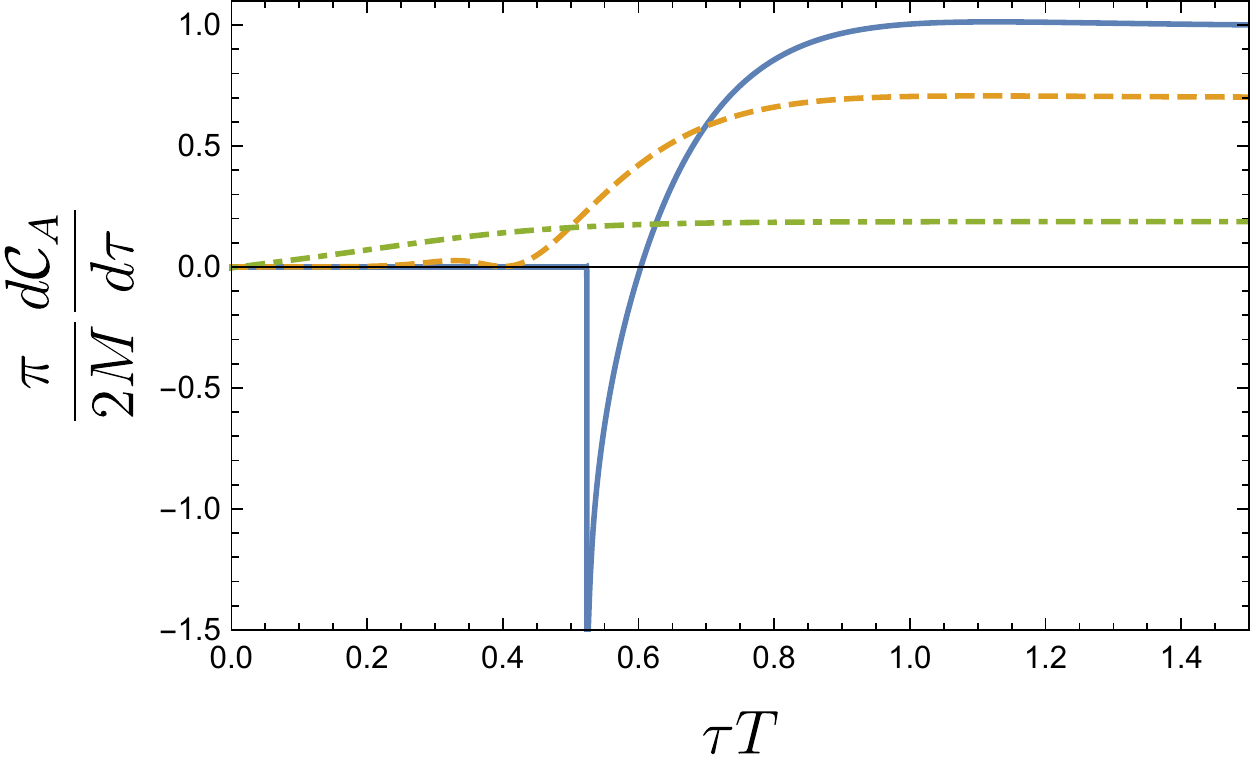}~~
\includegraphics[scale=0.5]{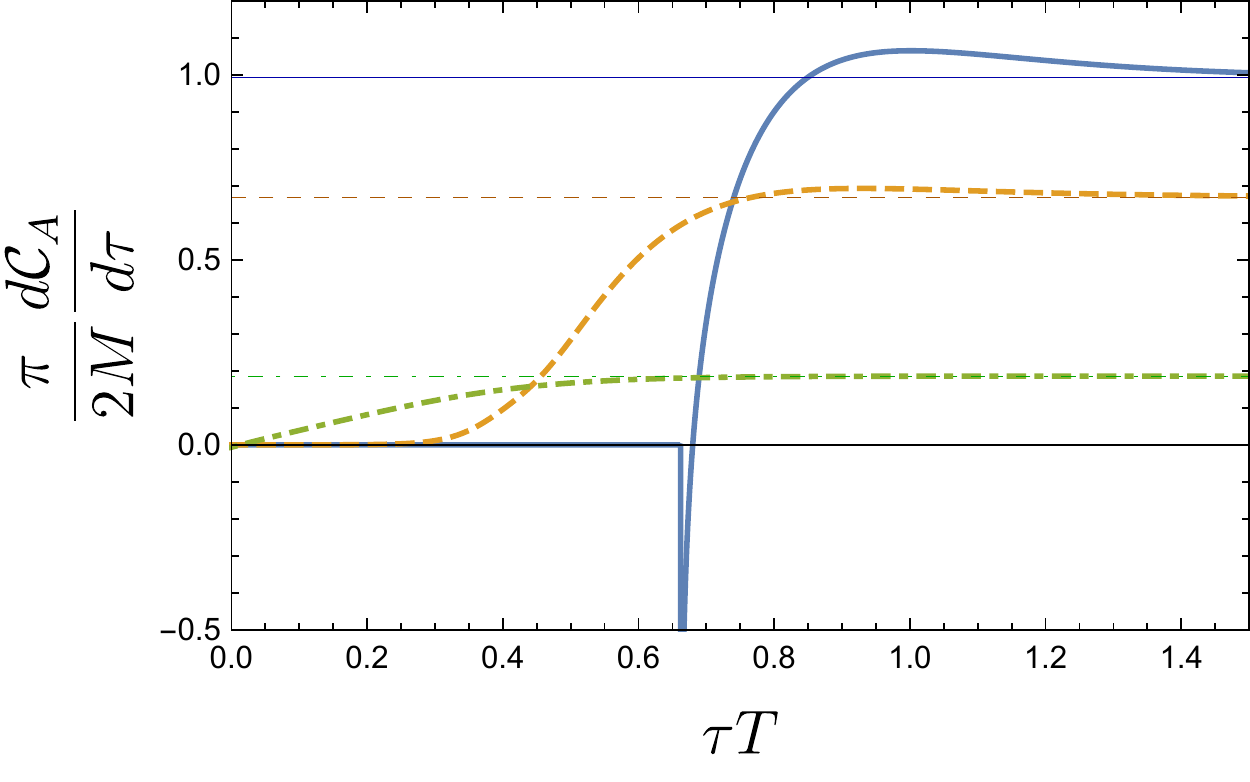}~~
\caption{The time derivative of complexity with $d=4$, $k=1$ and non-zero chemical potential, obtained by fixing the parameters in eq.~\eqref{eq:RTnuSphAdS5}. The various curves correspond to: $\nu = 0.1$ in blue (solid) , $\nu =1$ in orange (dashed) and $\nu=5$ in green (dot-dashed) for $T R =1$ (Left) and $T R = \frac{1}{2}$ (Right). In order to illustrate the violation of the bound, we explicitly show the late time limit from eq.~\eqref{ChargedActionLateTimeDiff} in the right figure.}
\label{fig:charged1}
\end{figure}

\begin{figure}
\centering
\includegraphics[scale=0.7]{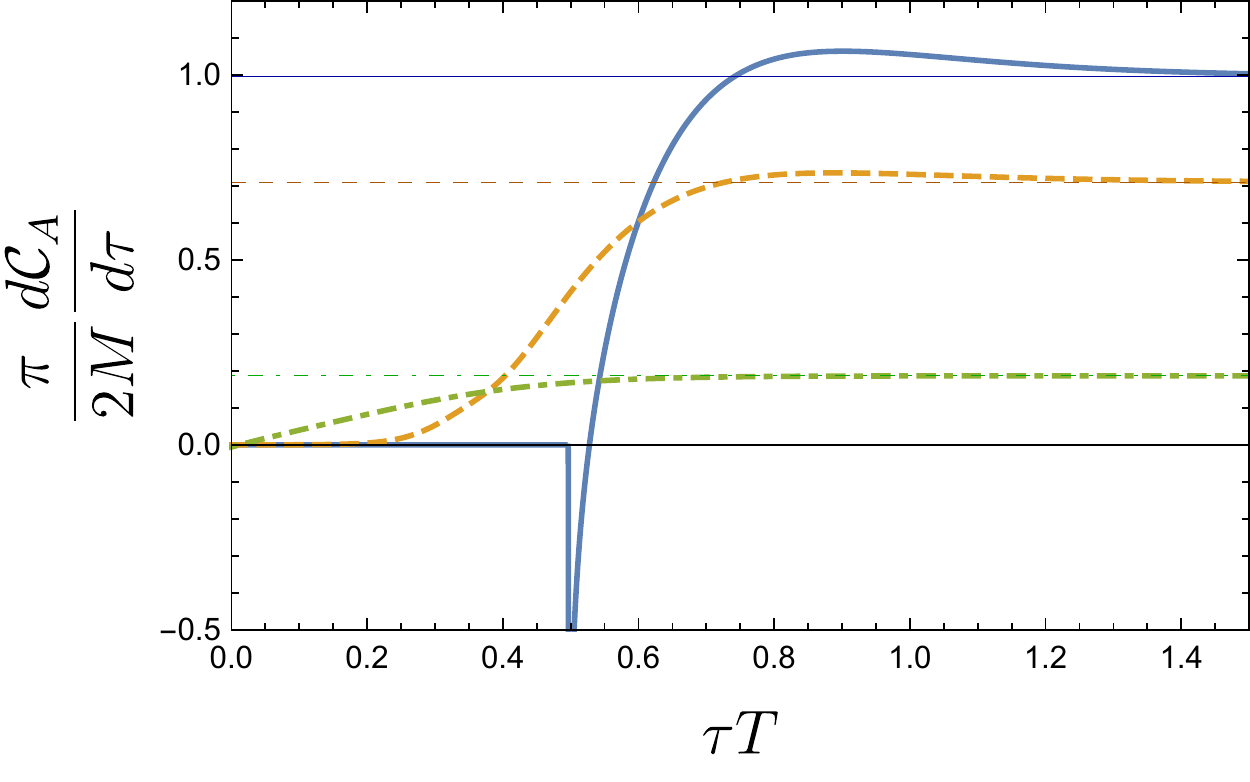}
\caption{The time derivative of complexity with $d=4$, $k=0$ and non-zero chemical potential, obtained by fixing the parameters in eq.~\eqref{eq:RTnuSphAdS5}. The various curves correspond to: $\nu = 0.1$ in blue (solid) , $\nu =1$ in orange (dashed)  and $\nu =5$ in green (dot-dashed). We varied the chemical potential while fixing the temperature as $T R = \frac{1}{2}$, where as before the scale $R$ in the planar geometry is related to an arbitrary scale in the boundary theory. }
\label{fig:charged12}
\end{figure}

Now it is straightforward to solve for the two meeting points numerically using eq.~\eqref{eq:meeting345} and then to evaluate the rate of change in complexity \reef{chargedEqAmazing}. To illustrate these results, we show $d \mathcal{C}_A/d \tau$ for $d=4$ in figures \ref{fig:charged1} and \ref{fig:charged12}.\footnote{As before, we set $\alpha=L/R$ for simplicity.} For these black holes, the boundary quantities $\nu$ and $RT$ in eq.~\eqref{eq:RTnu} can be obtained from the ratios $y$ and $z$ as
\begin{equation}\label{eq:RTnuSphAdS5}
\nu = \sqrt{\frac{C_J}{C_T}}\frac{\mu}{T} = \frac{3 \pi}{\sqrt{10}} \frac{y \sqrt{1+ y^2 +k z^2}}{(1-y^2)(2 + y^2 + k z^2)}\,, \qquad
RT = \frac{1}{2 \pi} \frac{(1-y^2) (2+ y^2 + k z^2)}{z}\,.
\end{equation}
In the figures, the rate of change in complexity is presented for fixed values of these boundary quantities.

\subsubsection{Comments}\label{subSec:CommentsCharged}

Let us make a number of observations about these results for the charged black holes. First, we note that in both figures, for very small charge (or small chemical potential), the rate of change in complexity develops a minimum at some finite time. This minimum becomes deeper and sharper for smaller charges, and so the behaviour smoothly approaches that of the neutral black holes ($\nu=0$), shown in figure \ref{d4Rate}. In particular, the pronounced minimum in $d \mathcal{C}_A/d \tau$ is centered around the neutral $\tau_{c}$, and its shape resembles closely the negative divergent rate of change observed right after $\tau_c$ in the neutral case, and as noted above, the late time limit approaches  $2M/\pi$, as expected for neutral AdS black holes.\footnote{In fact, one can easily show that eq.~\reef{tder44} is recovered in the zero charge limit analytically. The key observation is that $r_-$ vanishes as $r_-^{d-2} = q^2/\omega^{d-2}$ in this limit. Along with $r_m^1 \sim r_- $ and $r_m^2\simeq r^{\ssc (neutral)}_m$, eq.~\eqref{chargedEqAmazing} reduces to the neutral growth rate \reef{tder44} for $\tau>\tau_c$. We consider the early time behaviour in the zero charge limit below.}

Next, we might consider the extremal limit of the charged black holes where $T\to0$. It is straightforward to show $d \mathcal{C}_A/d \tau\simeq0$ in this limit. For example, from  eq.~\reef{eq:RTnuSphAdS5}, we see that this limit corresponds to $y\to1$ and this certainly produces a vanishing rate of change for the late time limit in eq.~\eqref{ChargedActionLateTimeDiff}. More generally, this limit corresponds to $r_-\to r_+$ and we find $r_m^1 \sim r_m^2$. The latter then produces a cancellation and vanishing  $d \mathcal{C}_A/d \tau\simeq0$ in eq.~\eqref{chargedEqAmazing}.

\paragraph{Late time expansion:}
In a very similar manner to the analysis of the late time limit in section \ref{tary}, we can obtain the late time limit of the growth rate of the holographic complexity for charged black holes. First, we decompose the inverse blackening factor as
\begin{equation}
\begin{split}
\frac{1}{f(r)} = \frac{1}{r_{+}-r_{-}}
\left(\frac{r_+}{F(r_+)r (r-r_+)}
-\frac{r_-}{F(r_-)r (r-r_-)} + H (r)
 \right)
\end{split}
\end{equation}
where we have defined:
\begin{equation}
f(r) \equiv F(r) (r-r_+)(r-r_-)
\end{equation}
and $F(r)$ is a strictly positive function. Further, we have defined
\begin{equation}
H (r) = \frac{F(r_+) r - F(r) r_+}{F(r_+)F(r) r (r-r_+)}
- \frac{F(r_-) r - F(r) r_-}{F(r_-)F(r) r (r-r_-)}\,,
\end{equation}
which is regular both at $r_+$ and at $r_-$ and decays at least as fast as $1/r^2$ when $r$ approaches infinity.
This leads to the tortoise coordinate:
\begin{equation}
r^*(r) =
\frac{\log \left(|r-r_+|/r \right)}{F(r_+)(r_{+}-r_{-})}
-\frac{\log \left(|r-r_-|/r \right)}{F(r_-)(r_{+}-r_{-})}
+ \frac{1}{r_+-r_-} \int^r H(\tilde r) d\tilde r.
\end{equation}
We have left the lower limit in the last integral implicit, as this choice does not influence the subtractions involved in the equations determining the meeting points.
Solving for the first subleading order in the late time limit of eq.~\eqref{eq:meeting345}, we obtain
\begin{equation}
r_m^1 = r_- \left(1+ c_-
e^{ -\frac{F(r_-)(r_+-r_-) }{2}\frac{L}{R}\tau}
\right), \qquad
r_m^2 = r_+ \left(1- c_+
e^{ -\frac{F(r_+)(r_+-r_-) }{2}\frac{L}{R}\tau}
\right)
\end{equation}
where $c_{+}$ and $c_{-}$ are positive constants given by
\begin{equation}
c_- = \left(\frac{r_+-r_-}{r_-}\right)^{\frac{F(r_-)}{F(r_+)}}e^{- F(r_-)\int_{r_-}^\infty H(\tilde r) d\tilde r} , \quad
c_+ = \left(\frac{r_+-r_-}{r_+}\right)^{\frac{F(r_+)}{F(r_-)}}e^{ F(r_+) \int_{r_+}^\infty H(\tilde r) d\tilde r}.
\end{equation}
From eq.~\eqref{chargedEqAmazing}, we can now demonstrate that
\begin{equation}
\begin{split}
& \frac{d \mathcal{C}_A}{d \tau} = \lim_{\tau \rightarrow\infty} \frac{d \mathcal{C}_A}{d \tau}  +
\frac{(r_+-r_{-})^2}{2}\frac{L^2}{R^2} \frac{\Omega_{d-1}(d-1)}{16 \pi^2 G_N }\tau
\\
&~~~~~~~~~~~~~~~\times  \left(
 c_{+} r_{+}^{d-1}F(r_+)^2  e^{-\frac{F(r_{+}) (r_+-r_-)}{2} \frac{L}{R}\tau }
 - c_{-} r_{-}^{d-1}F(r_-)^2 e^{-\frac{F(r_{-}) (r_+-r_-)}{2} \frac{L}{R}\tau} \right)
\end{split}
\end{equation}
where we have neglected terms that decay exponentially compared to those that decay as $\tau$ times an exponential above.
At very late times the exponent with smaller coefficient will dominate and will determine whether the limit is reached from above or from below.
We have checked the ratio $F(r_+)/F(r_-)=-f'(r_+)/f'(r_-)$ for a variety of dimensions and geometries and found that it is in general positive and smaller than one. As a consequence, $d \mathcal{C}_A/d \tau$ generally approaches the late time limit from above.

\paragraph{Early time behaviour:} We note that for the charged black holes, there is not a critical time before which the time derivative of the complexity is equal to zero. In the charged black hole, the past and future oriented joint terms (see the left panel in figure \ref{fig:charged}) start moving right away. However, we will show that for a small chemical potential, the time derivative of the complexity is exponentially suppressed at early times. In order to investigate this behaviour, we investigate the early time regime of the rate of change of complexity in an analytic expansion for small charges. To complete the picture, we also consider in this section the early time behaviour of the rate of change of complexity for near extremal black holes.

As we have already mentioned at the beginning of this subsection, in the limit in which the charge is small, the action does not change much for a certain period of time after $\tau=0$. In this situation, the future and past corner points (\ie $r_m^1$ and $r_m^2$ respectively, or $x_m^1$ and $x_m^2$ in terms of the dimensionless coordinate $x=r/r_+$) are exponentially close to the inner horizon $r_{-}$ at early times. For instance in $d=4$, we can derive the following expressions in a small charge expansion, \ie $y\rightarrow 0$,
\begin{align}
\begin{split}\label{eq.earlytimesmallcharge}
&x_m^1  = y \left(1 + \exp{ \left[ -\left(\frac{\pi (1 + k z^2)}{2 + k z^2} \, \frac{2 \tau T + \sqrt{1 + k z^2} }{y^3}\right) +\mathcal{O}\left( \frac{1}{y} \right) \right]} \right)  \, ,  \\
&x_m^2= y \left(1 + \exp{\left[ -\left(\frac{\pi (1 + k z^2)}{2 + k z^2} \, \frac{-2 \tau T + \sqrt{1 + k z^2} }{y^3}\right) + \mathcal{O}\left( \frac{1}{y} \right) \right]} \right)  \, .
\end{split}
\end{align}
This expansion demonstrates that the two corners remain exponentially close to $r_-$ at early times.
Given the above expression, it is clear that $r_m^1$ never leaves this regime and keeps approaching $r_{-}$.  However, in the second expression for $r_m^2$, the leading term in the exponent flips its sign at some $\tau=\tau_{c} = \frac{1}{2 T} \sqrt{1 + kz^2}$, which  is precisely the uncharged critical time given in eq.~\eqref{wop3}. Hence the rate of change of complexity given by eq.~\eqref{chargedEqAmazing} is exponentially suppressed as long as $\tau \lesssim \tau_{c}$.

Another case for which the early time behaviour can be studied in an analytic expansion is the near-extremal black holes. In this case, the inner and outer horizons are very close to each other as $y\rightarrow 1$. If we define $y=1-\epsilon$ where $\epsilon\ll1$, eq.~\eqref{eq:meeting345} yields at early times
\begin{align}
&x_m^1 = 1 - \frac{\epsilon}{2} \left( 1 + \pi \tau T  \right)+ \mathcal{O}(\epsilon \tau^3 T^3, \epsilon^2 \tau T , \epsilon^2 \log{\epsilon})  \, , \nonumber \\
&x_m^2 = 1 - \frac{\epsilon}{2} \left( 1 - \pi \tau T \right)+ \mathcal{O}(\epsilon \tau^3 T^3, \epsilon^2 \tau T , \epsilon^2 \log{\epsilon})  \, .
\end{align}
In general, the geometry and hence, the complexity are symmetric under $\tau\to-\tau$. Therefore only even derivatives of $\mathcal{C}_A$ are nonvanishing at $\tau=0$, \eg $\left. d\mathcal{C}_A/d \tau\right|_{\tau=0} =0$. We can evaluate the second derivative of $\mathcal{C}_A$ at $\tau=0$ using eqs.~\eqref{chargedEqAmazing} and  \eqref{eq:meeting3456}, and the expansion for $x_m\equiv x_m^1 = x_m^2$ at $\tau=0$ which reads
\small
\begin{align}
x_m=& 1-\frac{\epsilon }{2}+\frac{ \left(3 k z^2+7\right) \epsilon ^2 \log (\epsilon )}{4 \left(k z^2+3\right)} - \frac{\epsilon^2}{8 \left(k z^2+3\right)}\times \left( 16 \pi  \sqrt{\frac{1}{k z^2+2}}+3+28 \log (2)\right.
\\
& \left.+k z^2 \left(4 \pi  \left(k z^2+4\right) \sqrt{\frac{1}{k z^2+2}}+1+6 \log (4)\right)-8 \left(k z^2+2\right)^{3/2} \cot ^{-1}\left(\sqrt{k z^2+2}\right)\right)\, .\nonumber
\end{align}
\normalsize
Hence using the above results, the first nonvanishing derivative becomes
\begin{equation}
 \frac{d^2 \mathcal{C}_A}{d \tau^2}\bigg|_{\tau=0} = \frac{4   \left(k z^2+2\right)}{2 k z^2+3}\,\epsilon   \,   MT  + \mathcal{O}(\epsilon^3) \, .
\end{equation}
Note that the temperature here is of order $\epsilon$ and as a consequence the leading term in an $\epsilon$ expansion is in fact of order $\epsilon^2$. Despite being suppressed by the parameter $\epsilon$, the complexity grows quadratically (and the rate of change grows linearly) with $\tau$ at early times.

\paragraph{Lloyd's bound:} A generalization of Lloyd's bound for the case of charged black holes has been proposed in  \cite{Brown2} (see also \cite{Cai:2016xho}).
According to this suggestion, the natural bound for states at a finite chemical potential becomes
\begin{equation}\label{eq:ChargedBound}
\frac{d \mathcal{C}_{A}}{d t} \le \frac{2}{\pi} \left[ \left( M - \mu Q\right) - \left(M - \mu Q \right) \bigl|_{gs} \right] \, .
\end{equation}
This bound was inspired by the late time growth rate of holographic complexity for the charged black holes.
One important element of this proposed bound is that it involves the subtraction of certain thermodynamic quantities associated with the ground state (gs) of the system in question, which according to the proposal of \cite{Brown2} is the state minimizing $(M-\mu Q)$ for a given value of the chemical potential. For instance, for spherical black holes with $ \mu<  \frac{g L}{2 R \sqrt{2 \pi G}}\sqrt{\frac{d-1}{d-2}} $, the ground state is simply the vacuum solution ($M=Q=0$) with a constant gauge field, while for larger chemical potentials, the ground state is the extremal black hole with same chemical potential $\mu$ as the state of interest. However, it was also found in \cite{Brown2} that the proposed bound \eqref{eq:ChargedBound}  is violated for black holes which are intermediate or large compared to the AdS radius ($r_+ \gtrsim L$), while for small black holes the bound is exactly saturated. On the other hand, we showed earlier that the complexity calculated from the action always approaches its late time limit from above, and as a consequence we conclude that the bound in eq.~\eqref{eq:ChargedBound} is always violated.

\begin{figure}
\centering
\includegraphics[scale=0.45]{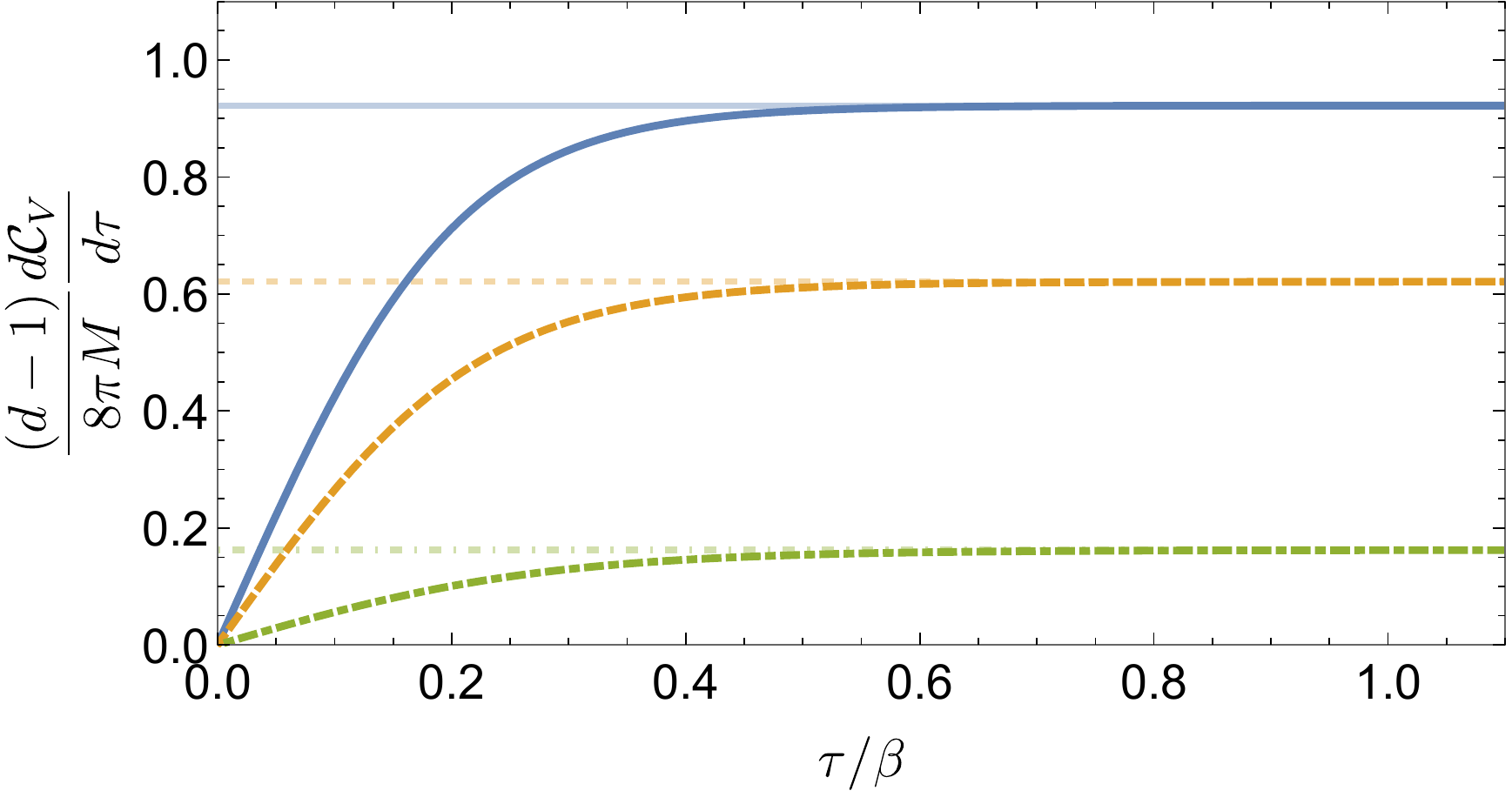}~~
\includegraphics[scale=0.45]{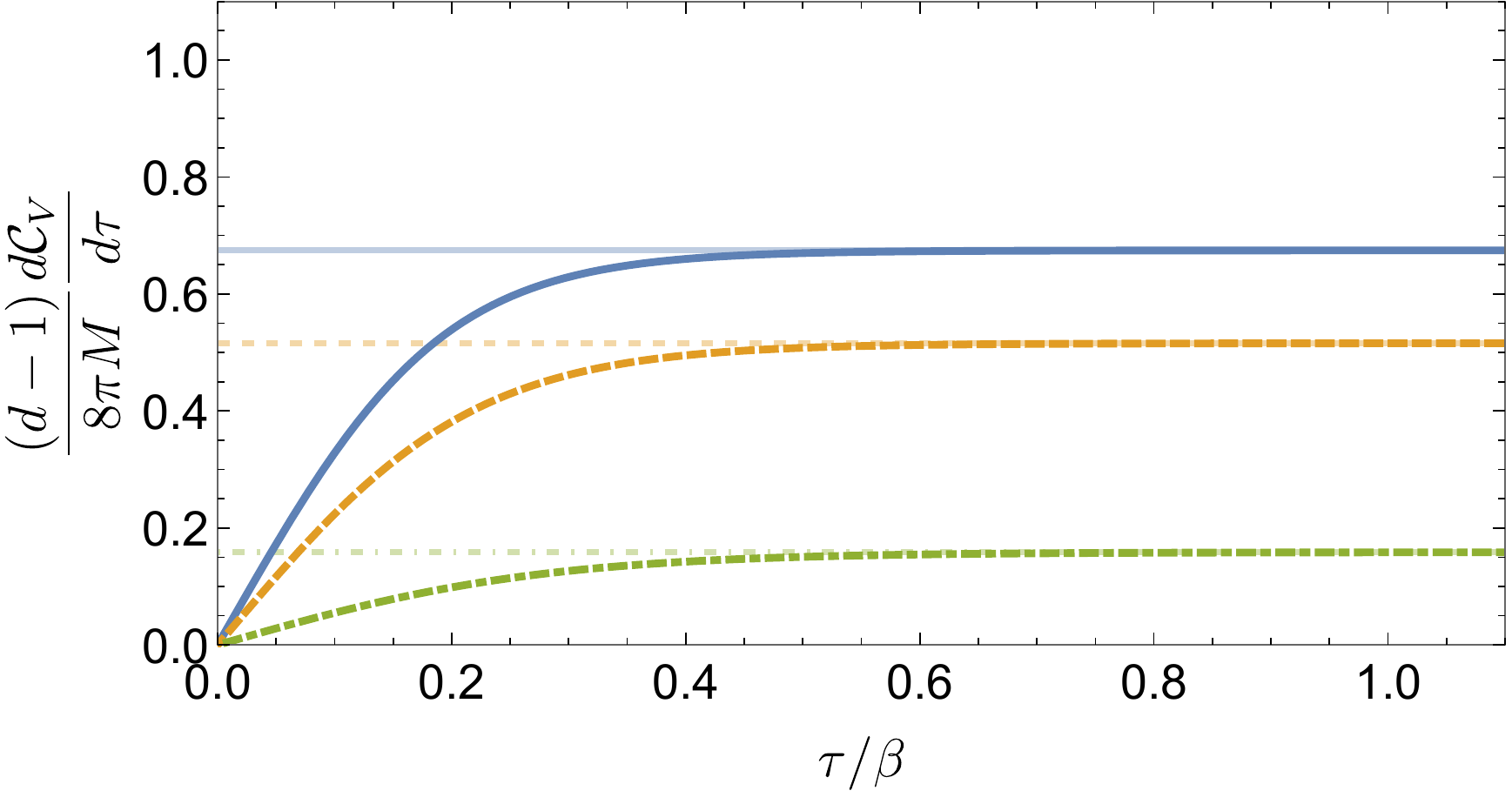}~~
\caption{The time derivative of complexity with $d=4$, $k=1$ with non-zero chemical potential, by fixing the parameters in eq.~\eqref{eq:RTnuSphAdS5}. The various curves correspond to: $\nu = 0.1$ in blue (solid), $\nu =1$ in orange (dashed) and $\nu=5$ in green (dot-dashed) for $T R =1$ (left) and $T R = \frac{1}{2}$ (right). Late time limits are obtained from eqs.~\eqref{r_m_eq}, \eqref{const_dv/dt} and are indicated by horizontal lines of the appropriate color.}
\label{fig:charged1vol}
\end{figure}

\subsection{Complexity=Volume}
We can also extend the analysis of section \ref{sec:EternalVolume} to evaluate the rate of change of complexity for the charged case using the CV proposal \reef{volver}. A maximal volume connecting the two boundaries anchored at $t_{L}$ and $t_{R}$ is depicted on the right side of figure \ref{fig:charged}. The analysis and the results are very similar to the uncharged case. For example, one still calculates the rate of change by computing $r_{min}$ (or the associated $E$) in eq.~\eqref{eq_rturn}, but now with the blackening factor for charged solutions in eq.~\eqref{ChargedMetric}. The growth rate can be evaluated as detailed in section \ref{sub_plot_CV}.

\begin{figure}
\centering
\includegraphics[scale=0.45]{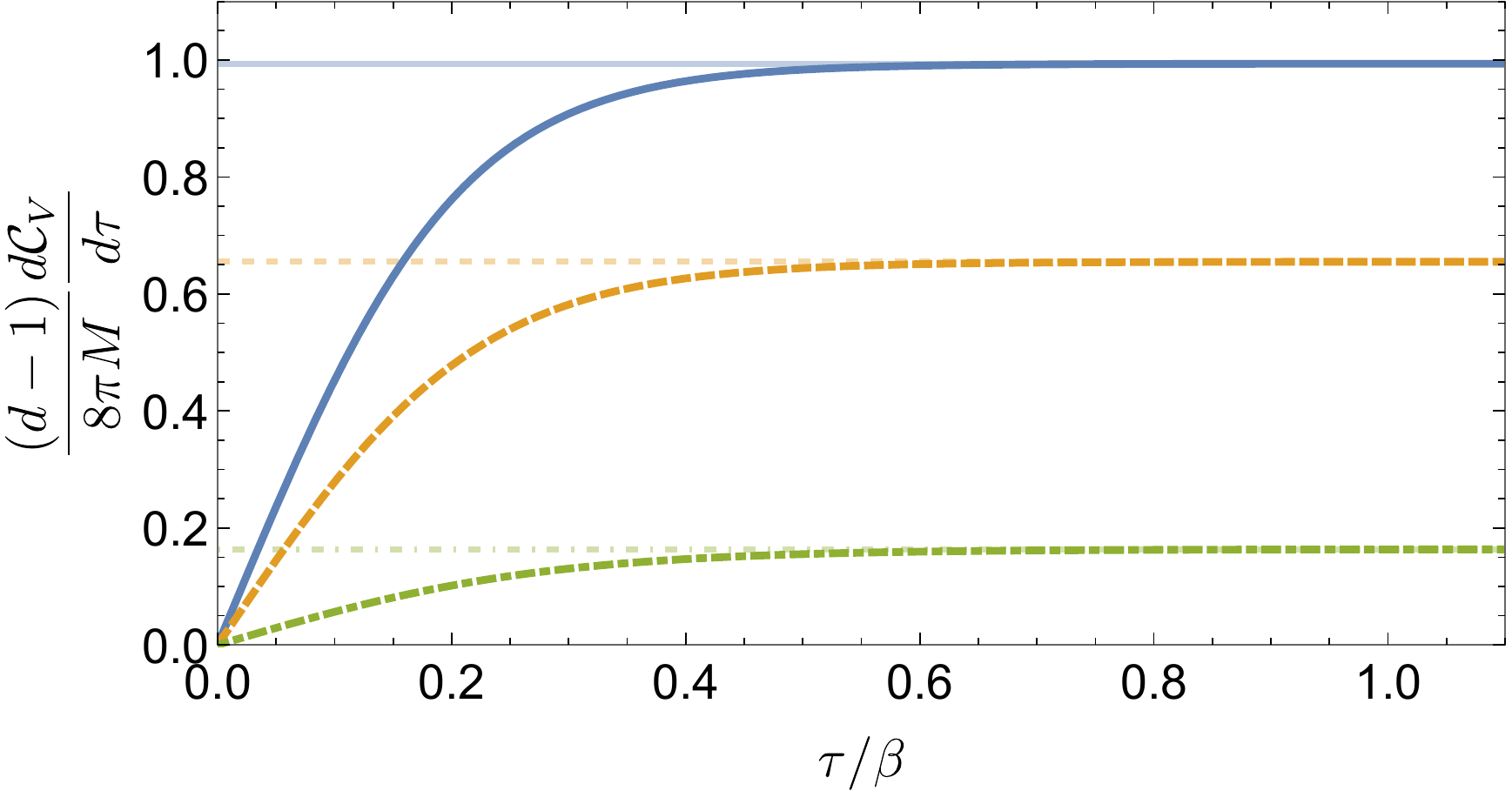}
\caption{The time derivative of complexity with $d=4$, $k=0$ with non-zero chemical potential, by fixing the parameters in eq.~\eqref{eq:RTnuSphAdS5}. The various curves correspond to: $\nu = 0.1$ in blue (solid), $\nu =1$ in orange (dashed)  and $\nu =5$ in green (dot-dashed). Curves are independent of $TR$ in eq. \eqref{eq:RTnu} as expected for the planar geometry. Late time limits are obtained from eqs.~\eqref{r_m_eq}, \eqref{const_dv/dt} and are indicated by horizontal lines of the appropriate color.}
\label{fig:charged12vol}
\end{figure}

We present some of the results in figures  \ref{fig:charged1vol} and \ref{fig:charged12vol}.  The growth rate depends on the charge parameter as expected, and it also approaches zero near the extremal limit, analogous to the previous results from CA. It smoothly approaches the neutral behaviour (\eg shown in figures \ref{vol_rates} and \ref{dvdt_d=4spherical}) in the limit $q\to0$.

\section{Discussion}\label{sec:Discussion}
In this paper, we computed the general time dependence of holographic complexity in various AdS black hole geometries.\footnote{Here, we have focused on eternal two-sided black holes and in a companion paper, we will also study one-sided black holes \cite{vaid}.} Further we examined the time dependence using both the complexity=action (CA) and the complexity=volume (CV) conjectures. Using the CV conjecture, the rate of change of complexity is a positive monotonically increasing function of time, and it saturates to a positive constant as $t \to \infty$.  In particular, for planar black holes, the limiting rate is given by eq.~\reef{planar_asympt_volume_rate},
\beq\tag{\ref{planar_asympt_volume_rate}}
\lim_{\tau\to \infty} \frac{d\mathcal{C}_V}{d \tau} = \frac{8 \pi M}{d-1}\,,
\eeq
as was first found in \cite{Stanford:2014jda}. When the boundary geometry is curved, this result is modified by various curvature corrections which become important when the temperature is of the same order as the curvature scale, \ie $RT\lesssim 1$.

Using the CA conjecture, the rate of change of the complexity shows some curious features. Of course, there is a universal late time rate of growth
\begin{equation}
\lim_{\tau\to \infty}\frac{d \mathcal{C}_A}{d \tau}=\frac{2M}{\pi}  \,,\label{XYZ2}
\end{equation}
as shown in eq.~\reef{latetimeSub}. This universal rate, discovered in \cite{Brown1,Brown2}, holds in any number of dimensions and is not affected by the boundary curvature. However, as also shown in eq.~\reef{latetimeSub}, $d \mathcal{C}_A/d \tau$ overshoots this late time limit at early times and approaches the final limit from above. Further $d \mathcal{C}_A/d \tau$ is initially zero and the complexity only begins to change after some critical time $\tau_c$ (for $d\ge3$). This initial phase of constant complexity was also observed in \cite{Brown1,Brown2}. In the bulk, the vanishing of $d \mathcal{C}_A/d \tau$ results because of the `boost' symmetry of the eternal black hole geometry and the fact that in this initial period of time the WDW patch touches both the past and future singularities, \eg see the left panel in figure \ref{PenroseBHa}. A third curious feature that we found is that immediately after $\tau=\tau_c$, $d \mathcal{C}_A/d \tau$ is divergent and negative, as shown in eq.~\reef{pop}\footnote{This negative spike (as well as the overshoot of the late time limit) in $d \mathcal{C}_A/d \tau$ also appears in different holographic settings, such as the holographic dual of non-commutative SYM theories \cite{NonComm}. We thank Josiah Couch for discussing this upcoming work with us.} --- see also figure \ref{d4Rate}.

We reiterate that the three features above only appear for the time rate of change evaluated with the CA proposal. None of these features appeared in the results found using the CV proposal in section \ref{sec:EternalVolume}. Further, when a chemical potential was introduced in section \ref{sec:ChargedEternal}, this washed out the unusual behaviour at early times, at least when the chemical potential was comparable to the temperature, as shown in figures \ref{fig:charged1} and \ref{fig:charged12}. Of course, as we discussed, the limit $q\to0$ was a smooth one and the curious behaviour found for the neutral black holes was recovered. So when the chemical potential was small but nonvanishing, $d \mathcal{C}_A/d \tau$ varied very little for an initial period and then quickly dipped to negative values before rising again. We can also add that with a chemical potential, $d \mathcal{C}_A/d \tau$ would still overshoot the late time limit but that the amount by which the limit was exceeded was much less pronounced when the chemical potential became large.

At this point, let us add that the curious behaviour found with the CA proposal also seems to be particular to the eternal black hole, \ie to the thermofield double state \reef{TFDx}. Analogous computations of the action for a one-sided black hole yield results more similar to those found here with the CV proposal \cite{vaid}. That is, in this context, $d \mathcal{C}_A/d \tau$ is a positive monotonically increasing function of time, which saturates to some positive constant in the late time limit.

In the above discussion, we commented that for higher dimensions (\ie $d\ge3$), the action (for neutral black holes) does not change at all for some period $-\tau_c\le\tau\le\tau_c$ and then changes very rapidly just after $\tau=\tau_c$. We observe that the time scale $\tau_c$ is of the order of the thermal time scale $\beta=1/T$, \eg see eq.~\eqref{wop3} for $d=4$. In particular, the latter equation demonstrates that the critical time is a physical quantity independent of the ambiguity  introduced by the normalization constant $\alpha$ of null normals. In contrast, the period of time over which $d \mathcal{C}_A/d \tau$ is negative, depends both on $\beta$ and on $\alpha$. For very small black holes, it is possible to obtain an estimate of this period by equating the RHS of eq.~\eqref{pop} with the constant term in the complexity $2M/\pi$ and we see that this period depends explicitly on the reference scale $\ell$ (as in $\alpha=L/\ell$) (\ie the spike lasts for $\delta t_0\sim \beta\, \left( \ell / \beta \right)^{2(d-1)/(d-2)}$). However, we might add that this negative spike can grow arbitrarily  wide\footnote{The growth rate is exceptionally slow with $\delta t_0\sim \beta\, \log\left[\log(\ell/\beta)\right]$ for very large values of $\ell$.} for extremely large values of $\ell$, or alternatively, for extremely small values of the parameter $\alpha$. While the latter remains a logical possibility, it also seems very unnatural for our complexity calculations, \eg see \cite{qft1,diverg}.

\begin{figure}
\centering
\includegraphics[scale=0.39]{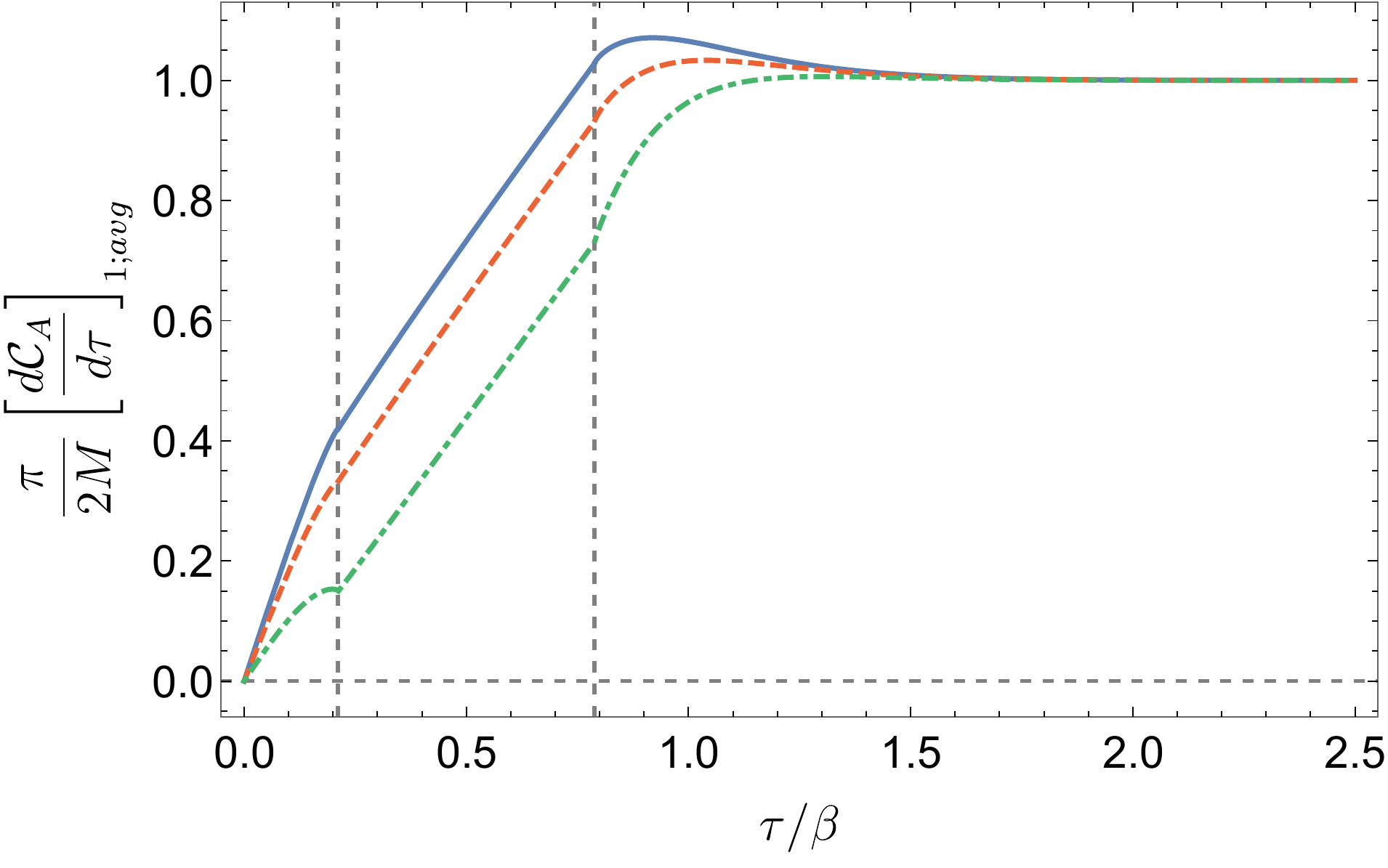}
\includegraphics[scale=0.36]{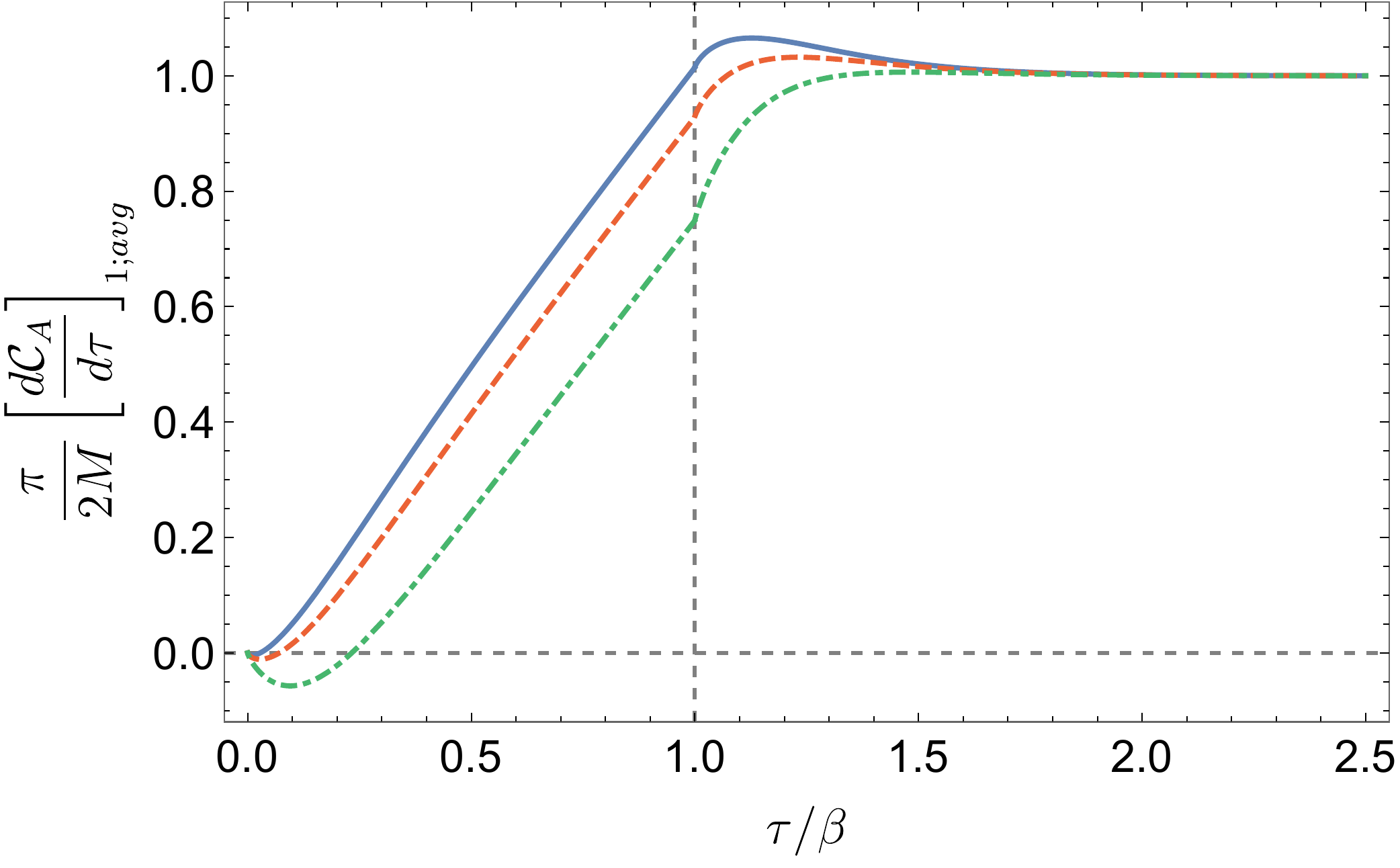}
\caption{The averaged rate of growth of complexity from eq.~\reef{averageC} (with $\gamma=1$) as a function of time for the $d=3$ planar black hole (left) and $d=4$ planar (right). Results are shown for several values of the horizon radius --- $r_h/L=1$ (blue), $r_h/L=1.5$ (dashed red) and $r_h/L=3.5$ (dot-dashed green). Note that, as in figures \ref{fig:BTZ} and \ref{d4Rate}, smaller black holes violate the Lloyd bound more strongly. Note also, that the averaged derivative is discontinuous at $|\tau/\beta \pm \frac{1}{2}|=\tau_c/\beta$, where for $d=3$, $\tau_c/\beta=\frac{1}{2\sqrt{3}}$ and for $d=4$,  $\tau_c/\beta=\frac{1}{2}$.}\label{fig:BTZaver}
\end{figure}

However, one might argue that the holographic definition of circuit complexity is not robust enough to consider time scales smaller than $\beta$ in the context of the eternal black hole.\footnote{We thank Lenny Susskind, Dan Roberts and Brian Swingle for correspondence on this point.} That is, we might only want to consider the behaviour of complexity over time scales which are longer than the thermal time scale. Therefore we defined an averaged version of $d \mathcal{C}_A/d \tau$ in eq.~\reef{averageC}, which is essentially a symmetric discrete time derivative with a time step $\Delta t=\gamma/T$. With a large enough $\gamma$, the complexity begins changing right away and the sharp negative spike in $d \mathcal{C}_A/d \tau$ is washed out by the averaging procedure.\footnote{This simply requires that $d \mathcal{C}_A/d \tau>0$ at $\tau=\gamma\beta/2$.} However, we note that this averaging does not remove the behaviour where the rate of change overshoots its late time limit.  This feature should not be associated with short times since in fact, the late time limit is being approached from above, as shown in eq.~\reef{latetimeSub}. Some examples of these averaged growth rates are shown in figure \ref{fig:BTZaver}.

Recall that \cite{Brown1,Brown2} suggested that the late time limit of $d\mathcal{C}_A/d\tau$ may be related to Lloyd's bound $2M/\pi$  for the rate of computation for a system of energy $M$ \cite{Lloyd}. These authors also proposed a generalization of Lloyd's bound that should apply for charged black holes --- see eq.~\eqref{eq:ChargedBound}. However, they also pointed out apparent violations of the latter bound for intermediate or large charge black holes (\ie $r_+ \gtrsim L$). However, our calculations of the rate of change of holographic complexity for general times showed that $d\mathcal{C}_A/d\tau$ always overshoots the late time limit. As a consequence, for every situation that we examined in sections \ref{tary} and \ref{subSec:CommentsCharged}, the corresponding bound on $d\mathcal{C}_A/d\tau$ was violated. This certainly calls into question these proposals or at least their interpretation (as we describe next).

Let us comment that similar violations are observed for the proposed bounds for the maximal rate of entanglement growth in relativistic systems \cite{tsun1,tsun2}.\footnote{We thank Mark Mezei for explaining this point  to us.} In this case, the proposal is that following a quantum quench, the rate of growth of the entanglement entropy for a large region will be bounded by
\beq
\frac{1}{s_{eq}\,A}\,\frac{dS_\mt{EE}}{dt} \le v_E\,
\eeq
where $s_{eq}$ is the equilibrium entropy density, $A$ is the area of the entangling surface, and $v_E (\le 1)$ is a universal velocity that depends on the dimension of the spacetime. In certain contexts, this bound can be proven but it requires considering a certain scaling regime where $\beta\ll t,R$ where $R$ is the characteristic size of the entangling region \cite{mark1}. In contrast, in numerical studies, one may find that the rate of growth actually overshoots the expected bound, \eg \cite{tsun2,mark2}. By analogy, it may be that one should only interpret the bounds on the growth of complexity in a particular scaling regime. For example, if we demand that $\beta\ll t$, then the corrections in eq.~\reef{latetimeSub} to the late time limit would be vanishingly small.  We might also point out that one needs to test carefully the validity of the assumptions entering in the derivation of Lloyd's bound in a holographic setup, in particular the use of orthogonalizing gates.\footnote{We thank William Cottrell and Miguel Montero for sharing their upcoming work \cite{SimpleComplexity} on this subject with us.}

We must also comment that the precise details of the manner in which $d\mathcal{C}_A/d\tau$ overshoots the late time limit depend on the normalization constant $\alpha$, which fixes the normal vectors on the null boundaries of the WDW patch. In our various plots, \eg figures \ref{fig:BTZ} and \ref{d4Rate}, we chose $\alpha=L/R$ for simplicity and as a result, the late time limit was only exceeded by a relatively small amount. However, by choosing  $\alpha$ to be very large, the amount by which $d\mathcal{C}_A/d\tau$ overshoots this limit can be made very large. This is easily demonstrated by examining  eq.~\eqref{tder44} evaluated for two different values of the normalization constant, \ie $\alpha_1$ and $\alpha_2$, but for the same time $\tau$ where $d\mathcal{C}_A/d\tau$ exceeds the late time limit for $\alpha_1$. Now we see in eq.~\eqref{tder44} shows that with $\alpha_2$,  $d\mathcal{C}_A/d\tau$ is the previous value plus a positive quantity multiplying $\log(\alpha_2/\alpha_1)$ and so by choosing $\alpha_2$ large enough, we can make the excess as large as we want.

We can also study the maximal rate of complexity growth analytically  when $\alpha$ is very large.  The simplest case to consider here is $d=2$ for which the maximum was calculated in appendix \ref{app:BTZnonsymmetric}. For example, if we choose $\alpha=L/\delta$, then eq.~\reef{gnu} yields
\beq
\left. \frac{d \mathcal{C}_A}{d \tau}\right|_{max}=\frac{2 M}{\pi} \left(1+\log\! \left[\frac{1}{2\pi\delta\, T}\right]\right)\,.
\label{gnu2}
\eeq
However, we should also remark that in this instance, the violation is an early time feature, \ie $d \mathcal{C}_A/d \tau$ peaks at precisely $\tau=0$ and the width of the peak is of order $\beta$. Hence the averaging discussed above will reduce the excess but it will still remain significant with this extreme choice of $\alpha$.
A similar result holds in higher dimensions. For instance, if we consider the planar uncharged black holes in section \ref{sec:EternalAction} with $\alpha=L/\delta$, then the limit  $\delta\to0$ yields
\begin{equation}\label{eq:PeakPland4}
\left. \frac{d \mathcal{C}_A}{d \tau}\right|_{max} = \frac{2 M}{\pi} \, \log \left( \frac{d}{4 \pi \delta \, T}  \right) + \mathcal{O}\left( \log{ \left( \log{\frac{1}{\delta \, T}} \right)} \right) \, ,
\end{equation}
for the leading behaviour of the peak of the growth rate. Note that this result reproduces the leading behaviour in eq.~\reef{gnu2} with $d=2$.

Having noted that the amount by which $d\mathcal{C}_A/d\tau$ exceeds that late time limit is controlled by $\alpha$, we might add that this produces a finite shift in the  complexity. That is comparing the complexity at late times for different choices of $\alpha$ has a rather simple expression
\beq
\Delta\CA(\alpha_1)-\Delta\CA(\alpha_2)=\frac{S}{2\pi^2}\,\log\left(\alpha_1^2/\alpha_2^2\right) \,.
\label{gnu3}
\eeq
That is, the total shift in the complexity caused by the overshoot scales with $S$, the entanglement entropy between the two CFTs in the thermofield double state \reef{TFDx}.
The $\Delta$ for the complexities in this difference indicates that we are subtracting two copies of the vacuum complexity. This subtraction removes the $\alpha$ dependence of the UV divergent contributions, which is not captured in the time derivative $d\mathcal{C}_A/d\tau$.\footnote{From \cite{diverg}, the leading UV behaviour is
$[\CA]_\mt{UV}(\alpha_1) -[\CA]_\mt{UV}(\alpha_2)\simeq-\frac{L^{d-1}}{4\pi^2G_N}\,\frac{V}{\delta^{d-1}} \log\left(\alpha_1^2/\alpha_2^2\right) $.}
Of course, we should also recall that the total holographic complexity diverges in this late time limit, since it is growing linearly with time.

As we first noted in eq.~\reef{tder45}, we should choose $\alpha = L/\ell$ in order that our general results for $d\mathcal{C}_A/d\tau$ can be fully expressed in terms of boundary quantities. That is, the argument of the logarithm in eq.~\reef{tder45} contains an errant factor of the AdS scale, which is not a quantity that the boundary CFT should know about, but this can be eliminated using our freedom in choosing $\alpha$. However, this choice for $\alpha$ also introduces some new scale $\ell$ in the boundary theory.  It is reassuring that precisely the same situation arises in the UV divergences of holographic complexity \cite{diverg}. That is, the contributions to the gravitational action coming from the joints where the null boundaries intersect the asymptotic cutoff surface also introduce logarithms where the argument contains the combination $L/\alpha$, as in eq.~\reef{tder45}. Of course, choosing $\alpha=L/\ell$ leaves us with the question of what the most appropriate choice for $\ell$ would be. While the ambiguity left in choosing $\ell$ may have originally seemed problematic, it was recently found that precisely the same ambiguity appears in complexity models for quantum field theory \cite{qft1,qft2} where the complexity of ground states of free scalar field theories were examined.\footnote{The complexity for a free scalar quantum field theory in the time-dependent thermofield double state, and the similarities and differences with the holographic results presented in this work, will be discussed in \cite{TimeDepQFTUs}.}
Further let us add that setting $\ell=e^{\sigma} \delta$, where $\sigma$ is some numerical factor and $\delta$ is the short-distance cutoff in the boundary theory, was a convenient choice because it removed an extra logarithmic factor in the leading UV divergence. However, our results show that with this choice, $d\mathcal{C}_A/d\tau$ would depend on the short-distance cutoff, \ie an apparently IR contribution to the complexity would now depend on the UV cutoff.

\begin{figure}[htbp]
\centering
\includegraphics[keepaspectratio, scale=0.3]{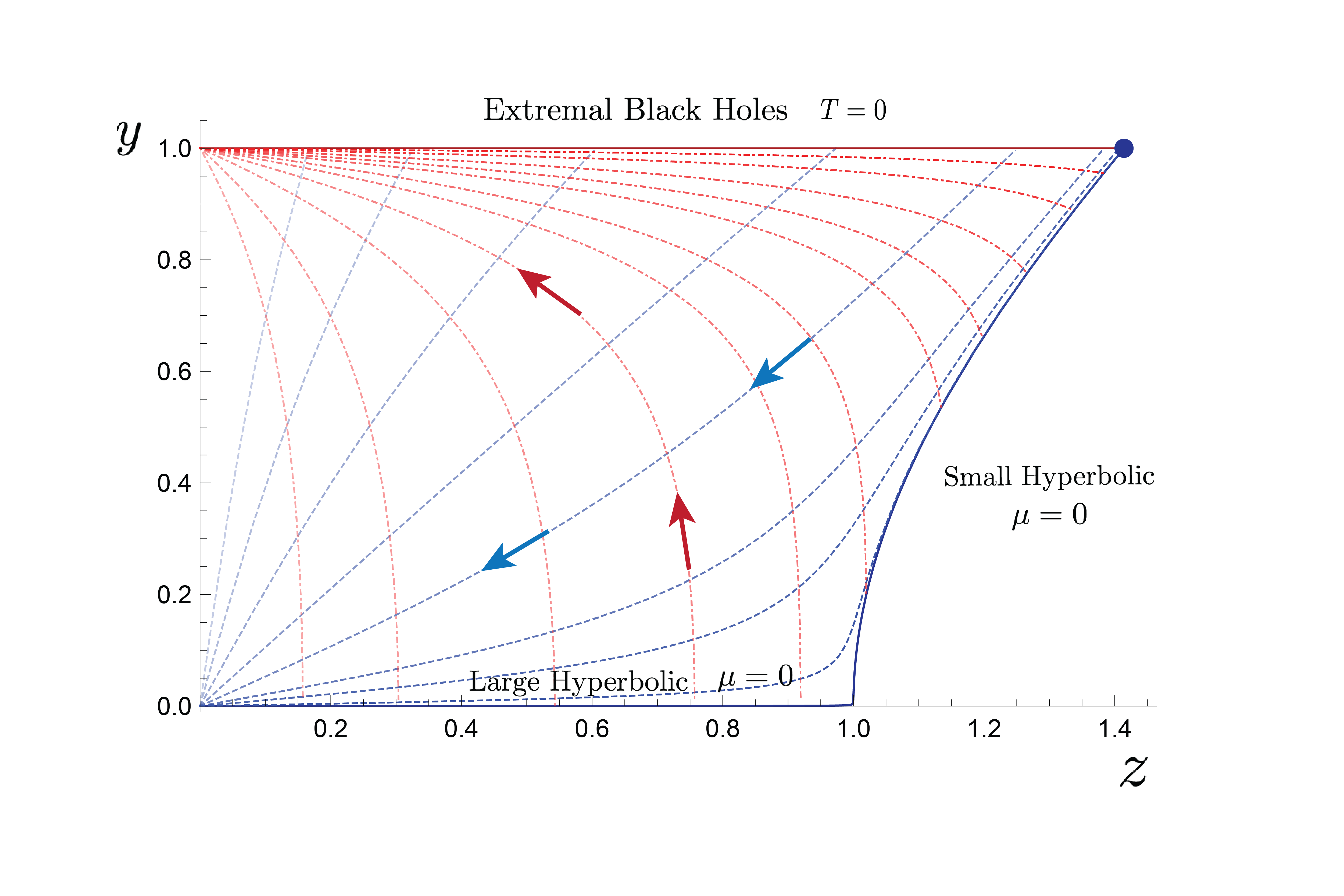}
	\caption{Lines of constant $\mu R$  (dashed blue) and constant $T R$ (dot-dashed red) for the hyperbolic black hole in $d=4$, with $y=\frac{r_-}{r_+}$ and $z=\frac{L}{r_+}$. The temperature and chemical potential increase as one moves towards the left, as indicated by the arrows. The line of extremal black holes at $y=1$ with finite chemical potential has states with infinite complexity. However, the extremal black hole represented by the blue dot with coordinates $y=1, z= \sqrt{2}$ is the small uncharged extremal hyperbolic black hole, with zero chemical potential and finite complexity (using the CA proposal). }
	\label{HyperbolicPhases}
\end{figure}

To close our discussion, we would like to return to our calculations of the complexity of charged AdS black holes. In particular, in section \ref{sec:ChargedEternal} (and appendix \ref{CformCharged}), we found that the complexity of formation diverged for extremal charged black holes. Both these results appeared using either the CA or CV conjectures. We stress that in the complexity of formation, there was still a cancellation of the UV divergences associated with the asymptotic boundary. Instead this divergence was a new IR divergence, associated with the infinitely long throat of the extremal black holes. Further, the results in section \ref{sec:ChargedEternal} indicate that the rate of change of the complexity vanishes for extremal black holes. If one considers the CA predictions, we find that extremal black holes with finite chemical potential has these IR divergences, while systems with zero chemical potential and zero temperature (\ie extremal hyperbolic black holes without any charge) have finite contributions to the complexity from the IR.\footnote{We might add that using the CV proposal actually yields a similar IR divergence for these black holes \cite{Formation}.} In order to illustrate these results, figure \ref{HyperbolicPhases} shows a schematic phase diagram for the hyperbolic black holes for $d=4$, in terms of the $y$ and $z$ variables introduced in eq.~\reef{eq:xyz}. There is a line of states at $y=1$ with finite chemical potential and zero temperature with infinite complexity, while the states with zero chemical potential ends in a point $y=1, z= \sqrt{2}$ with finite complexity.

Combining these results suggests a `Third Law of Complexity'.\footnote{We thank Henry Maxfield and Robie Hennigar for independently suggesting this connection.} That is, the corresponding `extremal' thermofield double states \reef{TFDq} at zero temperature and finite chemical potential are infinitely complex compared to the finite temperature states. Hence no physical process should be able to produce the extremal states in a finite amount of time. It would be interesting to further test this idea by examining the complexity of extremal spinning black holes \cite{rot1,rot2}.

\section*{Acknowledgments}
We would like to thank Alice Bernamonti, Adam Brown, William Cottrell, Josiah Couch, Bartek Czech, Lorenzo Di-Pietro, Federico Galli, Robie Hennigar, Javier Martinez, Henry Maxfield, Mark Mezei, Miguel Montero, Djordje Radicevic, Dan Roberts, Jamie Sully, Lenny Susskind, Todd Sierens, Brian Swingle, Tadashi Takayanagi and Ying Zhao for useful comments and discussions. We also would like to thank Ipsita Mandal for collaboration on the initial stages of this project. Research at Perimeter Institute is supported by the Government of Canada through Industry Canada and by the Province of Ontario through the Ministry of Research \& Innovation. This research was supported in part by the National Science Foundation under Grant No. NSF PHY11-25915. SC acknowledges support from an Israeli Women in Science Fellowship from the Israeli Council of Higher Education. RCM is supported by funding from the Natural Sciences and Engineering Research Council of Canada, from the Canadian Institute for Advanced Research and from the Simons Foundation through the ``It from Qubit'' collaboration.
SS thanks Perimeter Institute for their hospitality during this project.
The work of SS is supported in part by the Grant-in-Aid for JSPS Research Fellow, Grant Number JP16J01004.

\appendix

\section{Details of Complexity=Action for BTZ Black Holes} \label{app:BTZnonsymmetric}
In this appendix, we add some more details of the holographic complexity for the BTZ holes, using the complexity=action proposal.
Much of these results are already summarized in section \ref{btzztb}. The new results here include the derivation of our results for non-symmetric boundary times $\tau_L\neq \tau_R$ and their generalization for negative times.

\subsection{General Boundary Times}\label{sec_nonsym_BTZ}
\begin{figure}[htbp]
	\vspace{2ex}
	\begin{minipage}[b]{0.5\hsize}
		\centering
		\includegraphics[keepaspectratio, scale = 0.05]{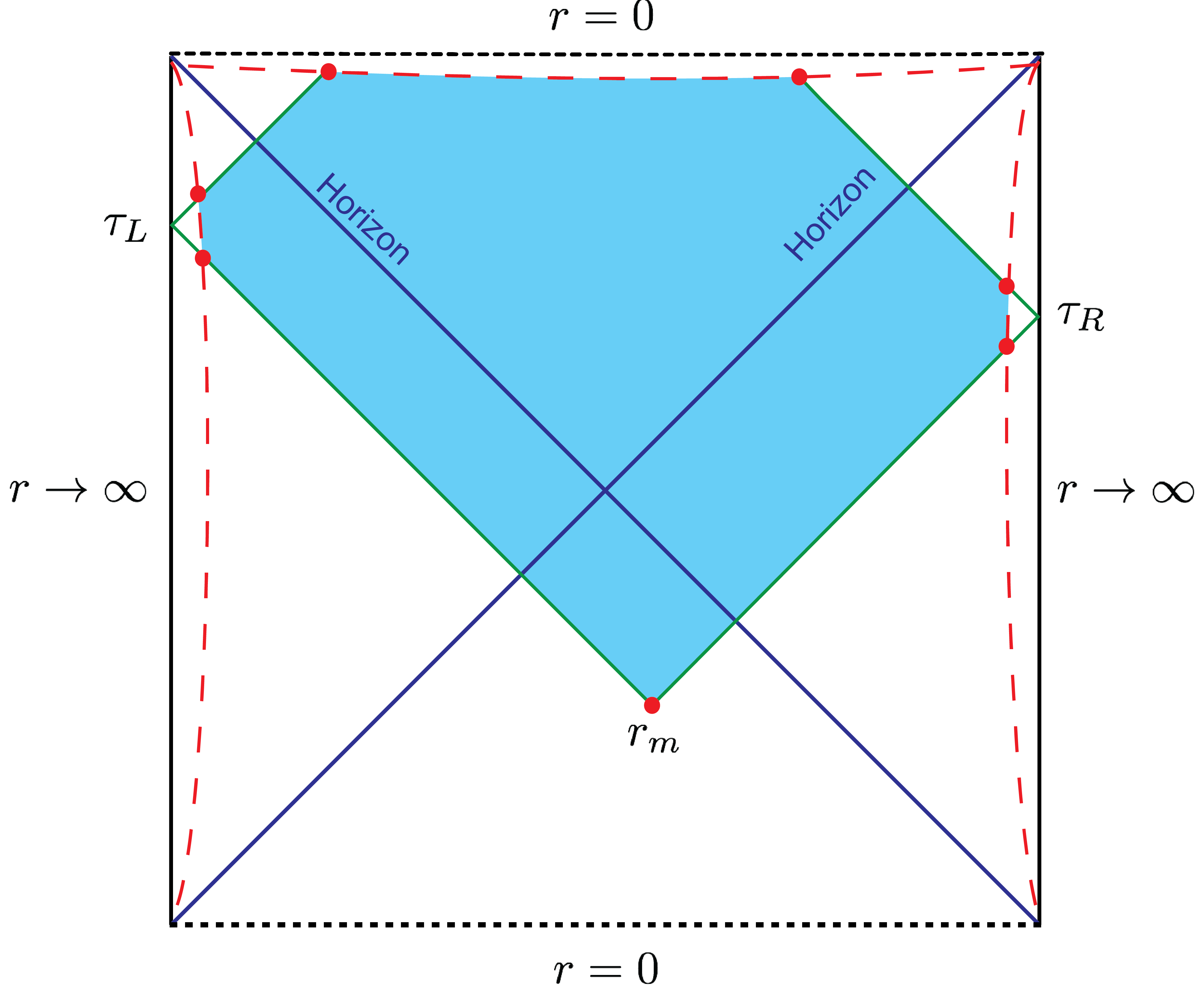}
		\subcaption{$\tau=\tau_L+\tau_R>0$}\label{BTZ_WDW_positive_t}
	\end{minipage}
	\begin{minipage}[b]{0.5\hsize}
		\centering
		\includegraphics[keepaspectratio, scale=0.05]{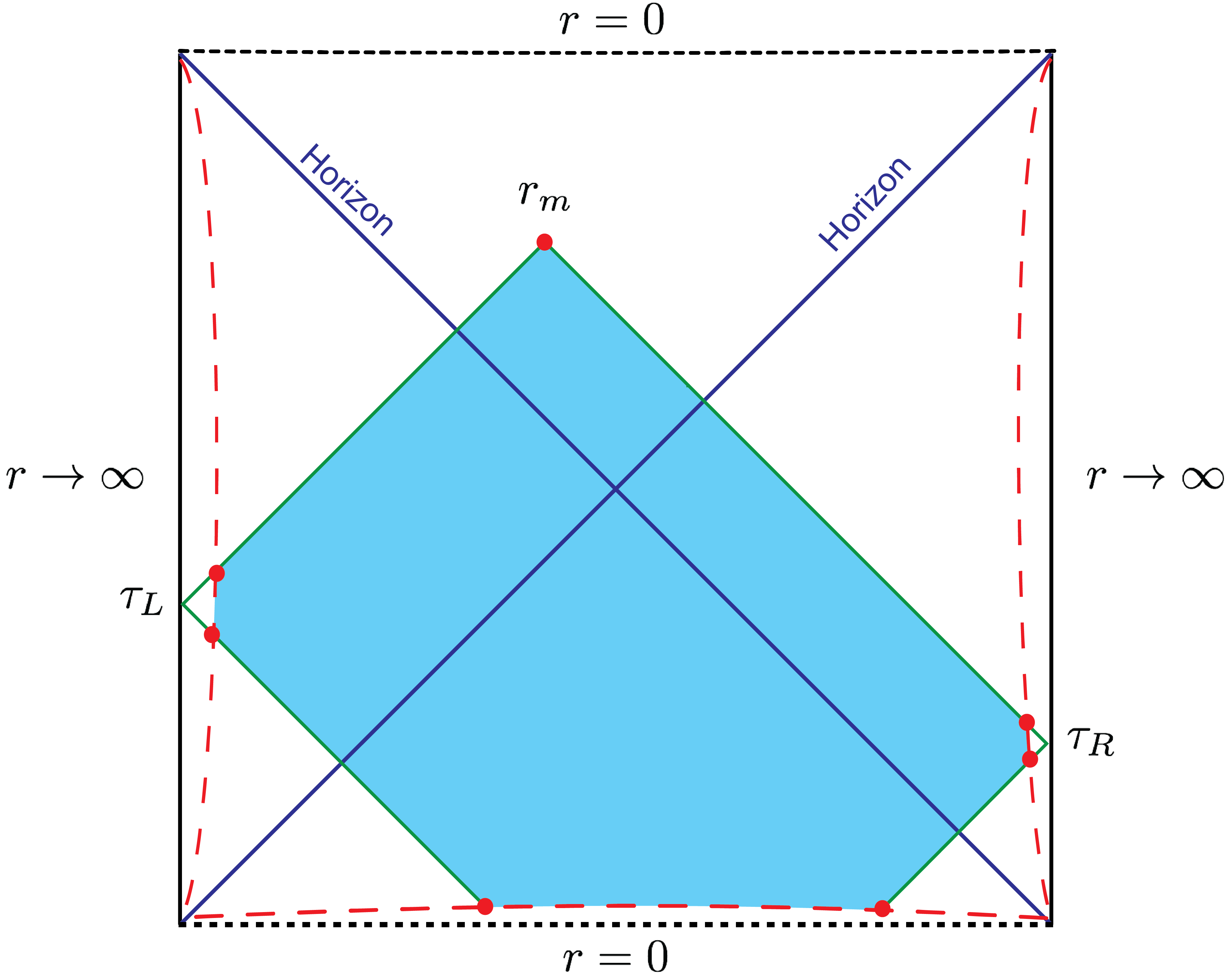}
		\subcaption{$\tau=\tau_L+\tau_R<0$}\label{figures/fig_btz/BTZ_WDW_negative_t}
	\end{minipage}
	\caption{
		The WDW patches in the BTZ black hole background. The dashed lines represent the cutoff surfaces.
		The left (right) panel illustrates the case in which $\tau=\tau_L+\tau_R>0$ $(\tau=\tau_L+\tau_R<0)$. }
	\label{fig_btz_wdw}
\end{figure}

We consider the BTZ metric, given in eq.~\reef{btz2}:
\begin{align}
d s^2 = -f(r) \frac{L^2}{R^2 }d \tau^2 + \frac{dr^2}{f(r)}+ r^2 d \phi^2\,,
\quad \quad {\rm with}\ \ f(r) =\frac{r^2-r_h^2}{L^2}\,.
\end{align}
The boundary metric takes the form given in eq.~\reef{bdyBTZ} and so a constant time slice is simply a circle with circumference $2\pi R$.  For general boundary times $(\tau_L, \tau_R)$,\footnote{Here, we do not assume $\tau_L=\tau_R$, but, of course, the result depends only on the total time $\tau=\tau_L+\tau_R$ due to the symmetry of the background.}
the WDW patch takes the form depicted in figure   \ref{fig_btz_wdw}.
When the total time $\tau=\tau_L+\tau_R$ is positive $\tau>0$ (or negative $\tau<0$),
the WDW patch does not reach the past (future) singularity
and there is a past (future) corner represented by the dot in figure \ref{fig_btz_wdw}.
The radial coordinate $r_m$ of this joint is given by
\begin{align}
r_m(\tau_L, \tau_R)= r_h \tanh \frac{r_h |\tau_L+\tau_R|}{2L R}
=r_h \tanh \frac{r_h |\tau|}{2L R} \,.
\label{rmeet_btz}
\end{align}

The action for the WDW patch consists of a bulk term, surface terms and joint terms, as described in the main text:
\begin{align}
I_{BTZ}= I_{\bulk}+ I_{\surf} + I_{\jnt} .
\end{align}
The bulk term $I_{\bulk}$ is given by
\begin{align}
I_{\bulk}  = -\frac{2L R}{G_N \delta} - \frac{r_h^2 |\tau|}{4 G_N L R} +  \frac{r_{m}}{2 G_N},
\end{align}
where we used the cutoff $r=r_{\max} = LR/\delta + r_h^2 \delta/(4LR)$ corresponding to the UV regulator $z=\delta$ in a Fefferman-Graham expansion (see \cite{Formation}).

As in the text, we choose here an affine parametrization for the null generators. For this reason, the null surface terms vanish. Therefore the only nonvanishing surface contributions come from the surface at the future singularity and the UV cutoff surfaces.
The contribution from the singularity is given by
\begin{align}
I_{\surf,sing}=\frac{r_h^2 |\tau|}{4 G_N LR}.
\end{align}
The contribution from the cutoff surfaces is UV-divergent:
\begin{align}\label{BTZcutsurf}
I_{\surf,cut}=\frac{2 LR}{G_N \delta}.
\end{align}
Thus, the total surface term is given by
\begin{align}
I_{\surf} = \frac{2 LR}{G_N \delta} + \frac{r_h^2 |\tau|}{4 G_N L R}.
\end{align}
The normalization of null vectors ${\bf k}_L$ and ${\bf k}_R$ are set to be the same as in the main text:
\begin{align}
{\bf k}_L\cdot \hat{\tau}_L = {\bf k}_R\cdot \hat{\tau}_R= \pm \alpha
\label{null_norm1}
\end{align}
where $\hat{\tau}_L = \partial_{\tau_L}$ and $\hat{\tau}_R = \partial_{\tau_R}$, and the sign is chosen as $+$ $(-)$ for future (past) null surfaces.
The joint contributions come from joints at $r=r_{m}$ and at $r=r_{\max}$ and are given by
\begin{align}\label{BTZcutcorn}
I_{\jnt,cut}=&-\frac{LR}{G_N \delta} \log \frac{\alpha \delta}{L}. \\
I_{\jnt,r_m}=&-\frac{r_m}{4 G_N} \log \left|\frac{L^2 f(r_m)}{\alpha^2 R^2}\right|  =\frac{r_h}{2 G_N} \tanh \frac{r_h |\tau|}{2LR}
\log\left(\frac{\alpha R}{r_h}\cosh\frac{r_h \tau}{2 LR}\right).
\end{align}
Therefore, the total action reads
\begin{align}
I_{BTZ} &= I_{\bulk}+ I_{\surf} + I_{\jnt} \nonumber\\
&= \frac{r_h}{2 G_N}  \tanh \frac{r_h |\tau|}{2LR}
\left[1+\log\left(\frac{\alpha R}{r_h}\cosh\frac{r_h \tau}{2 LR}\right)
\right]
-\frac{LR}{G_N \delta} \log \frac{\alpha \delta}{L}\,.
\label{unreg_action_btz}
\end{align}
We can regularize it by subtracting twice the action of the WDW patch in the vacuum AdS space, following \cite{Formation}. If we consider the Neveu--Schwarz vacuum of the boundary theory \cite{couscous}, \ie with the metric $f_0(r) = r^2/L^2+1$, the action of  vacuum AdS space is given by a sum of  eq.~(4.5) of \cite{Formation} and eqs.~\eqref{BTZcutsurf} and \eqref{BTZcutcorn} which remain the same for the empty AdS background (but need to be multiplied by a factor of a half if we consider a single copy of empty AdS). We therefore obtain:
\begin{align}
I_{AdS}=\frac{\pi L}{4G_N}  -\frac{L R}{2 G_N \delta} \log \frac{\alpha \delta}{L}.
\label{pureAdS}
\end{align}
The regularized action is then given after the subtraction by:
\begin{align}
I_{reg}(\tau_L,\tau_R) &= I_{BTZ}(\tau_L,\tau_R) - I_{AdS}(\tau_L) -I_{AdS}(\tau_R) \\
&=-\frac{\pi L}{2G_N}+\frac{r_h}{2 G_N}
\tanh \frac{r_h |\tau|}{2LR}\,
\left[1+\log\left(\frac{\alpha R}{r_h}\cosh\frac{r_h \tau}{2 LR}\right)\right] .
\label{btz_regular_action}
\end{align}
The finite part of the holographic complexity from the CA conjecture is thus
\begin{align}
\Delta\mathcal{C}_A(\tau_L,\tau_R) = \frac{I_{reg}}{\pi} =
-\frac{L}{2G_N}+\frac{r_h}{2 \pi G_N}
\tanh \frac{r_h |\tau|}{2LR}\,
\left[1+\log\left(\frac{\alpha R}{r_h}\cosh\frac{r_h \tau}{2 LR}\right)\right] \,.
\label{CA_general_R}
\end{align}
This result can also be written as
\begin{align}
\Delta\mathcal{C}_A(\tau_L,\tau_R)=-\frac{c}{3}
+ \frac{2 M}{\pi^2 T}  \tanh \left(\pi T|\tau|\right)\,
\Bigl(1+\log\left[\frac{\alpha}{2\pi L T}\cosh\left(\pi T \tau\right)\right]\Bigr)\, ,
\label{CA_general_R_bdry}
\end{align}
where $c$ is the central charge of the boundary CFT, given by $c=3L/(2G_N)$, $M$ is the mass of the BTZ black hole $M=r_h^2/(8G_NLR)$,   and $T$ is the temperature $T=r_h/(2\pi LR)$. We can think of this result as the complexity of formation of the thermofield double state, with general times $\tau_L$, $\tau_R$. Note that the temperature should satisfy $T>1/(2\pi R)$ so that the BTZ black hole is the dominant saddle point for the gravitational theory in the bulk.
In order to express $\Delta\mathcal{C}_A$ solely in terms of boundary quantities,
choose the normalization constant $\alpha=L/\ell$, where $\ell$ is a new length scale in the boundary theory, as discussed in section \ref{tary} --- see also \cite{diverg}.

The holographic complexity of the AdS vacuum is independent of time and hence taking the derivative of eq.~\reef{CA_general_R_bdry} with respect to time $\tau=\tau_L+\tau_R$, yields the rate of growth appearing in eq.~\eqref{btznew2}
\beq
\frac{d\mathcal{C}_A}{d\tau} =\frac{2 M}{\pi} \left(1+\text{sech}^2\left(\pi T \tau\right)\, \log\! \left[\frac{\alpha}{2\pi L T}\,\cosh \left(\pi T \tau\right)\right]\right) \,.
\eeq
We are assuming $\tau>0$ here.

Unlike the higher dimensional case, $d\mathcal{C}_A/d \tau$ is finite at $\tau=0$.\footnote{Recall the discussion for higher dimensional black holes around eq.~\reef{pop}.}
In fact, we have
\begin{align}
\frac{d \mathcal{C}_A}{d \tau}(\tau\to0^+) = \frac{2M}{\pi} \Bigl(1+\log \frac{\alpha}{2\pi LT}\Bigr)\,.
\end{align}
At late times,
we have
\begin{align}
\frac{d \mathcal{C}_A}{d \tau}(\tau \rightarrow \infty) \sim \frac{2M}{\pi} \Big[1+4\Big(\pi T \tau+ \log \frac{\alpha}{4\pi L T}\Big)e^{-2 \pi T \tau}
+\cdots\Big] \,.
\end{align}
Noting the coefficient of the exponential is positive, we find that it approaches $2M/\pi$ from above. In figure \ref{fig:BTZ}, we see that $\frac{d \mathcal{C}_A}{d \tau}$ has a maximum at some time $\tau_{peak}$. We can determine the latter by evaluating
$\frac{d^2 \mathcal{C}_A}{d \tau^2}(\tau_{peak})=0$
and we find
\begin{align}
\tau_{peak} = \frac{1}{\pi T}\, \cosh^{-1} \frac{\sqrt{e} \,2\pi L T}{\alpha}\,.
\end{align}
At that time, $\frac{d \mathcal{C}_A}{d \tau}$ is greater than $2M/\pi$ with,
\begin{align}
\frac{d \mathcal{C}_A}{d \tau}(\tau_{peak}) = \frac{2M}{\pi} \Bigl[1+ \frac{1}{2 e} \Bigl(\frac{\alpha}{2 \pi L T}\Bigr)^2 \Bigr] > \frac{2M}{\pi}\, .
\label{joker}
\end{align}
Hence $d\CA/d\tau$ always exceeds the Lloyd bound and further the violation increases for smaller black holes, \ie smaller temperatures. Substituting the minimum temperature, $T=1/(2\pi R)$, into eq.~\reef{joker} yields
\beq
\frac{d \mathcal{C}_A}{d \tau}(\tau_{peak})\bigg|_{T=\frac1{2\pi R}} = \frac{2M}{\pi} \Bigl[1+ \frac{1}{2 e} \Bigl(\frac{\alpha\,R}{L }\Bigr)^2 \Bigr] \, .
\label{joker2}
\eeq
Note that implicitly the above expressions require $2\pi L T\ge \alpha/\sqrt{e}$. Otherwise the maximum occurs at $\tau=0$, \ie
\beq
\left. \frac{d \mathcal{C}_A}{d \tau}\right|_{max}=\frac{d \mathcal{C}_A}{d \tau}(\tau=0)=\frac{2 M}{\pi} \left(1+\log\! \left[\frac{\alpha}{2\pi L T}\right]\right)\qquad{\rm for}\ \ 2\pi L T< \alpha/\sqrt{e}\,.
\label{gnu}
\eeq
We observe, however, that the details of  the violation of Lloyd's bound depend on the normalization constant $\alpha$, \ie whether or not the violation is large depends crucially on the choice of $\alpha$.

\subsection{Boundary Counterterm}
\label{sec_ct}
\begin{figure}[htbp]
	\begin{center}
		\includegraphics[keepaspectratio, scale = 0.8]{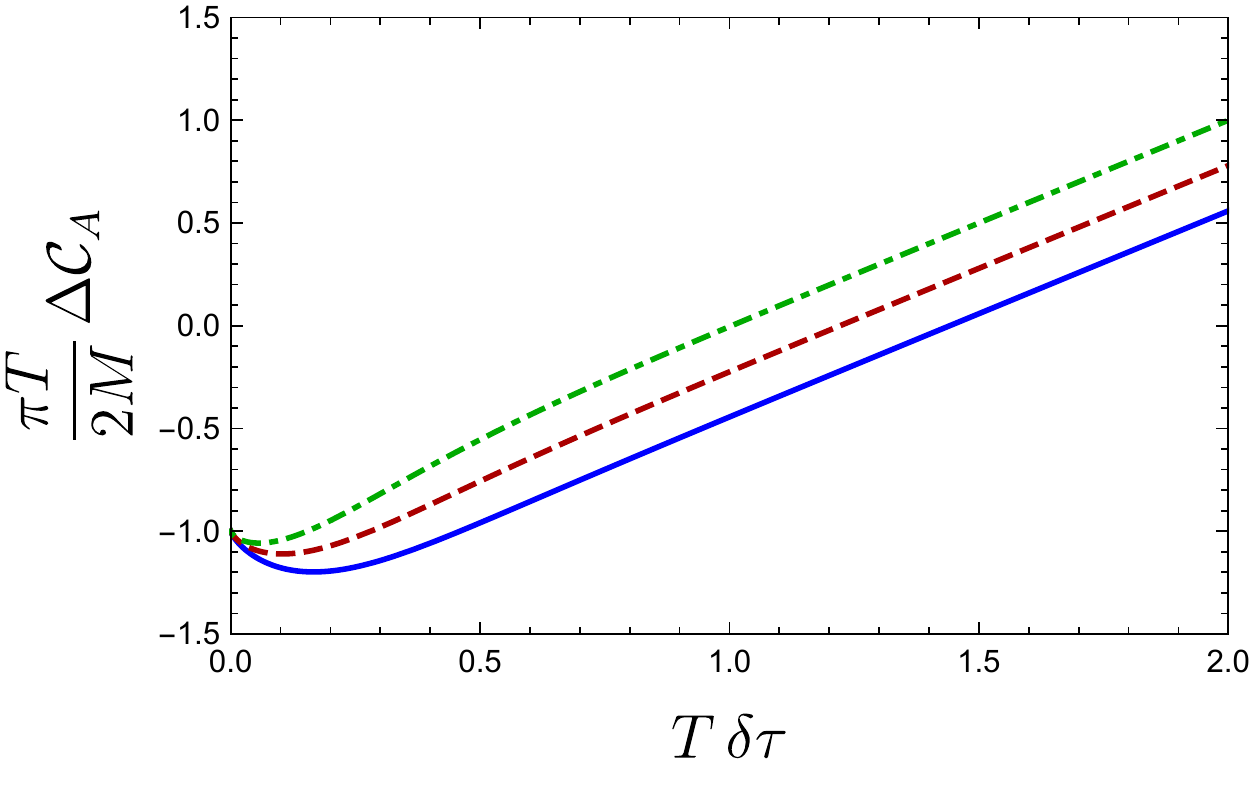}
		\caption{
			Plot of $[\pi/(2M\beta)] \mathcal{C}_A(\tau)$ with $T R =\frac{1}{2 \pi}$
			for $R/\tilde{L}=0.5$ (solid blue), $R/\tilde{L}=1.0$ (dashed red) and $R/\tilde{L}=2.0$ (dot-dashed green).
	}
		\label{action-time_plot}
		\vspace{-3ex}
	\end{center}
\end{figure}

We will now add the boundary counterterm to the action, which was introduced in \cite{RobLuis} to make the action invariant under the reparametrizations of null boundaries of the WDW patch.
As we see in appendix \ref{walk}, the counterterm for the affine parametrization $\lambda=r/\alpha$, which corresponds to the normalization of ${\bf k}_L$ and ${\bf k}_R$ in the previous subsection \ref{sec_nonsym_BTZ}, is\footnote{This expression holds for the general boundary size $2\pi R$.}
\begin{align}
\Delta I^{BTZ}_{\Sigma}
&= -\frac{1}{G_N} r_{max}
\Bigl(\log \frac{r_{max}}{\alpha\tilde{L}}
-1
\Bigr)
+\frac{1}{2 G_N} r_{m}
\Bigl(\log \frac{r_{m}}{\alpha\tilde{L}}
-1
\Bigr),
\label{BTZ_ct}
\end{align}
where $\tilde{L}$ is an arbitrary constant.
Similarly the counter term for pure AdS$_3$ is given by
\begin{align}
\Delta I^{AdS}_{\Sigma}
&= -\frac{1}{2G_N} r^{AdS}_{max}
\Bigl(\log \frac{r^{AdS}_{max}}{\alpha\tilde{L}}
-1
\Bigr),
\end{align}
where we assume that the arbitrary constant $\tilde{L}$ is the same as that in BTZ.
Subtracting this from eq.~\eqref{BTZ_ct}, we obtain the regularized counter term
\begin{align}
\Delta I^{reg} & = \Delta I^{BTZ}_{\Sigma} - 2 \Delta I^{AdS}_{\Sigma}
= \frac{1}{2 G_N} r_{m}
\Bigl(\log \frac{r_{m}}{\alpha\tilde{L}}
-1
\Bigr)
\nonumber \\
&= \frac{r_h}{2 G_N} \tanh \frac{r_h |\tau|}{2L R}
\Bigl[\log \Bigl(\frac{r_h}{\alpha\tilde{L}} \tanh \frac{r_h |\tau|}{2L R}\Bigr)
-1
\Bigr].
\end{align}
Adding this result to eq.~\eqref{btz_regular_action},
the regularized BTZ action with the counter term is given by
\begin{align}
I_{BTZ}&=-\frac{\pi L}{2G_N}+\frac{r_h}{2 G_N}
\tanh \frac{r_h |\tau|}{2LR}\,
\left[\log\left(\frac{R}{\tilde{L}}\sinh\frac{r_h |\tau|}{2 LR}\right)\right] .
\end{align}
Note that $\alpha$-dependence cancels out.
We thus obtain the holographic complexity
\begin{align}
\Delta\mathcal{C}_A(\tau_L,\tau_R) &= \frac{I_{BTZ}}{\pi}
= -\frac{L}{2G_N}+\frac{r_h}{2 \pi G_N}
\tanh \frac{r_h |\tau|}{2LR}\,
\left[\log\left(\frac{R}{\tilde{L}}\sinh\frac{r_h |\tau|}{2 LR}\right)\right]
\\
&= -\frac{c}{3}
+ \frac{2 M}{\pi^2 T}  \tanh \left(\pi T |\tau| \right) \,
\left(\log\left[\frac{R}{\tilde{L}}\sinh\left(\pi T|\tau|\right)\right]\right).
\label{CA_BTZ_counter}
\end{align}
Of course, the boundary counterterm introduces a new arbitrary length scale $\tilde{L}$. Hence we again encounter an ambiguity of the choice of the arbitrary length scale like in the choice of $\alpha$ without the counterterm or the ambiguous factor in the CV conjecture \reef{volver}. The plots of eq.~\eqref{CA_BTZ_counter} for various $R/\tilde{L}$ are shown in figure \ref{action-time_plot}.

\begin{figure}[htbp]
	\begin{center}
		\includegraphics[keepaspectratio, scale = 0.8]{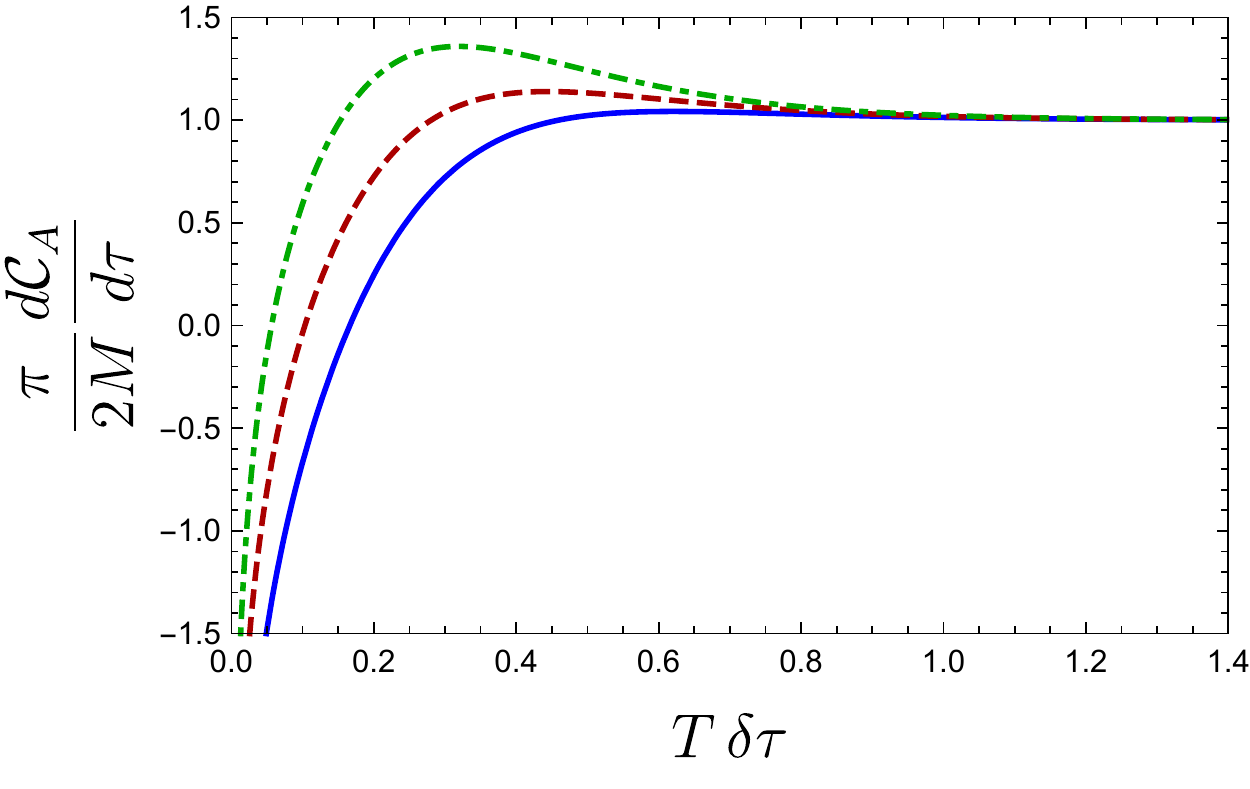}
		\caption{
			Plot of $[\pi/(2M)] d\mathcal{C}_A/d\tau$
			for $R/\tilde{L}=0.5$ (solid blue), $R/\tilde{L}=1.0$ (dashed red) and $R/\tilde{L}=2.0$ (dot-dashed green).
			The curves diverge at $\tau=0$ and approach to 1 from above at late times.
		}
		\label{BTZdCdt_counter}
		\vspace{-3ex}
	\end{center}
\end{figure}

The time derivative of the holographic complexity for $\tau>0$ is
\begin{align}
\frac{d \mathcal{C}_A}{d\tau} =
\frac{2M}{\pi} \left[1
+\frac{\log(\frac{R}{\tilde{L}}\sinh(\pi T \tau))}{\cosh^2(\pi T \tau)}
\right].
\end{align}
We show the plots for various choices of $\tilde{L}$ in Fig.~\ref{BTZdCdt_counter}.
Unlike the case without the counter term,
$d \mathcal{C}_A/d\tau$ is divergent at $\tau=0$ with
\begin{align}
\frac{d \mathcal{C}_A}{dt} \sim
\frac{2M}{\pi} \log\left[\pi T \tau \right]\qquad
{\rm for}\ \ 0<T \tau\ll1\,.
\end{align}
This divergence might be comparable to that found for higher dimensional black holes at $t=t_c$, \ie see eq.~\reef{pop}. However, the complexity of formation \eqref{CA_BTZ_counter} still has a finite value at $\tau=0$, as
\begin{align}
\Delta\mathcal{C}_A(0) = -c/3\,.
\end{align}
This matches the complexity of formation for Neveu-Schwarz vacuum found in \cite{Formation}.
At late times, $d\mathcal{C}_A/d\tau$ behaves as
\begin{align}
\frac{d \mathcal{C}_A}{d\tau} \sim
\frac{2M}{\pi} \left[1 +
4\left(\pi T \tau+\log\frac{R}{2 \tilde{L}}\right)e^{-2\pi T\tau}
\right].
\end{align}
Thus, the rate of growth still approaches the universal limit $2M/\pi$ from above, for any choices of $\tilde{L}$.

\section{Additional Examples of  Time Dependence of Complexity} \label{app:MoreAction}

In eq.~\reef{tder1}, we provided a general expression for the time rate of change of the holographic complexity of (neutral) AdS black holes using the CA conjecture. We examined some specific examples in section \ref{samples} for boundary CFTs with $d=2$ and 4 --- see also appendix \ref{app:BTZnonsymmetric}. Further, in eq.~\eqref{dvdt_general}, together with eqs.~(\ref{eq_rturn}) and \reef{t_r_E}, we provided an expression for the rate of change of complexity based on the CV conjecture, and examined numerically the cases of $d=2$, and planar geometry with $d=3,4$ in subsection \ref{sub_plot_CV}. In this appendix, we provide further examples of the time dependence of holographic complexity.  We show that qualitatively the holographic complexity behaves in the same way in a  different  (odd) dimension, namely $d=3$, using the CA conjecture.  We also explore the influence of the choice of horizon geometry on the results of the CV conjecture in $d=3$ and $d=4$.

\subsection{CA Results in $d=3$}
\begin{figure}
\begin{center}
\includegraphics[scale=0.6]{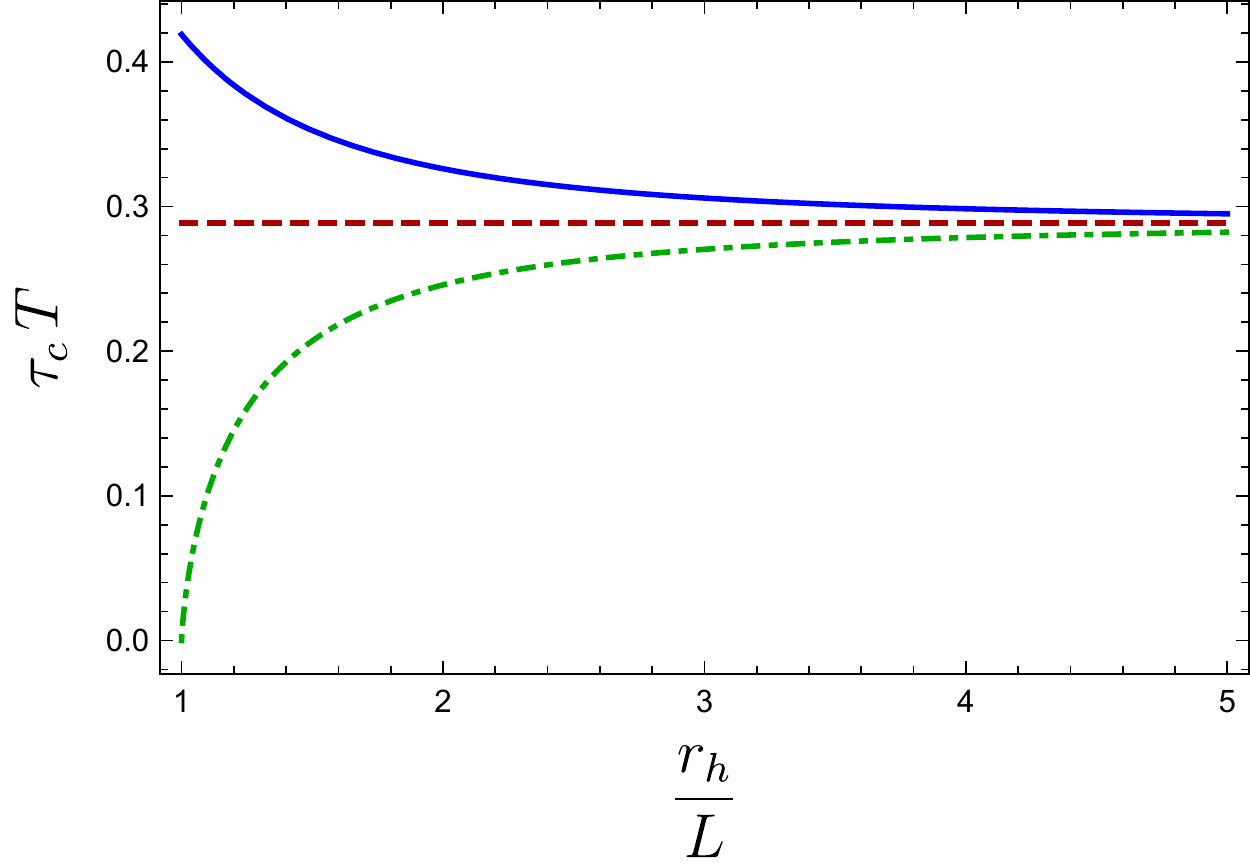}
\caption{Critical time as a function of the horizon radius for $d=3$ for the various geometries -- spherical $k=1$ (blue, solid), planar $k=0$ (red, dashed) and large hyperbolic $k=-1$, $r_h>L$ (green, dot-dashed). }\label{fig:d3ActionEternaltcOrh}
\end{center}
\end{figure}

For the case of $d=3$, we have the dimensionless tortoise coordinate $x^{*} (x, R T) = \frac{r_h}{L^2} \, r^{*}(r)$, where we have used the definition $x\equiv\frac{r}{r_h}$. This leads to
\small
\begin{equation}\label{rstarEqD3}
\hspace{-5pt}x^{*}(x, R T) = \frac{1}{ \frac{k L^2}{r_h^2}+3} \left[ \log \left[\frac{|x-1|}{\sqrt{\frac{k L^2}{r_h^2}+x^2+x+1}}\right] +\frac{\left(\frac{2 k L^2}{r_h^2}+3\right) }{ \sqrt{\frac{4 k L^2}{r_h^2}+3}} \tan ^{-1}\left[\frac{2 x+1}{\sqrt{\frac{4 k L^2}{r_h^2}+3}}\right]  \right]  ,
\end{equation}
\normalsize
and
\begin{equation}\label{vinfD3}
x^*_{\infty} = \frac{\pi  \left(\frac{2 k L^2}{r_h^2}+3\right)}{2 \left(\frac{k L^2}{r_h^2}+3\right) \sqrt{\frac{4 k L^2}{r_h^2}+3}}\, .
\end{equation}
We can evaluate the critical time $\tau_{c}$ using eq.~\eqref{eq:criticalTaume}. This leads to
\small
\begin{align}
\tau_c = \frac{1}{4 \pi  T \sqrt{\frac{4 k L^2}{r_h^2}+3}} \left[ \sqrt{\frac{4 k L^2}{r_h^2}+3} \log \left(\frac{k L^2}{r_h^2}+1\right)+\left(\frac{4 k L^2}{r_h^2}+6\right) \tan ^{-1}\left(\sqrt{\frac{4 k L^2}{r_h^2}+3}\right) \right] \, .
\end{align}
\normalsize
We can apply these results to evaluate the rate of change of holographic complexity for spherical, planar and large hyperbolic black holes. By large hyperbolic black holes, we mean that $r_h/L\ge1$ which implies that the mass is positive. Actually, we assumed here that $r_h>{2L}/{\sqrt{3}}$ for the hyperbolic case with $k=-1$. In the regime $L\le r_h\le{2L}/{\sqrt{3}}$, $f(r)$ has two additional negative real roots. While these do not indicate the existence of additional horizons, the tortoise coordinate is modified in this case and takes the form
\small
\begin{align}
\begin{split}
x^{*}(x, R T) =  & \frac{1}{3-\frac{L^2}{r_h^2} }   \log \left(\frac{|x-1|}{\sqrt{-\frac{L^2}{r_h^2}+x^2+x+1}}\right)
 -  \frac{\left(3-\frac{2 L^2}{r_h^2}\right)}{\left(3-\frac{L^2}{r_h^2}\right) \sqrt{\frac{4 L^2}{r_h^2}-3}}   \coth ^{-1}\left(\frac{2 x+1}{\sqrt{\frac{4 L^2}{r_h^2}-3}}\right) \,
\end{split}
\end{align}
\normalsize
 and the critical time for the hyperbolic black holes in this mass range reads
\begin{equation}
\tau_c  = \frac{1}{4 \pi T \sqrt{\frac{4 L^2}{r_h^2}-3}} \left( \sqrt{\frac{4 L^2}{r_h^2}-3} \log \left(1-\frac{L^2}{r_h^2}\right)+\left(6-\frac{4 L^2}{r_h^2}\right) \tanh ^{-1}\left(\sqrt{\frac{4 L^2}{r_h^2}-3}\right) \right) \, .
\end{equation}
We present a plot of $\tau_c \, T$ as a function of the horizon radius in figure \ref{fig:d3ActionEternaltcOrh}.

\begin{figure}
\begin{center}
\includegraphics[scale=0.6]{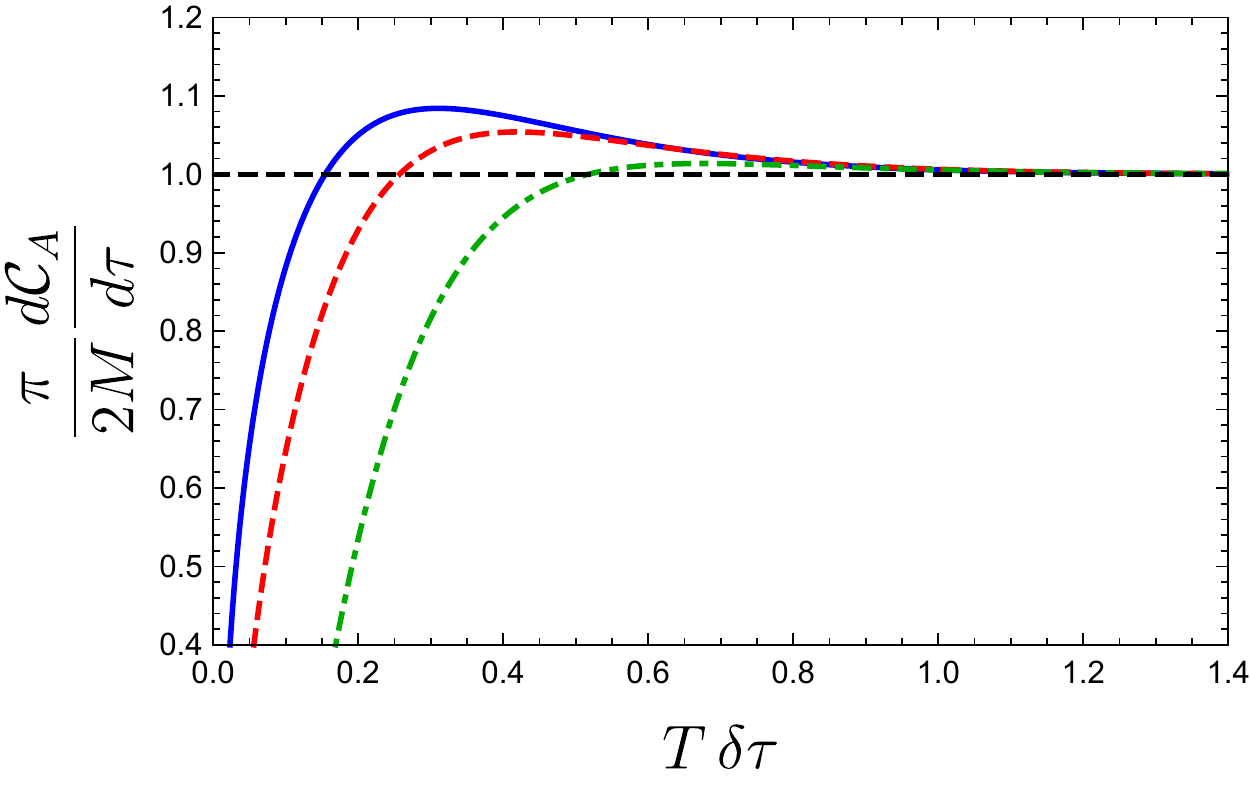}
\includegraphics[scale=0.6]{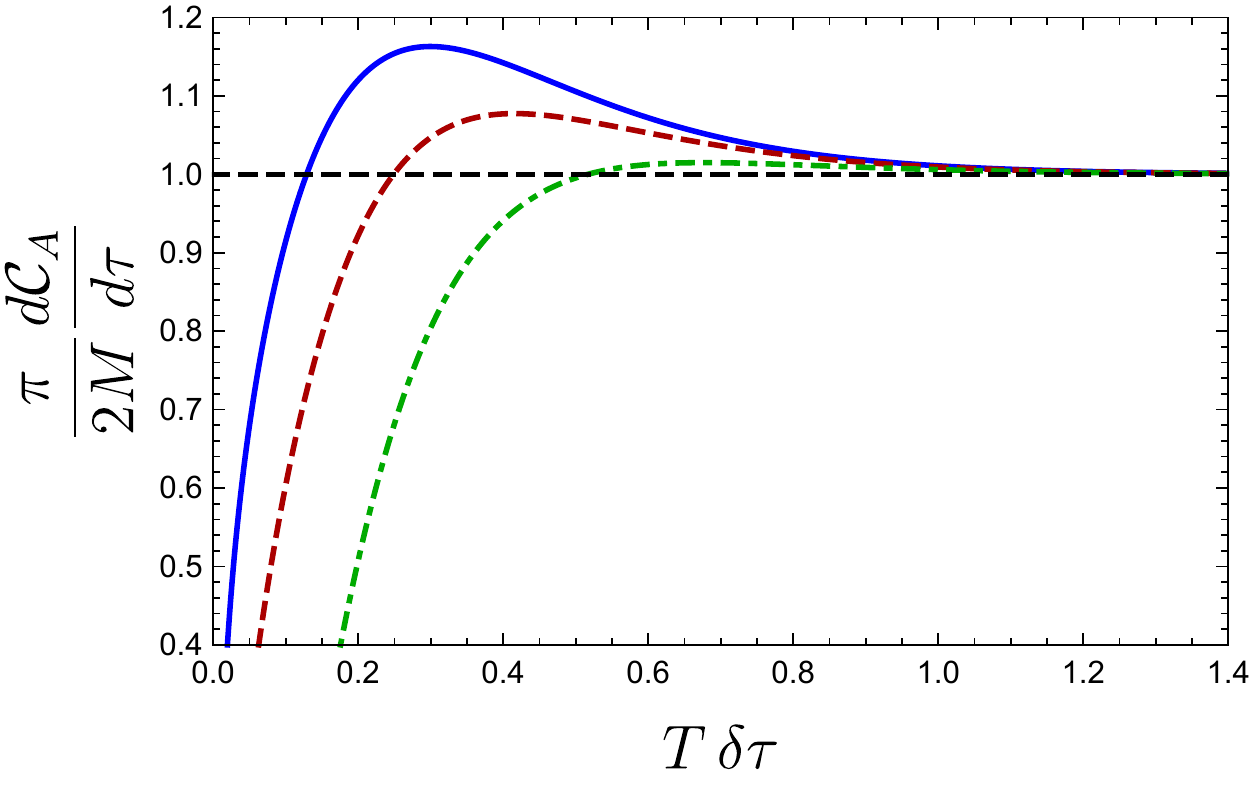}
\caption{Time derivative of complexity as a function of time for spherical (left) and planar (right) geometries in $d=3$ boundary dimensions for various values of the horizon radius -- $r_h= L$ (solid blue),  $r_h=1.5 L$ (dashed red),  $r_h= 3.5 L$ (dot-dashed green). We present the plot as a function of the time coordinate in units of the thermal scale $\delta \tau \, T= (\tau -\tau_c) \, T$. We stress again that the complexity starts changing at $\tau_c$ and each of the curves presented has a different value of $\tau_c$. For these parameters, the violation of the late time bound is clearly manifest.}\label{fig:d3ActionEternal}
\end{center}
\end{figure}

\begin{figure}
\begin{center}
\includegraphics[scale=0.7]{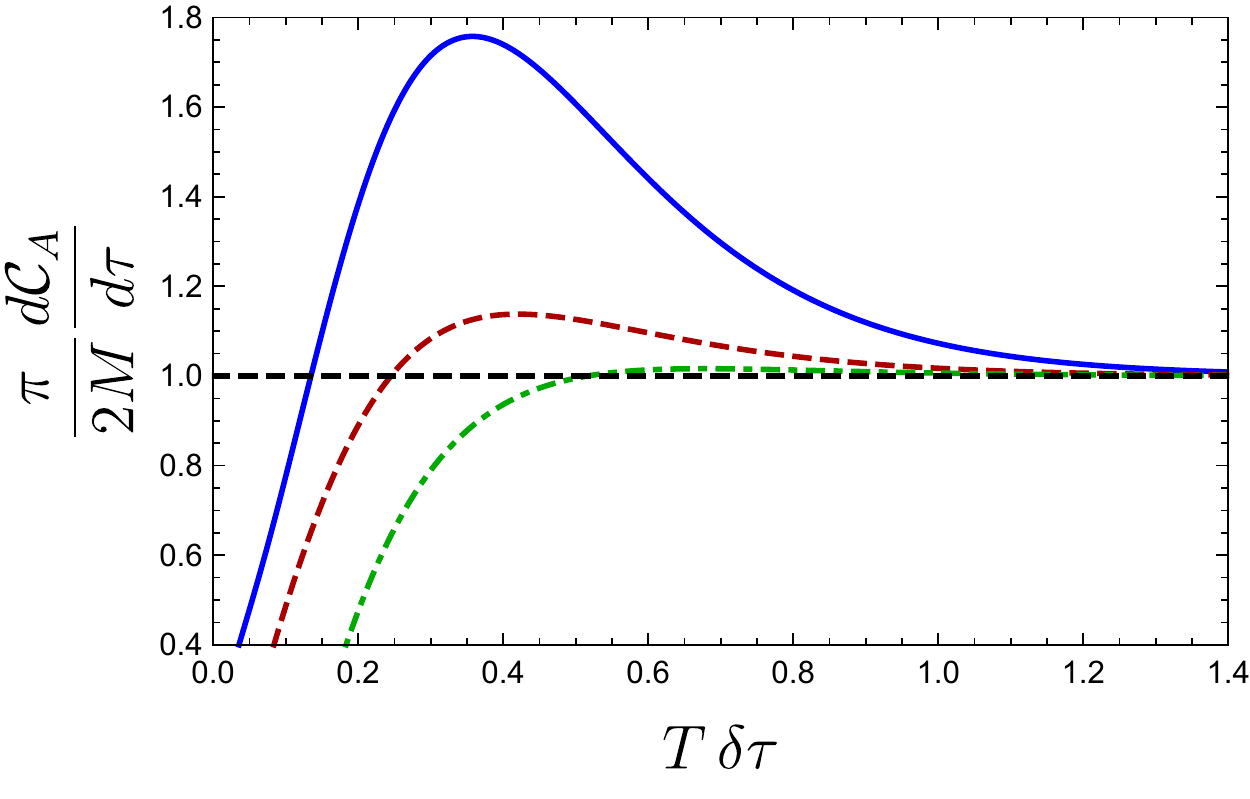}
\caption{Time derivative of complexity as a function of time for large hyperbolic black holes ($r_h > L$) in $d=3$ boundary dimensions for various values of the horizon radius --  $r_h=1.1L$ (blue), $r_h=1.5L$ (dashed red),  $r_h=3.5L$ (dot-dashed green). We present the plot as a function of the time coordinate in units of the thermal scale $\delta \tau \, T= (\tau -\tau_c) \, T$. We stress again that the complexity starts changing at $\tau_c$ and each of the curves presented has a different value of $\tau_c$. For these parameters, the violation of the late time bound is clearly manifested.}\label{fig:d3ActionEternalLHyper}
\end{center}
\end{figure}

\begin{figure}
\begin{center}
\includegraphics[scale=0.39]{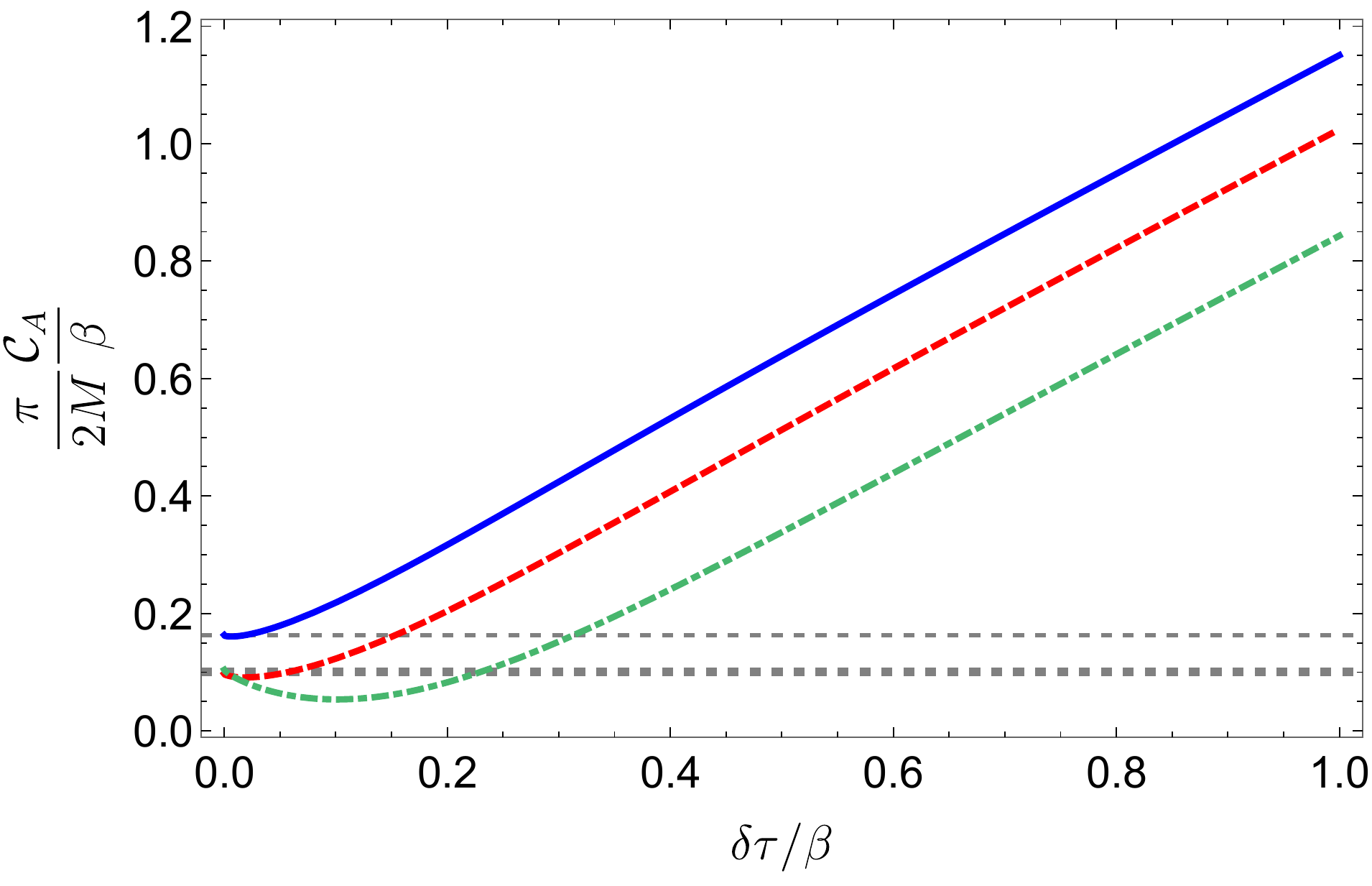}
\includegraphics[scale=0.39]{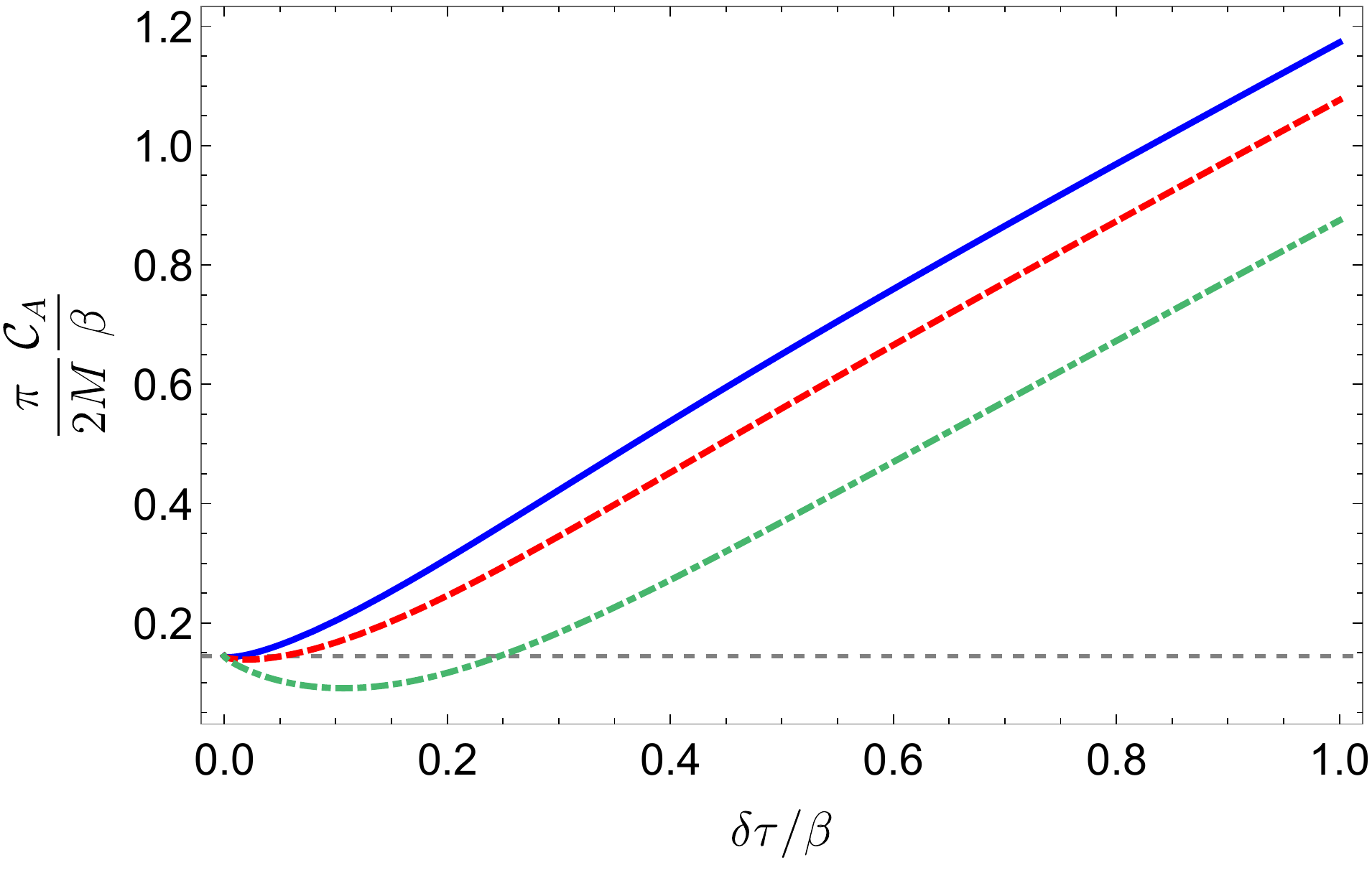}
\caption{
Integrated complexity as a function of time for spherical (left) and planar (right) geometries in $d=3$ boundary dimensions for various values of the horizon radius -- $r_h= L$ (solid blue),  $r_h=1.5 L$ (dashed red),  $r_h= 3.5 L$ (dot-dashed green). We see that it does not diverge at $\tau=\tau_c$ ($\delta \tau =0$). The value at $\delta\tau=0$ has been set according to the complexity of formation, see \cite{Formation}.}\label{fig:d3ActionEternalInt}
\end{center}
\end{figure}

\begin{figure}
\begin{center}
\includegraphics[scale=0.4]{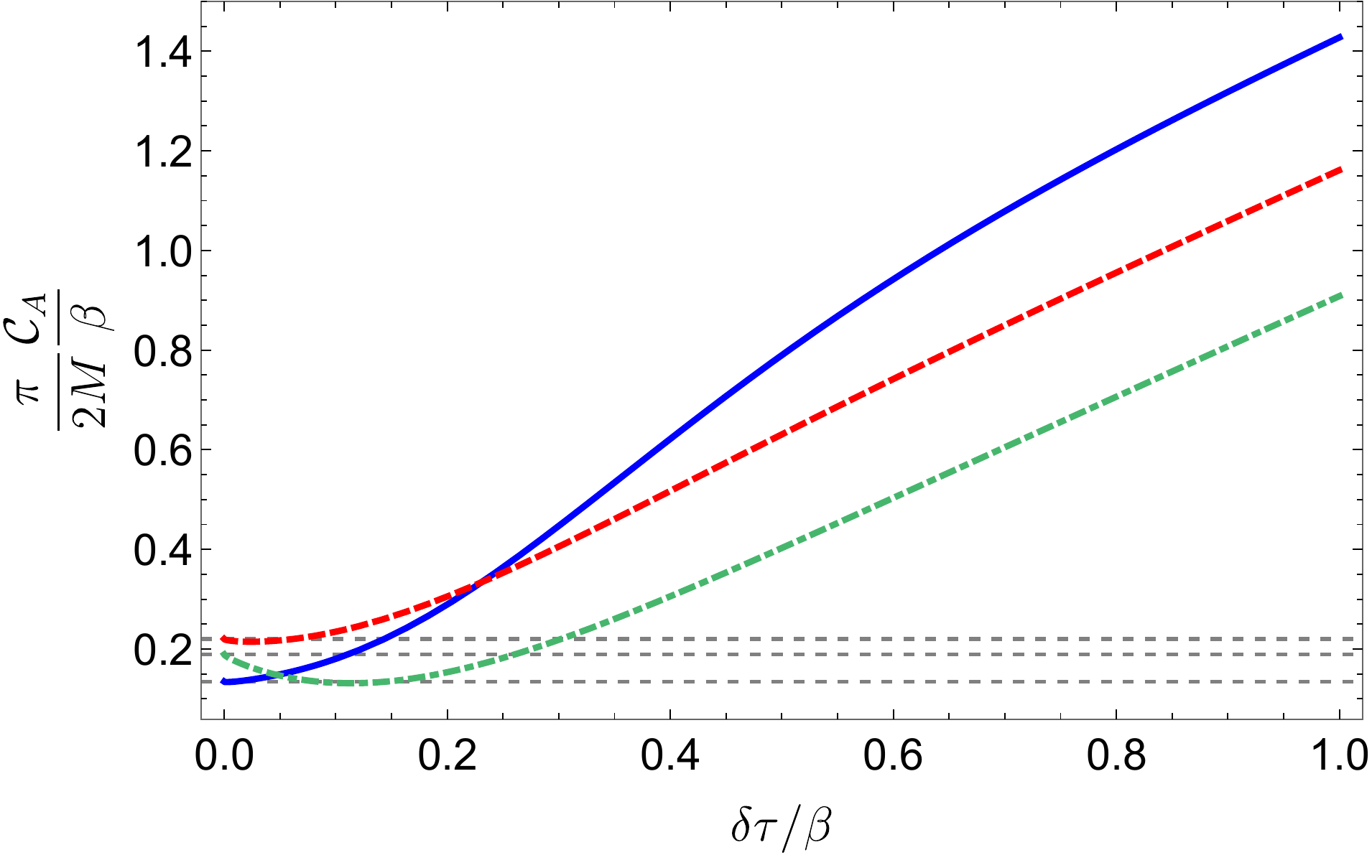}
\caption{Integrated complexity as a function of time for large hyperbolic black holes in $d=3$ boundary dimensions for various values of the horizon radius --  $r_h=1.1L$ (blue), $r_h=1.5L$ (dashed red),  $r_h=3.5L$ (dot-dashed green). We see that it does not diverge at $\tau=\tau_c$ ($\delta \tau =0$). The value at $\delta\tau=0$ has been set according to the complexity of formation, see \cite{Formation}.}\label{fig:d3ActionEternalLHyperInt}
\end{center}
\end{figure}

After solving numerically for $x_m$, the results are presented in figure \ref{fig:d3ActionEternal} for $k=0,1$, and in figure \ref{fig:d3ActionEternalLHyper} for $k=-1$. The overall behaviour of the rate of change of complexity is very similar to the results shown in figure \ref{d4Rate} for spherical and planar black holes in $d=4$.
We also present the integrated complexity in figures \ref{fig:d3ActionEternalInt} and \ref{fig:d3ActionEternalLHyperInt} to demonstrate that there is no divergence near $\tau=\tau_c$. That is, these figures show ${\mathcal C}_A(\tau)- {\mathcal C}_A(\tau_c)= \int_{\tau_c}^\tau d\tau \frac{d {\mathcal C}_A}{d\tau}$. Even though $d {\mathcal C}_A/d\tau$ diverges at the critical time (see eq.~\reef{pop}), it is an integrable singularity and the complexity itself only shows a mild variation at this point. We have also included as the integration constant ${\mathcal C}_A(\tau_c)$ the complexity of formation, see \cite{Formation}. Recall that the complexity of formation is given by the complexity of the thermofield double state minus twice that of  the vacuum state of the CFT, and so presents a natural finite value for the complexity for $|\tau|<\tau_c$. Again we found similar results for $d=4$, although we do not explicitly show the corresponding figures here.

\subsection{CV Results for Other Geometries}
\begin{figure}[htbp]
	\vspace{2ex}
	\begin{minipage}[b]{0.5\hsize}
		\centering
		\includegraphics[keepaspectratio, scale=0.43]{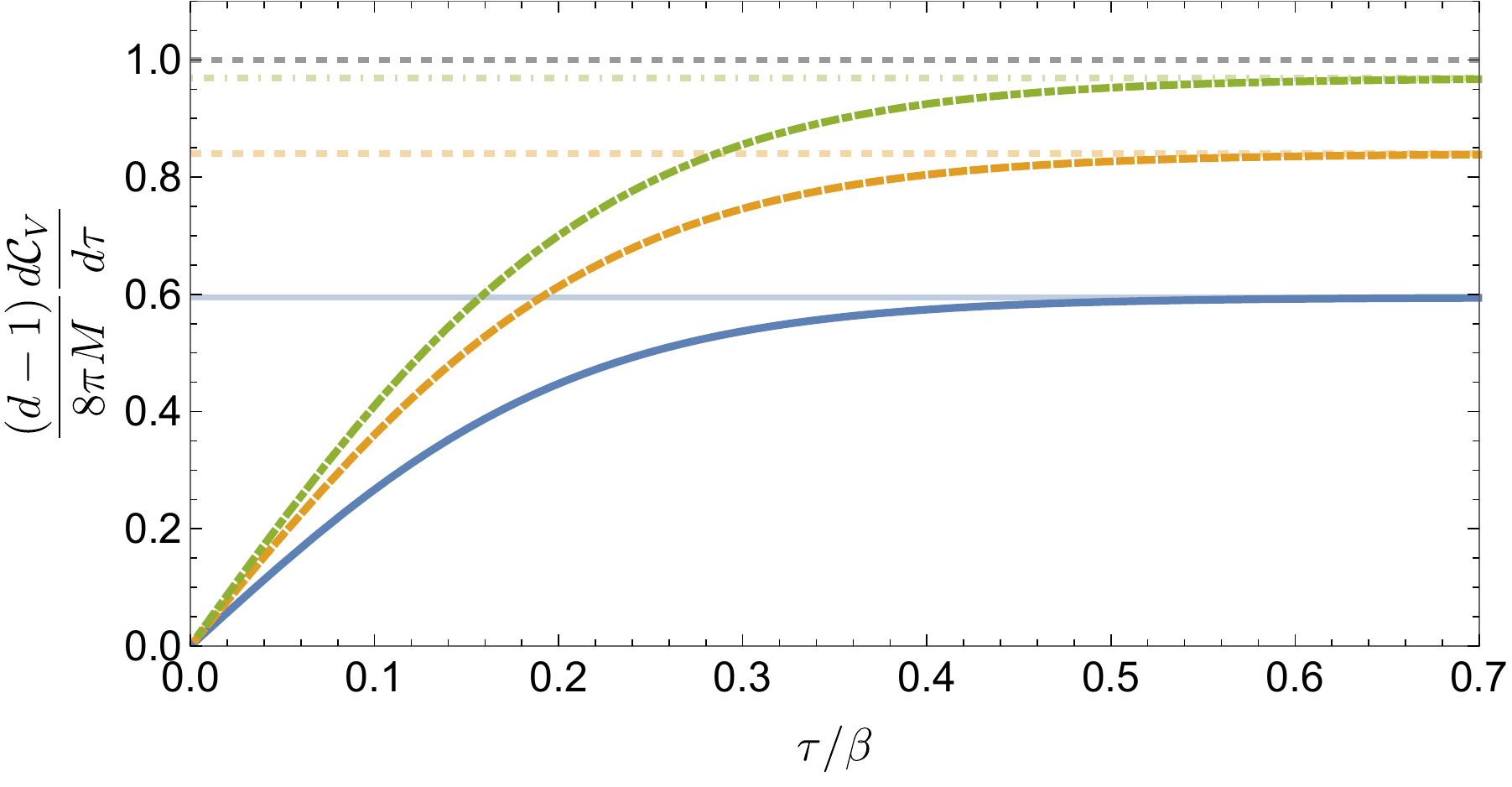}
		\subcaption{$d=3$}\label{dvdt_d=3spherical}
	\end{minipage}
	\begin{minipage}[b]{0.5\hsize}
		\centering
		\includegraphics[keepaspectratio, scale=0.39]{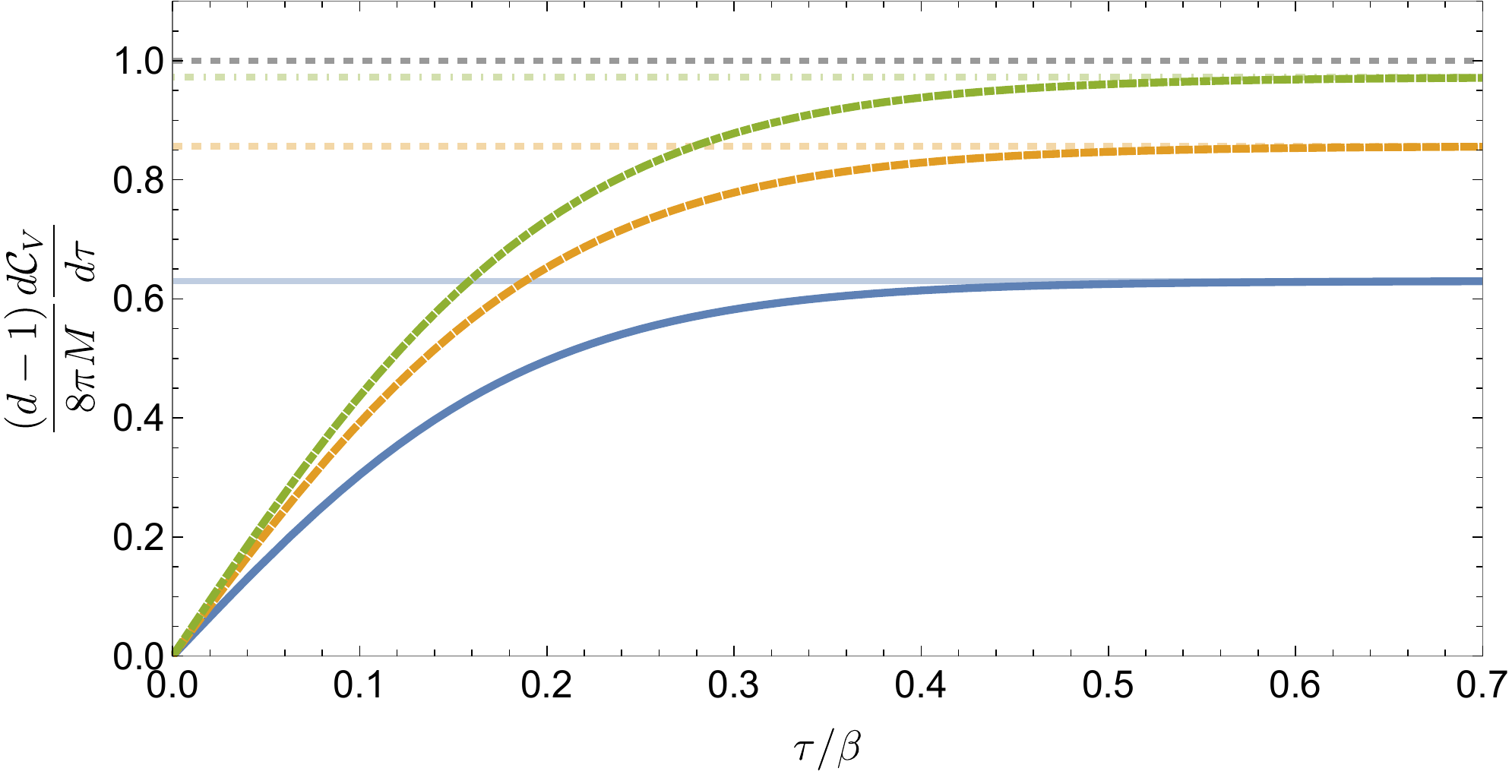}
		\subcaption{$d=4$}\label{dvdt_d=4spherical}
	\end{minipage}
	\caption{
		Plots of $\frac{d-1}{8\pi M}  d\mathcal{C}_V/dt$ for spherical black holes ($k=1$) for various values of the horizon radius -- $r_h/L=1$ (blue), $r_h/L=2$ (yellow), $r_h/L=5$ (green). At late times, they approach to the asymptotic values indicated in figure \ref{asympto_volume_spherical}. The asymptotic value at late times is always smaller than 1 and approaches to 1 for large black holes. We can also see that the asymptotic value is always approached from below.}
	\label{fig_dvdt_spherical}
\end{figure}

\begin{figure}[htbp]
	\vspace{2ex}
	\begin{minipage}[b]{0.5\hsize}
		\centering
		\includegraphics[keepaspectratio, scale=0.47]{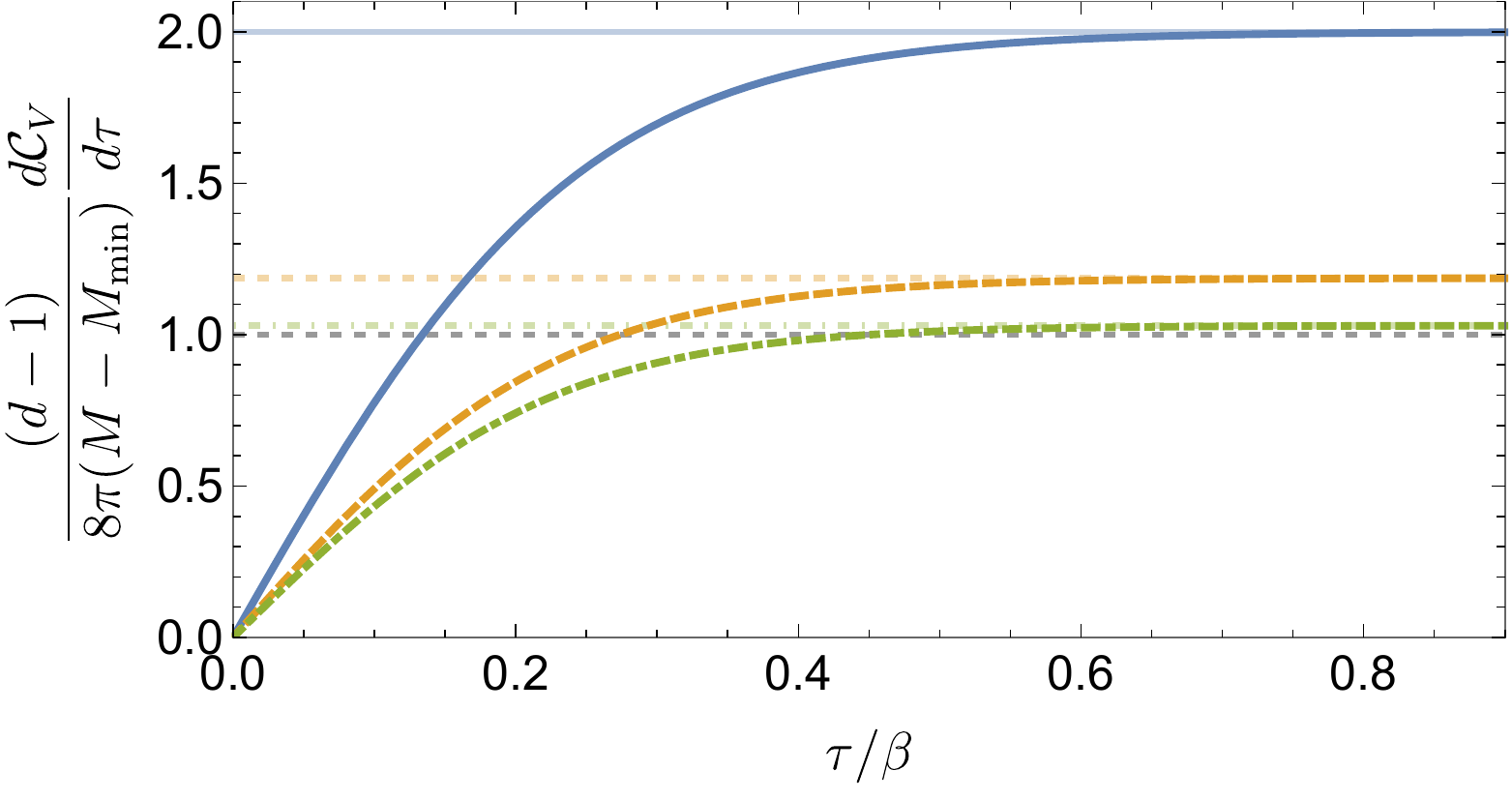}
		\subcaption{$d=3$}\label{dvdt_d=3hyper}
	\end{minipage}
	\begin{minipage}[b]{0.5\hsize}
		\centering
		\includegraphics[keepaspectratio, scale=0.47]{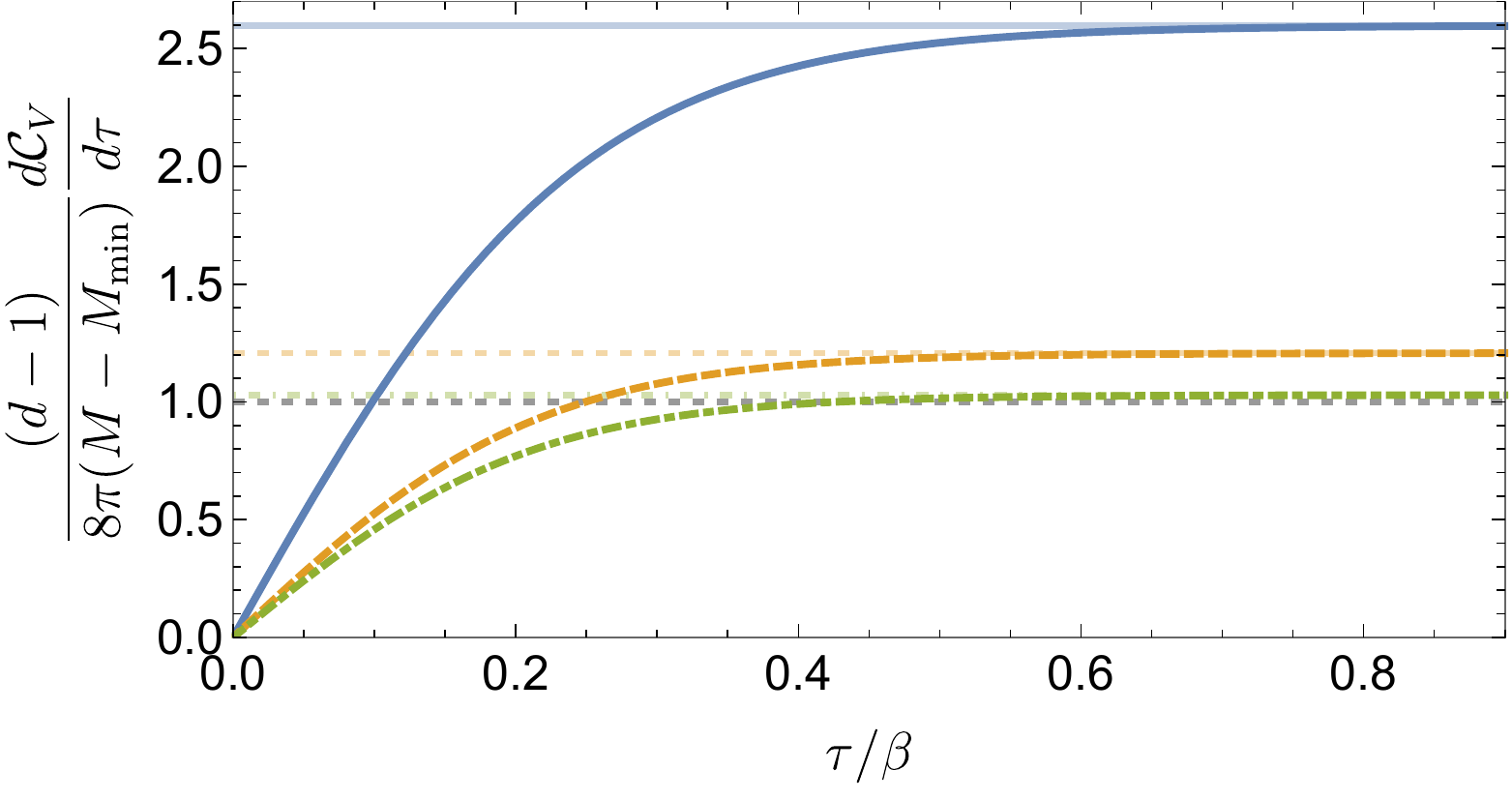}
		\subcaption{$d=4$}\label{dvdt_d=4hyper}
	\end{minipage}
	\caption{
		Plots of $\frac{d-1}{8\pi (M-M_{min})}  d\mathcal{C}_V/dt$ for hyperbolic black holes ($k=-1$) for various values of the horizon radius -- $r_h/L=1$ (blue), $r_h/L=2$ (yellow), $r_h/L=5$ (green).  Recall that $M_{min}$ was introduced to avoid divergences as the mass takes both positive and negative values, see eq.~\eqref{Mminhyper} and the explanation above it.
		The asymptotic values at late times are greater than 1 for small black holes and
		approach to 1 for large black holes. The asymptotic value is always approached from below.}
	\label{fig_dvdt_hyper}
\end{figure}

To complete the picture of the time dependence, we also give some examples of the results of the complexity=volume conjecture for the other geometries (\ie spherical and hyperbolic horizons) in $d=3$ and $d=4$ in figures \ref{fig_dvdt_spherical} and \ref{fig_dvdt_hyper}.

\section{Late Time Behaviour for the CV Proposal} \label{app:MoreVolume}
In this appendix, we provide further details with regards to the late time growth of the holographic complexity, using the CV proposal. In particular, we will determine the leading correction of the late time behaviour of $d\mathcal{C}_V/dt$ given in eq.~\reef{planar_asympt_volume_rate}.

Eq.~\eqref{t_r_E} determines $r_{min}$ as a function of $t$.
Using the function $W(r)$ in eq.~\eqref{w(r)}, eq. \eqref{t_r_E} can be written as
\begin{align}
\frac{t}{2} = -\int^{\infty}_{r_{min}} \!\!\! dr
\frac{W(r_{min})}{f(r)\sqrt{-W(r)^2+W(r_{min})^2}}.
\label{rturn_int_W}
\end{align}
Noting that $\frac{d}{dr}[W(r)^2-W(r_{min})^2]_{r=r_{min}}$ vanishes at $r_{min}=\tilde{r}_{min}$, we introduce a function $Y(r;r_t,\tilde{r}_{min})$ defined as\footnote{This decomposition makes transparent the fact that the denominator of the integral \eqref{rturn_int_W} has generally an order one root, while for $r_{min}=\tilde{r}_{min}$, it has a root of order 2.}
\begin{align}
W(r_{min})-W(r) \equiv (r-r_{min})(r- 2\tilde{r}_{min}+r_{min}) Y(r;r_{min},\tilde{r}_{min}).
\end{align}
We then have
\begin{align}
Y(r_{min};r_{min},\tilde{r}_{min})=-\frac{W'(r_{min})}{2(r_{min}-\tilde{r}_{min})},
\quad  Y(\tilde{r}_{min};\tilde{r}_{min},\tilde{r}_{min})
=- \frac12 W''(\tilde{r}_{min}) .
\end{align}
Separating the integrand in eq.~\eqref{rturn_int_W} as follows
\begin{align}
&-\frac{W(r_{min})}{f(r)\sqrt{-W(r)^2+W(r_{min})^2}}
\nonumber\\
&=\frac{-\sqrt{W(r_{min})}r_{min}}{f(r_{min}) r\sqrt{2(r-r_{min})(r-2\tilde{r}_{min}+r_{min}) Y(r_{min};r_{min},\tilde{r}_{min})}}
+ j(r;r_{min},\tilde{r}_{min}) ,
\end{align}
where
\footnotesize
\begin{align}
&j(r;r_{min},\tilde{r}_{min}) \equiv -\sqrt{W(r_{min})} \nonumber\\
&\times
\left(\frac{
	f(r_{min})\sqrt{2W(r_{min})Y(r_{min};r_{min},\tilde{r}_{min})}r
	-f(r)\sqrt{[W(r)+W(r_{min})]Y(r;r_{min},\tilde{r}_{min})}r_{min}}{f(r_{min})f(r) r
	\sqrt{2\left[W(r)+W(r_{min})\right](r-r_{min})(r-2\tilde{r}_{min}+r_{min}) Y(r_{min};r_{min},\tilde{r}_{min})Y(r;r_{min},\tilde{r}_{min})}}\right)
\end{align}
\normalsize
eq.~\eqref{rturn_int_W} becomes\footnote{Again, we note that the second integral would have a pole at $r=r_h$ and what is actually meant by it is to subtract this pole as described in section \ref{sec:EternalVolume}.}
\begin{align}
\frac{t}{2}
&=\frac{r_{min}^{2d-3/2}\log \Bigl(\frac{\tilde{r}_{min}+\sqrt{r_{min}(2m-r_{min})}}{r_{min}-\tilde{r}_{min}}\Bigr)}
{W(r_{min})^{\frac32}\sqrt{2(2\tilde{r}_{min}-r_{min})  Y(r_{min};r_{min},\tilde{r}_{min})}}
+ \int^{\infty}_{r_{min}} \!\!\!dr j(r;r_{min},\tilde{r}_{min}).
\label{t_rturn_rm}
\end{align}
We then set
\begin{align}
\frac{r_{min}}{\tilde{r}_{min}} = 1+ e^{-\frac{\sqrt{-W(\tilde{r}_{min})^3 W''(\tilde{r}_{min})}}{ \tilde{r}_{min}^{2d-2}}\frac{t}{2}}[c_1 + \epsilon(t)],
\end{align}
where $c_1$ is a constant and $\epsilon(t)$ is a function which goes to zero as $t\to\infty$.
Inserting this form into \eqref{t_rturn_rm} and taking the limit $t\to \infty$,
we obtain
\begin{align}
0=\frac{\tilde{r}_{min}^{2d-3/2}}{\sqrt{-W(\tilde{r}_{min})^3 W''(\tilde{r}_{min})}} \log \frac{2}{c_1}
+ J(\tilde{r}_{min}) ,
\end{align}
where $J(\tilde{r}_{min})$ is a finite function
\begin{align}
J(\tilde{r}_{min})\equiv \int^{\infty}_{\tilde{r}_{min}} \!\!\!dr j(r;\tilde{r}_{min},\tilde{r}_{min}) .
\label{int_j}
\end{align}
In fact, the integrand $j(r;\tilde{r}_{min},\tilde{r}_{min})$ is harmless around $r\sim \tilde{r}_{min}$ and $r\sim \infty$ because it behaves as
\begin{align}
j(r;\tilde{r}_{min},\tilde{r}_{min}) &\sim \tilde{r}_{min}^{2d-1} \frac{[-6(2d-1)W''(\tilde{r}_{min})+W'''(\tilde{r}_{min}) \tilde{r}_{min} ]}
{6[-W(\tilde{r}_{min})W''(\tilde{r}_{min})]^{\frac32}}
+\mathcal{O}(r-\tilde{r}_{min}), \\
j(r;\tilde{r}_{min},\tilde{r}_{min}) &\sim - \frac{\tilde{r}_{min}^{2d-1}}
{\sqrt{-W(\tilde{r}_{min})^3 W''(\tilde{r}_{min})}} \frac{1}{r^2}
+\mathcal{O}(r^{-3}),
\end{align}
and thus $J(\tilde{r}_{min})$ is finite.\footnote{The integrand also has a pole at $r=r_h$ but it can be cured as discussed in section \ref{sec:EternalVolume}.}
Therefore, the late time behaviour of $r_{min}$ is given by
\begin{align}
r_{min}= \tilde{r}_{min} \Bigl[
1+ 2 e^{-\frac{\sqrt{-W(\tilde{r}_{min})^3 W''(\tilde{r}_{min})}}{ 2 \tilde{r}_{min}^{2d-2}}(t-2J(\tilde{r}_{min})) } +\epsilon(t)
\Bigr].
\label{rturn_late}
\end{align}
Using eq.~ \eqref{r_m_eq} the coefficient of $t$ is computed as
\begin{align}
-\frac{\sqrt{-W(\tilde{r}_{min})^3 W''(\tilde{r}_{min})}}{ 2 \tilde{r}_{min}^{2d-2}}
= - \frac{\sqrt{-f(\tilde{r}_{min})\left[(d-1)(d-2)k+d \frac{\tilde{r}_{min}^2}{L^2}\right]}}{2 \tilde{r}_{min}}.
\end{align}
In particular for \emph{planar black holes} (\ie $k=0$),
inserting the analytical expression of $\tilde{r}_{min}$ in eq.~ \eqref{rtild} the coefficient is given by
\begin{align}
-\frac{\sqrt{-W(\tilde{r}_{min})^3 W''(\tilde{r}_{min})}}{ 2 \tilde{r}_{min}^{2d-2}} =- \frac{d\, r_h}{2^{1+\frac{1}{d}} L^2}
=-2^{-\frac{1}{d}} \frac{2\pi}{\beta},
\end{align}
and the rate of change in complexity follows from eq.~\eqref{latetime_dcv/dt} and eq.~\eqref{rturn_late}. One can find that the late time behaviour is given by
\begin{equation}\label{decayrateCVlatetime}
\frac{d-1}{8 \pi M} \frac{d\mathcal{C}_V}{dt} = 1- 2 d^2  e^{-2^{1-\frac{1}{d}} \frac{2\pi}{\beta}
	(t-2J(\tilde{r}_{min}))} + \cdots,
\end{equation}
where the dots stand for corrections which decay faster at late times than the leading exponential in eq.~\eqref{decayrateCVlatetime}.

\section{Complexity of Formation for Charged Black Holes} \label{CformCharged}

In this appendix, we evaluate the complexity of formation for charged black holes. The complexity of formation for uncharged black holes was examined in detail in \cite{Formation}. There, the complexity of formation is defined as the additional complexity involved in preparing two copies of the boundary CFT in the entangled thermofield double state \reef{TFDx} (evaluated at $t_L=t_R=0$) compared to preparing each of the CFTs in their vacuum state. Using the CA proposal,\footnote{Of course, an analogous calculation can also be performed using the CV proposal --- see below.} the bulk calculation consists of evaluating the gravitational action for the WDW patch (anchored at $t_L=t_R=0$) in the (neutral) AdS black hole background and subtracting twice the action for the WDW in an appropriate vacuum of AdS space. A key aspect of this subtraction is that all of the UV (large $r$) divergences cancel, which as a consequence leaves a UV finite result.

Hence in the present charged case, the first question to settle is what is the appropriate reference state to compare to the charged thermofield double state \reef{TFDq}. Here we recall that it was shown in \cite{Chamblin:1999tk} that at zero temperature and with a spherical boundary, the ground state for the fixed chemical potential ensemble is pure AdS for $ \mu<  \frac{g L}{2R \sqrt{2\pi G}} \sqrt{\frac{(d-1)}{(d-2)}}$  and an extremal black hole of the same chemical potential for $ \mu>\frac{g L}{2R \sqrt{2\pi G}} \sqrt{\frac{(d-1)}{(d-2)}}$. It was also noted there  that this extremal black hole may be unstable and decay by the emission of charged particles. For the planar boundary geometry (\ie $k=0$) and the hyperbolic one (\ie $k=-1$), the ground state is always the extremal black hole.

Hence in evaluating the complexity of formation for the charged thermofield double state, one suggestion is to subtract the holographic complexity corresponding to an extremal black hole with the same chemical potential \cite{Brown2}. However, we find that the holographic complexity for an extremal black hole contains an additional infrared divergence and hence a meaningful comparison cannot be achieved by comparing a charged black hole to the corresponding extremal one. We will see that this IR divergence appears for both the CA and the CV conjectures. Therefore, we simply choose the uncharged vacuum ($\omega = q = 0$) as our reference state, \ie we subtract the holographic complexity of two copies of the corresponding AdS vacuum.

As in section \ref{sec:ChargedEternal}, it is convenient to work with the dimensionless variables introduced in eq.~\reef{eq:xyz}. Recall
\begin{equation}\tag{\ref{eq:xyz}}
x\equiv\frac{r}{r_+}, \qquad
y\equiv\frac{r_-}{r_+}, \qquad
z\equiv\frac{L}{r_+}.
\end{equation}
The first is a dimensionless radial coordinate, while the latter two can be defined in terms of boundary quantities, as in eq.~\reef{eq:RTnu}. Further, in the following, we will focus on the case of $d=4$, where that latter expressions are explicitly given in eq.~\reef{eq:RTnuSphAdS5}. In principle then, we can invert these formula to write our results in terms of the boundary quantities, $\nu=\sqrt{C_J/C_T}\,\mu/T$ and $RT$. In the planar geometry, \ie $k=0$, for $d=4$ eq.~\reef{eq:RTnuSphAdS5} reads
\begin{equation}
\nu = \sqrt {\frac{C_J}{C_T}} \frac{\mu}{T} = \frac{3 \pi}{\sqrt{10}}\,\frac{ y \sqrt{y^2+1}}{ (1-y^2)(2+y^2)}\,,
\qquad
RT = \frac{1}{2 \pi} \frac{(1-y^2) (2+ y^2)}{z}\,.
\label{razzle}
\end{equation}
Then the first of these equations can be inverted  to obtain
\begin{equation}\label{yofnu}
y^2 =\frac{\sqrt{3} \sqrt{15 \nu^2-\pi  \sqrt{80 \nu^2+9 \pi ^2}+3 \pi ^2}}{2\sqrt{5}\,\nu}-\frac12
\end{equation}
and for the second, we may write
\beq
z = \frac{1}{2 \pi} \frac{(1-y^2(\nu)) (2+ y^2(\nu))}{RT}\,.
\eeq

\subsection{Complexity=Action}\label{sec:CAForm}

Using the CA proposal, the complexity of formation is given by:
\begin{equation}\label{CformAM}
\Delta \mathcal C_{A} = \frac{1}{\pi} \left[\Delta I_{\bulk} + I_{\jnt}\right]
\end{equation}
where
\begin{equation}\label{CformAM2}
\begin{split}
\Delta I_{\bulk} = & \frac{\Omega_{k,d-1}}{2\pi  G_N}
\int_{r_m}^{r_{\max}}
\left(-\frac{d}{L^2} +\frac{ q^2 (d-2)}{r^{2(d-1)}}\right) r^{d-1} \left(
r^*_{\infty}-r^*(r)\right)dr \\
& +
\frac{d\Omega_{k,d-1}}{2 \pi G_N  L^2}
\int_0^{r_{\max}^{\vac}} r^{d-1}
\left(r^{*}_{\infty,\text{vac}}-r^*_{\text{vac}}(r)
\right)dr
\end{split}
\end{equation}
and
\begin{equation}\label{CformAM3}
I_{\jnt} = -\frac{\Omega_{k,d-1}}{4\pi G_N} r_m^{d-1} \log \frac{L^2 |f(r_m)|}{R^2 \alpha^2}.
\end{equation}
The meeting point $r_m$ is obtained by numerically solving \eqref{eq:meeting345} for $\tau=0$, \ie
\begin{equation}
r^*(r_m)= r^*_{\infty}\,. \label{bumb9}
\end{equation}
Note that here the future and past meeting points are at the same value of the radial coordinate, \ie $r_m^1 = r_m^2 = r_m$. Further, $r_{\max}$ corresponds to the UV cutoff $z=\delta$ in the Fefferman-Graham expansion of the respective metric.

\subsubsection{Planar $d=4$}

We proceed by analyzing charged planar black holes. Recall that with $q=0$, the planar black holes produced $\Delta \mathcal C_{A} =S/(2\pi)$ where $S$ is the entanglement entropy of the thermofield double state \reef{TFDx} \cite{Formation}. For the curved horizons, there were curvature corrections to this simple result, proportional to inverse powers of $RT$. Below, we will find that this expression receives corrections even with $k=0$ in the charged case. Since the curvature vanishes, all of the nontrivial behaviour comes from the finite chemical potential.

As before, we redefine the tortoise coordinate \reef{tort2} in terms of dimensionless variables
\begin{equation}\label{blackxyz}
\begin{split}
\tilde f(x,y) \equiv & \, z^2 f(r) = \frac{\left(x^2-1\right) (x-y) (x+y) \left(x^2+y^2+1\right)}{x^4}
\\
x^*(x,y) \equiv &  \frac{r^*(r)}{z^2 r_+} = \int^x \frac{dx}{\tilde f(x,y)} = \frac{y^3}{4 y^4-2 y^2-2} \log \frac{|x-y|}{x+y}
\\
&
-\frac{1}{2 \left(y^4+y^2-2\right)} \log \frac{|x-1|}{x+1} +
\frac{\left(y^2+1\right)^{3/2}}{2 y^4+5 y^2+2}
\tan ^{-1}\left(\frac{x}{\sqrt{y^2+1}}\right).
\end{split}
\end{equation}
This allows us to rewrite eq.~\reef{bumb9} for the meeting points as
\begin{equation}
x^*(x_m,y)=x^*_\infty = \frac{\pi  \left(y^2+1\right)^{3/2}}{4 y^4+10 y^2+4}
\end{equation}
where $x_m\equiv r_m/r_+$. Given eq.~\reef{yofnu}, we see that $x_m$ is a function of $\nu$ only.

There is a subtlety in numerically solving for the meeting point for small values of the charge. The reason is that $r_-$ approaches zero as $r_-^{d-2} = q^2/\omega^{d-2}$ and the tortoise coordinate peaks very sharply around $r_-$. The meeting point equation $r^*_\infty = r^*(r_m)$ solves for the point in which the asymptotic value of the tortoise coordinate intersects back with the curve. As a consequence of the special form of the curve for small values of $r_-$, this happens very close to $r_-$. In fact, in the limit that $r_-$ (or equivalently $y$) approaches zero, the meeting point can be approximated by (see eq. \eqref{eq.earlytimesmallcharge} with $\tau=0$ and $k=0$):
\begin{equation}
 x_m =   y \left(1+\exp\left({-\frac{\pi }{2 y^3 }+\mathcal{O}\left(\frac{1}{y}\right) }\right)\right)\, .
\end{equation}
This means that the corner contribution is nonvanishing  in the $r_-\rightarrow 0$ limit despite the fact that $r_m$ approaches zero. In our plots, we have used similar approximations for the cases of small $\nu$.

\begin{figure}[htbp]
\centering
\includegraphics[keepaspectratio, scale=0.6]{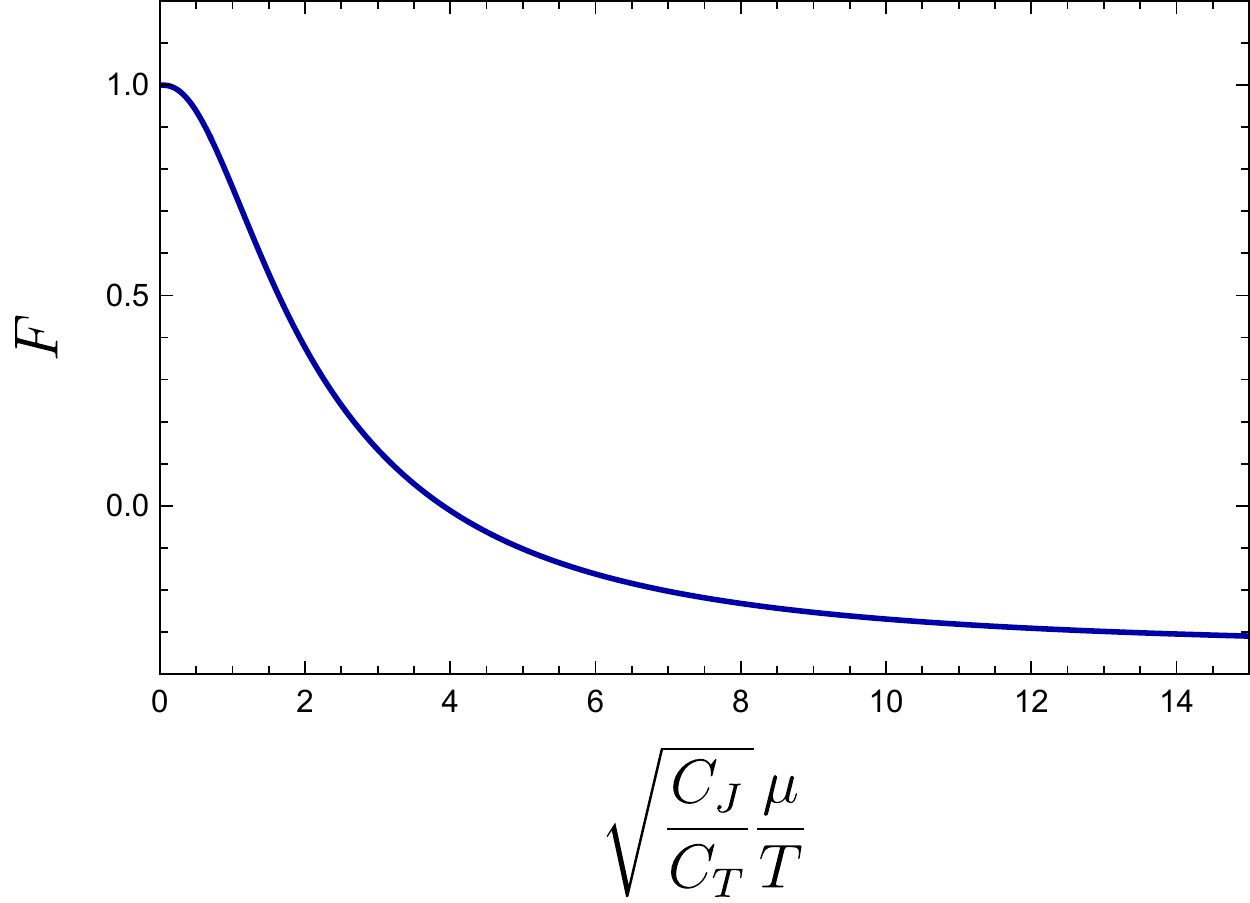}
\includegraphics[keepaspectratio, scale=0.6]{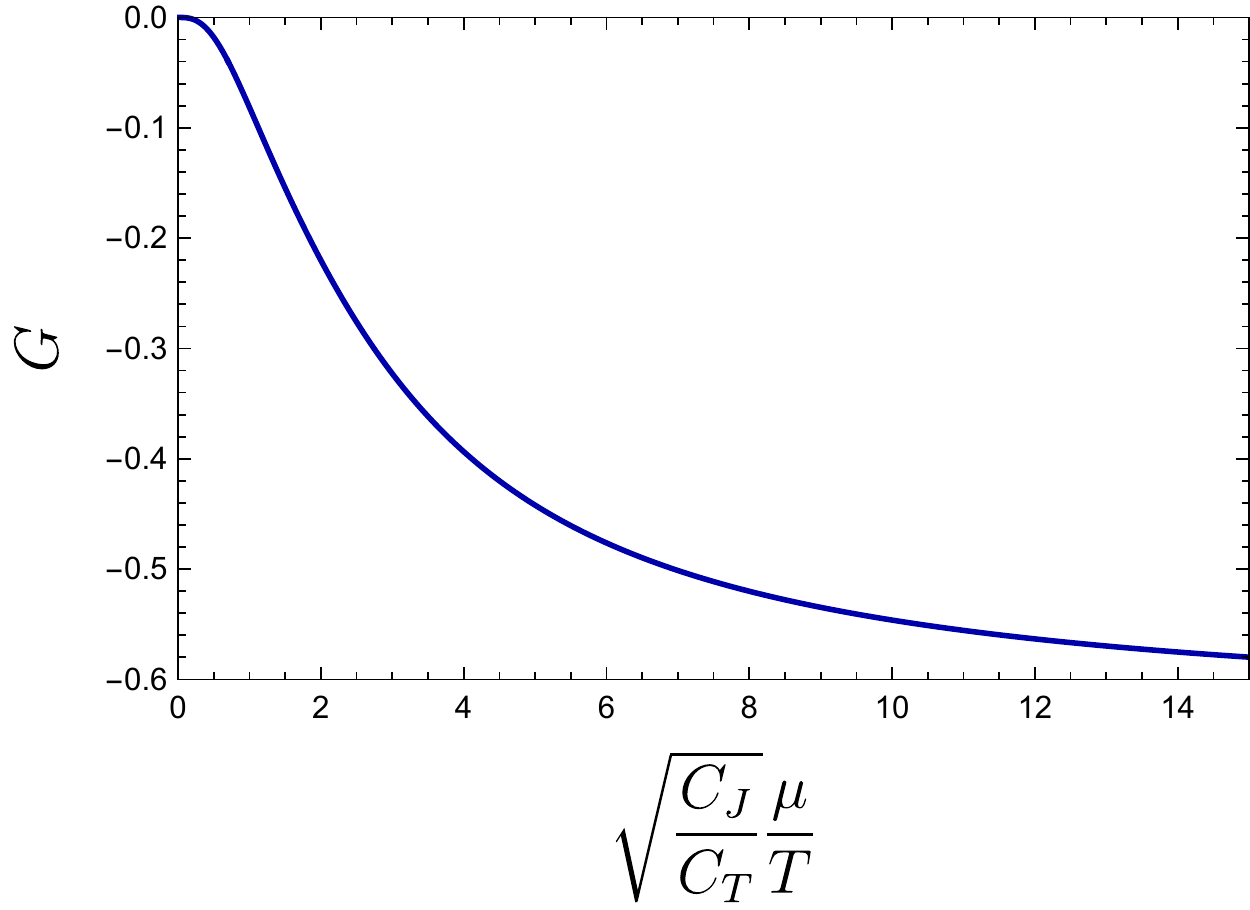}
	\caption{The functions $F(\nu)$ and $G(\nu)$ defined in eq.~\reef{dazz2} which appear in the complexity of formation \reef{ComplexityChargedFormT} for charged planar AdS$_{5}$ black holes as a function of $\nu \equiv \sqrt{\frac{C_J}{C_T}}\frac{\mu}{T}$.}
	\label{ChargedFormPlanarAdS51}
\end{figure}

Motivated by the results of \cite{Formation} for the neutral case, we will be interested in evaluating the ratio of complexity of formation over entropy. Using eq.~\eqref{CformAM} we find
\begin{equation}
\frac{\Delta \mathcal{C}_{\text{form}}}{S}  = \frac{1}{ \pi} \left[\frac{\Delta I_{\bulk}}{S} + \frac{I_{\jnt}}{S}\right]
\end{equation}
where
\begin{equation}
\begin{split}
\frac{\Delta I_{\bulk}}{S} = &
\frac{8}{\pi } \frac{x_{\max}^3}{3} +
\int_{x_m}^{x_{\max}}
\frac{4 }{\pi  x^3}\left(-2 x^6+y^4+y^2\right)  \left(
x^*_{\infty}-x^*(x,y)\right)dx
\end{split}
\end{equation}
and
\begin{equation}
\frac{I_{\jnt}}{S} = -\frac{ x_m^{d-1}}{\pi } \log \frac{r_+^2 |\tilde f(x_m,y)|}{R^2 \alpha^2}
=
-\frac{ x_m^{d-1}}{\pi } \log \left|\frac{g_2(x_m,y) L^2 T^2}{\alpha^2}\right|
,
\end{equation}
where we have defined
\begin{equation}
g_2(x,y) =
\frac{4 \pi ^2 \left(x^2-1\right) \left(x^2-y^2\right) \left(x^2+y^2+1\right)}{ x^4 \left(y^2-1\right)^2 \left(y^2+2\right)^2}
\end{equation}
and the planar black hole complexity of formation is regularized at infinity by subtracting two copies of the vacuum \cite{Formation}. A meaningful comparison between the two spacetimes is achieved by placing the cutoff at $x_{\text{max}}\equiv r_{\text{max}}/r_+$ corresponding to $z=\delta$ in the Fefferman-Graham expansion of the respective metric (see \eg appendix A of \cite{Formation}).
We see that the complexity of formation can be naturally split into a sum of two functions
\begin{equation}\label{ComplexityChargedFormT}
\Delta \mathcal{C}_{A} \equiv \frac{S}{2\pi} \left(F (\nu) + G(\nu) \log \left( \frac{T^2 L^2}{\alpha^2} \right) \right) \, ,
\end{equation}
where $S$ is the entropy of the charged AdS black hole, given in eq.~\reef{charge22}; and $F(\nu)$ and $G(\nu)$ are universal functions that depend only on the ratio $\nu$ through their dependence on $y$ as follows
\begin{equation}
\begin{split}
G(\nu)  = G(y) = & -\frac{2 }{\pi}\,x_m^{d-1} \,,
\\
F(\nu) =  F(y) = & -\frac{2 }{\pi}\,x_m^{d-1} \log 	|g_2(x_m,y)|  +
\frac{16}{\pi } \frac{x_{\max}^3}{3}
\\
&
\quad -
\int_{x_m}^{x_{\max}}
\frac{8 }{\pi  x^3}\left(2 x^6-y^4-y^2\right)  \left(
x^*_{\infty}-x^*(x,y)\right)dx \,.
\end{split}\label{dazz2}
\end{equation}
We note that our result for the complexity of formation depends on the arbitrary parameter $\alpha$ associated to the normalization of null normals.
The two functions $G(\nu)$ and $F(\nu)$ are shown in figure \ref{ChargedFormPlanarAdS51} as a function of $\nu=\sqrt{\frac{C_{J}}{C_{T}}} \frac{\mu}{T}$. Note that
in the limit $\nu \rightarrow 0$, the complexity of formation agrees with the uncharged result found in \cite{Formation}, \ie
$F(\nu\to0)\to1$ and $G(\nu\to0)\to0$.

As we showed in section \ref{sec:ChargedEternal}, we can write an expansion of the complexity of formation for small charge as an expansion in the parameter $y$, which reads
\begin{equation}
\tag{\ref{eq:SeriesCoFPlan}}
\Delta \mathcal{C}_A = \frac{S}{2 \pi}\left(1 + \left( \frac{20}{3 \pi} +\frac{4}{\pi} \log\!\left[ \frac{y z}{2} \frac{\alpha R}{L}\right] \right) y^3
 + \cdots\right) \, .
 \end{equation}
In order to probe the limit of extremal black holes, \ie $T \rightarrow 0$ with $\mu$ finite, we investigate eq.~\eqref{ComplexityChargedFormT} in this limit. The result is divergent in the $T \rightarrow 0$ limit. To see this we use the expansion for $x_m$ near extremality
\begin{equation}\label{eq.extd4k011}
x_m=1-\frac{\epsilon }{2}+\frac{7}{12} \epsilon ^2 \log\epsilon -\frac{8 \sqrt{2} \pi +3+28 \log (2)-16 \sqrt{2}\, \cot^{-1}\!\sqrt{2}}{24} \,\epsilon ^2  + \cdots \, ,
\end{equation}
where we have defined $y\equiv 1-\epsilon$, and evaluate the complexity of formation
\begin{equation}\label{eq.extd4k022}
\Delta \mathcal{C}_{A}  =
\frac{2S}{\pi^2} \left(
\log \left( \frac{\alpha}{L T} \right) + \frac{1}{3}-\log \left(\frac{\pi  }{\sqrt{3}  }\right)
+ \mathcal{O} \left( R T \, \log \, R T \right) \right)\, .
\end{equation}
Note that the limit $RT\rightarrow 0$ corresponds to the limit $\nu\rightarrow \infty$, so the correction, where we have left implicit a function of $z$, is in fact a function of $\nu$ only.
We find that the result diverges logarithmically at low temperatures and the coefficient of the logarithmic divergence is proportional to the entanglement entropy of the system. The result also depends on the arbitrary length scale $\ell\equiv L/\alpha$ associated to the normalization of null normals. We will see in the next subsection that a similar divergence at low temperatures appears using the CV conjecture.

\subsubsection{Spherical $d=4$}

\begin{figure}
\centering
\includegraphics[scale=0.5]{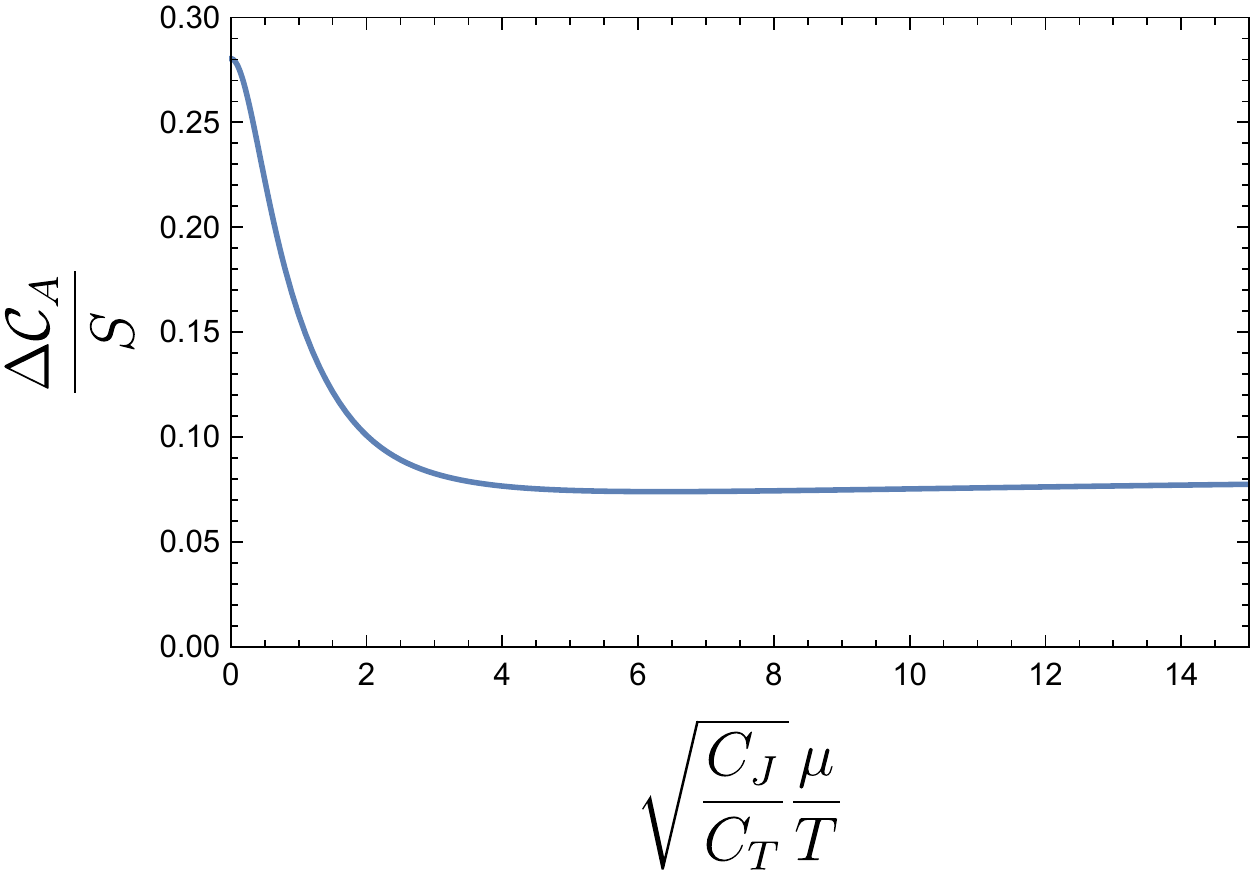}~~
\includegraphics[scale=0.55]{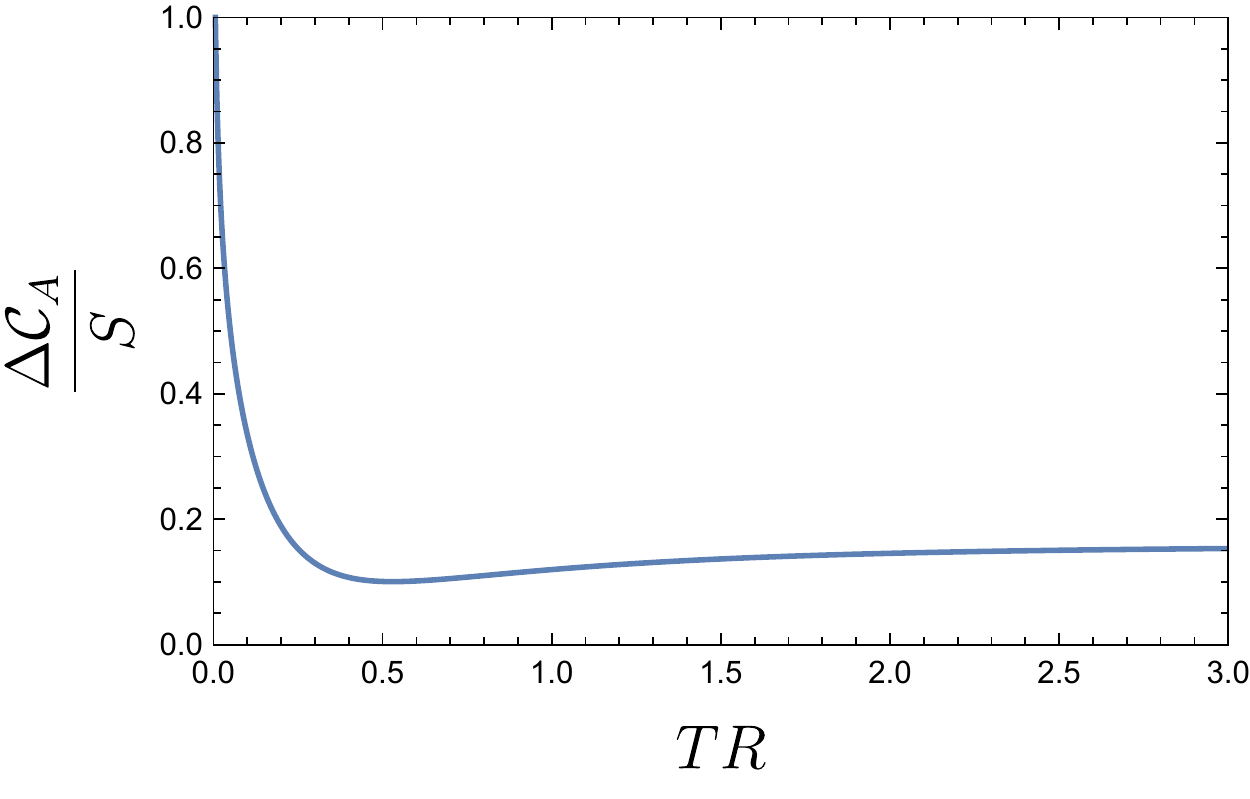}~~
\caption{Complexity of formation for spherical charged black holes in $d=4$. In the left panel,  we fix $R T= \frac{1}{2}$ and we show the dependence on the dimensionless boundary quantity $\nu$. In the right panel, we fix the quantity $\sqrt{\frac{C_J}{C_T}} \mu R=\nu RT = 1$ and show the dependence on $RT$.}
\label{fig:CFormationSph}
\end{figure}

The calculation of the complexity of formation for spherical charged black holes follows closely the one of the planar case. However, the two contributions from eq.~\eqref{CformAM} need to be evaluated using the appropriate blackening factor \eqref{ChargedMetric} with $k=1$. We show the results for $d=4$ in figure \ref{fig:CFormationSph} and note that again the complexity of formation diverges in the low temperature (near extremal) limit.

As in the planar case, we can find the leading behaviour when  $R T$ is small. The expansion for the meeting point reads
\small
\begin{align}
x_m = &1-\frac{\epsilon }{2}+ \frac{\left(3 z^2+7\right) \epsilon ^2 \log (\epsilon )}{4 z^2+12}-
\\
& -\frac{\epsilon ^2 \left(z^2 (1+12 \log (2))+8 \left(z^2+2\right)^{3/2} \tan ^{-1}\left(\sqrt{z^2+2}\right)+3+28 \log (2)\right)}{8 \left(z^2+3\right)} +\mathcal{O}(\epsilon^3 \log\epsilon)\nonumber
\end{align}
\normalsize
and that of the complexity of formation
\small
\begin{align}
&\Delta \mathcal{C}_{A}  =
\frac{S}{3 \pi^2 (3 +z^2)} \left(
-9 \left(z^2+2\right) \log \left(\frac{\pi  R T z}{z^2+3}\right)-3 \left(z^2+3\right) \log \left(\frac{L^2 \left(z^2+3\right)}{\alpha ^2 R^2 z^2}\right)+z^2 \log 64 \right.\nonumber \\
&~~~\left.+\left(z^2+3\right) \left(3 (\pi  z-2) z^2+2\right)-6 z^2 \left(z^2+2\right)^{3/2} \tan ^{-1}\left(\sqrt{z^2+2}\right) +\mathcal{O} \left( R T \, \log \, R T \right) \right) \, .
\end{align}
\normalsize
Notice that as $z \rightarrow 0$, we recover the planar result in eq.~\eqref{eq.extd4k022}.
However, unlike in the planar case,
now the overall coefficient that controls the divergence for small temperatures depends on $z$, which in turn depends on the product of the boundary size and the chemical potential.
The exact relation is obtained from eq.~\eqref{eq:RTnuSphAdS5}, which leads to the relation
\begin{equation}
z  = \frac{3 \sqrt{2}}{\sqrt{40 \left(\sqrt{\frac{C_J}{C_T}} \mu  R\right)^2-9}}\, .
\end{equation}
where $C_J$ and $C_T$ are the coefficients in the two point function of stress tensors or currents, respectively, see eq.~\eqref{cjct22}.
The value of chemical potential for which $z$ becomes imaginary in this expression exactly matches the value for which the extremal black holes cease to exist (see discussion at the beginning of this appendix).
We stress once more that the conclusion that the complexity of formation diverges in the zero temperature limit holds also in the spherical geometry.

It is also interesting to write the first few terms in a small charge (small $y$) expansion. In fact, we will also expand our results for small $z$  (large temperatures). In order to compare the results for charged black holes to those of neutral black holes found in \cite{Formation}, we express  the result of \cite{Formation} for spherical neutral black holes in $d=4$, as an expansion in small $z$, (large horizon radius)
\begin{equation}\label{FormUnchargedExp}
\frac{\Delta \mathcal{C}_A}{S} \bigg|_{\mu =0} = \frac{1}{2 \pi} +  \frac{z^3}{\pi} - \frac{9 \, z^4}{16 \pi} + \mathcal{O}(z^6) = \frac{1}{2 \pi}  + \frac{1}{\pi^4} \frac{1}{(T R)^3}-\frac{9}{16 \pi^5} \frac{1}{(T R)^4} + \mathcal{O}\left(\frac{1}{(T R)^6}\right) \, .
\end{equation}
The dependence on $z^3$ in the expansion comes from the vacuum contribution to the complexity of formation for the spherical geometry, as can be seen from the $L^3 \, \delta_{k,1}$ dependence in equation (3.14) in \cite{Formation}.
For charged black holes, a double expansion in $y$ and $z$ reads
\begin{align}
\frac{\Delta \mathcal{C}_A}{S}  &= \left( \frac{1}{2 \pi} +  \frac{z^3}{\pi} - \frac{9 \, z^4}{16 \pi}  \right) - \left( \frac{9 z^2}{8 \pi} - \frac{3 \, z^4}{16 \pi}  \right) y^2   \nonumber \\
&+\left( \frac{2}{3 \pi^2} \left(5 + 3 \log \frac{R \alpha y z}{2 L} \right) - \frac{z^2}{\pi^2} + \frac{z^4}{2 \pi^2} \right) y^3+ \mathcal{O}(z^5, y^4) \, .
\end{align}
We see by comparing this expression to \eqref{FormUnchargedExp} that the neutral limit is recovered in the zero charge limit $y \rightarrow 0$.

\subsection{Complexity=Volume}
In this subsection, we examine the complexity of formation evaluated using the CV conjecture. For simplicity, we will only consider planar (\ie $k=0$) charged black holes in $d=4$. The complexity of formation is then given by the following integrals:
\begin{equation}\label{VolChargedForm1}
\Delta \mathcal{C}_V = \frac{2 \Omega_{0,3}}{G_N L}
\left[
\int_{r_+}^{r_{\max}}
\frac{r^{3}dr }{\sqrt{f(r)}}
- \int_{0}^{r_{\max}}
\frac{r^{3}dr }{\sqrt{f_0(r)}}
\right],
\end{equation}
where $f(r)$ is the blackening factor \eqref{ChargedMetric} with $k=0$, and $f_0(r)=r^2/L^2$ is the corresponding `blackening' factor for empty AdS space.
Using eqs.~\eqref{eq:xyz} and \eqref{blackxyz} we can  perform a change of variables in \eqref{VolChargedForm1} to the dimensionless coordinate $x$ and then decompose the integration region for $x<1$ and $x>1$ which reads
\begin{equation}\label{CFormVol11}
\Delta \mathcal{C}_V = 8\, S \left[ \int_1^{\infty}\left(\frac{x}{\sqrt{\tilde f(x,y)}}-1\right) x^2 dx -\frac{1}{3}\right].
\end{equation}
To evaluate the remaining integral, it is useful to perform the following change of variables:
\begin{equation}
x^2=\frac{1}{u} + 1
\end{equation}
and the integral can be evaluated explicitly yielding
\begin{equation}\label{CVFinal2}
\Delta \mathcal{C}_V = \frac{8 \left(y^4+y^2+1\right) }{3 \sqrt{y^2+2}}\,S \,K\!\left(\frac{2 y^2+1}{y^2+2}\right)
\end{equation}
where $K$ is the complete elliptic integral of the first kind and $y$ can be expressed in terms of the boundary quantity $\nu$ using eq.~\eqref{yofnu}.

There are two interesting limits to explore. The small charge limit $\nu \rightarrow 0$ and the near extremal limit $\nu \rightarrow \infty$. In the small charge limit, an expansion of eq.~\eqref{CVFinal2} reads:
\begin{equation}
\begin{split}
\Delta \mathcal{C}_V =
S &\left(
\frac{2 \sqrt{\pi }\Gamma \left(-\frac{3}{4}\right)}{\Gamma \left(-\frac{1}{4}\right)}
+\frac{5 }{9 \pi ^{3/2}} \left(\frac{4 \Gamma \left(\frac{1}{4}\right)}{\Gamma \left(\frac{3}{4}\right)}-\frac{\Gamma \left(-\frac{1}{4}\right)}{\Gamma \left(\frac{5}{4}\right)}\right)\nu^2\right.
\\
&~~~~\left.
+
\frac{100  \left(7 \sqrt{2} \Gamma \left(-\frac{1}{4}\right) \Gamma \left(\frac{3}{4}\right) \Gamma \left(\frac{7}{4}\right)-12 \pi  \Gamma \left(\frac{5}{4}\right)\right)}{81 \pi ^{9/2} \Gamma \left(\frac{7}{4}\right)} \nu^4\right)\,.
\end{split}
\end{equation}
Note that the leading term above matches the complexity of formation given in eq.~(5.8) of \cite{Formation} for a planar neutral black hole in $d=4$.
A near extremal (small temperature, or equivalently in the planar case, large $\nu \equiv \sqrt{\frac{C_J}{C_T}} \frac{\mu}{T}$) expansion reads:
\begin{equation}
\Delta \mathcal{C}_V = \frac{4 S}{\sqrt{3}} \log \left(\frac{48 \sqrt{5} \nu}{\pi }\right)
-\frac{\pi  S }{3 \sqrt{15} \nu}
\left(1+9 \log \left(\frac{48\sqrt{5} \nu}{\pi }\right)\right)+\cdots\, ,
\end{equation}
and we see that the complexity of formation is logarithmically divergent at extremality. This is similar to what we found using the action conjecture, see eq.~\eqref{eq.extd4k022}. The reason for the divergence is easily understood looking back at the integral in eq.~\eqref{CFormVol11} and the definition of
$\tilde f(x,y)$ in eq. \eqref{blackxyz}. In the near extremal limit $\nu \rightarrow \infty$, the function $\tilde f(x,y)$ has two zeros in the neighborhood of $x=1$ namely
\begin{equation}
x_1=1, \qquad x_2=y= 1-\frac{\pi }{2 \sqrt{5} \nu} +\dots.
\end{equation}
and so we are approximately integrating $1/(x-1)$ all the way to $x=1$.

\begin{figure}[htbp]
\centering
\includegraphics[keepaspectratio, scale=0.43]{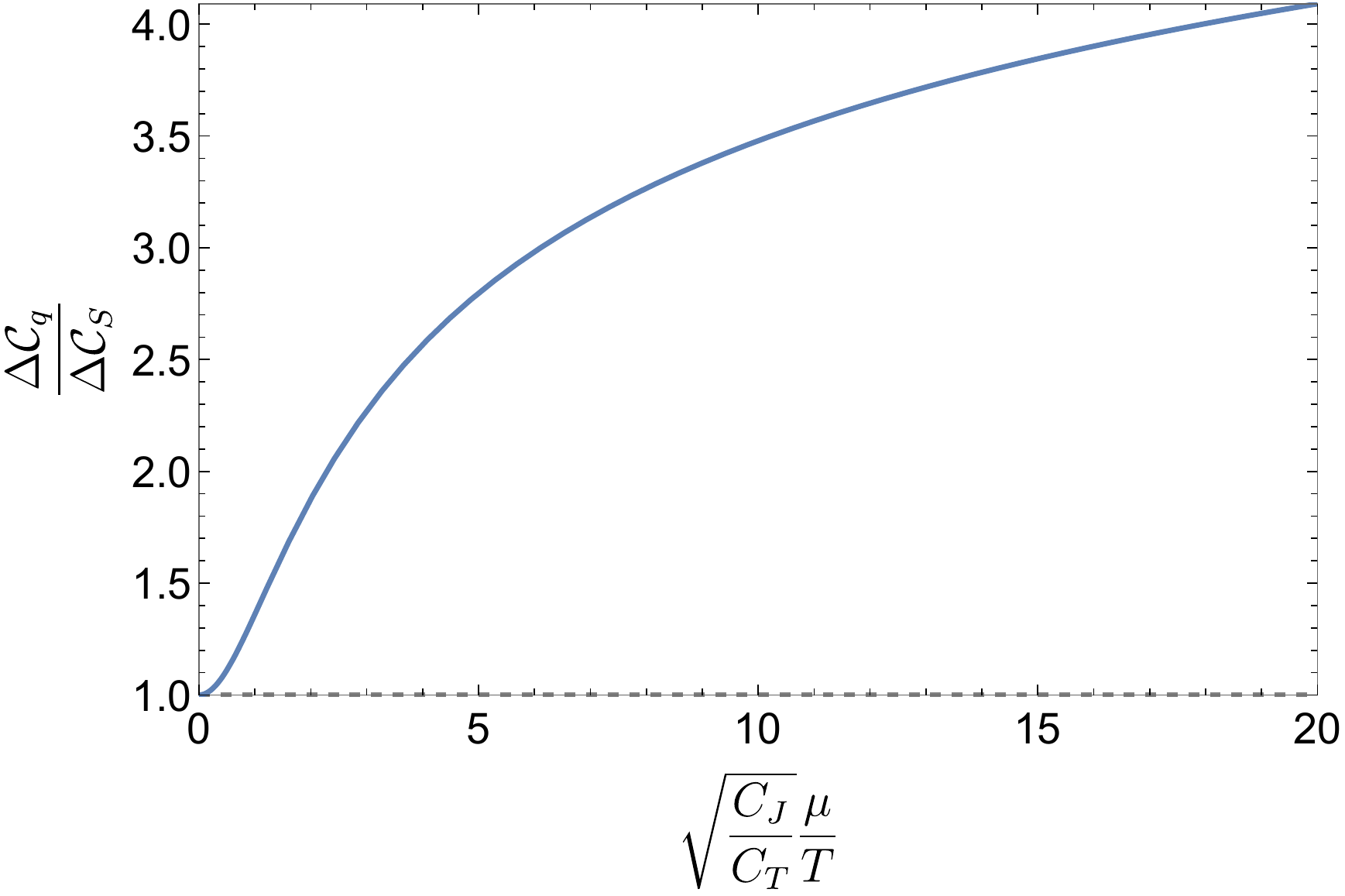}
\includegraphics[keepaspectratio, scale=0.43]{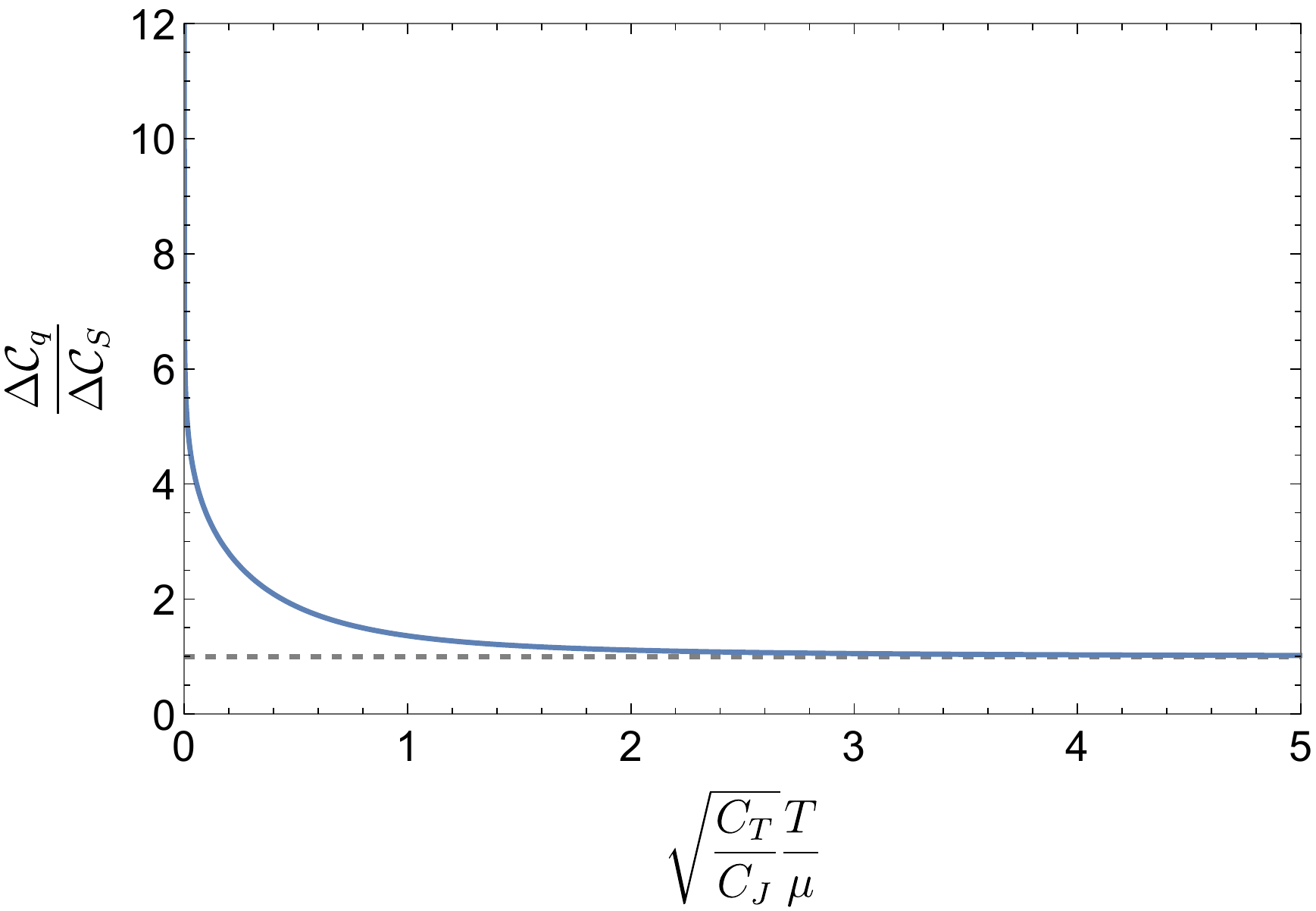}
	\caption{
		Complexity of formation from the CV conjecture normalized by $\Delta \mathcal{C}_S$ for planar ($k=0$) charged black holes in $d=4$ as a function of the dimensionless ratio of boundary quantities $\nu\equiv\sqrt{\frac{C_{J}}{C_{T}}} \frac{\mu}{T}$ and its inverse. Extremal black holes ($T=0$) have divergent complexity of formation also using the CV conjecture.}
	\label{CV:ChargedFormPlanarAdS5}
\end{figure}

\begin{figure}[htbp]
\centering
\includegraphics[keepaspectratio, scale=0.43]{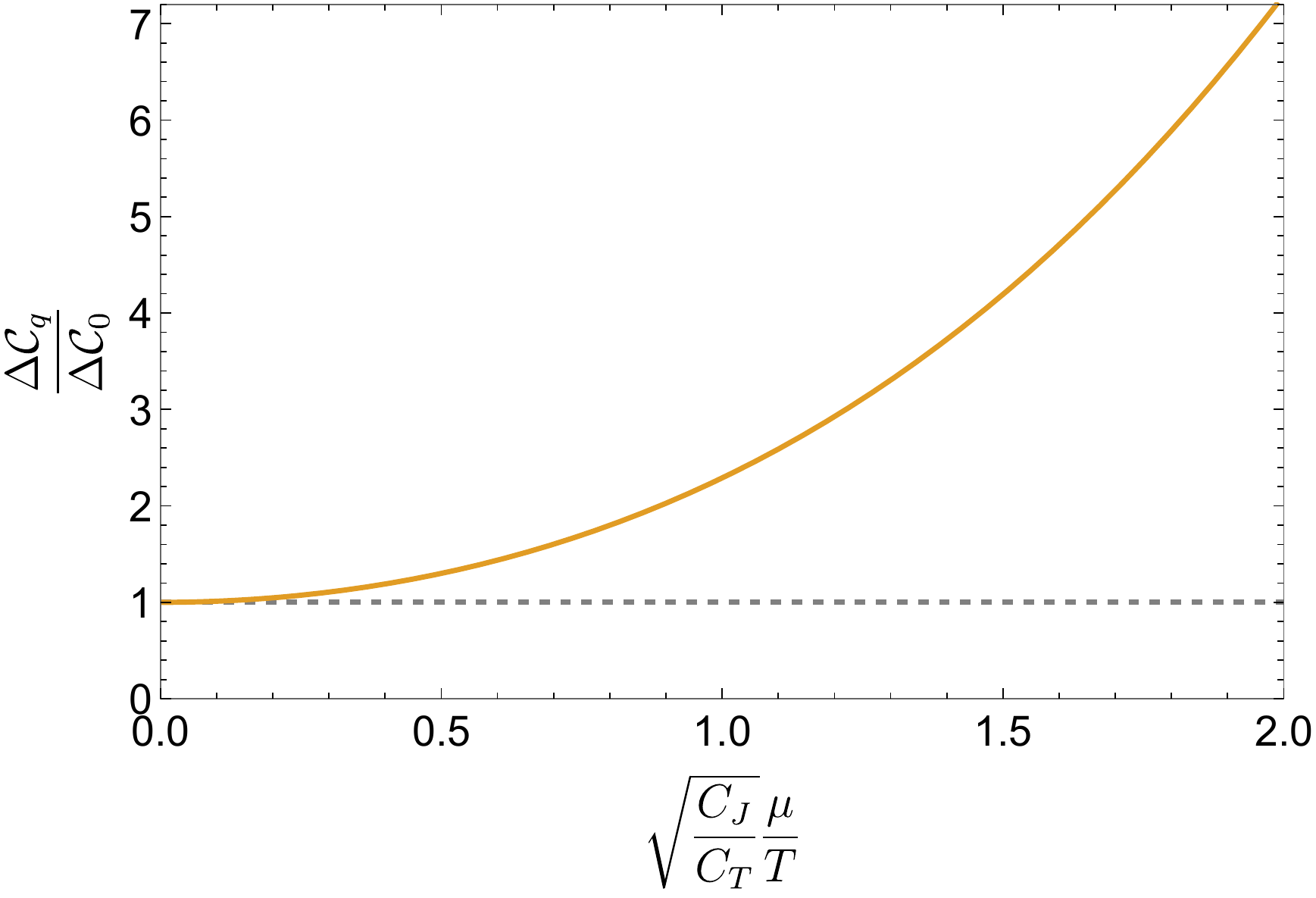}
\includegraphics[keepaspectratio, scale=0.43]{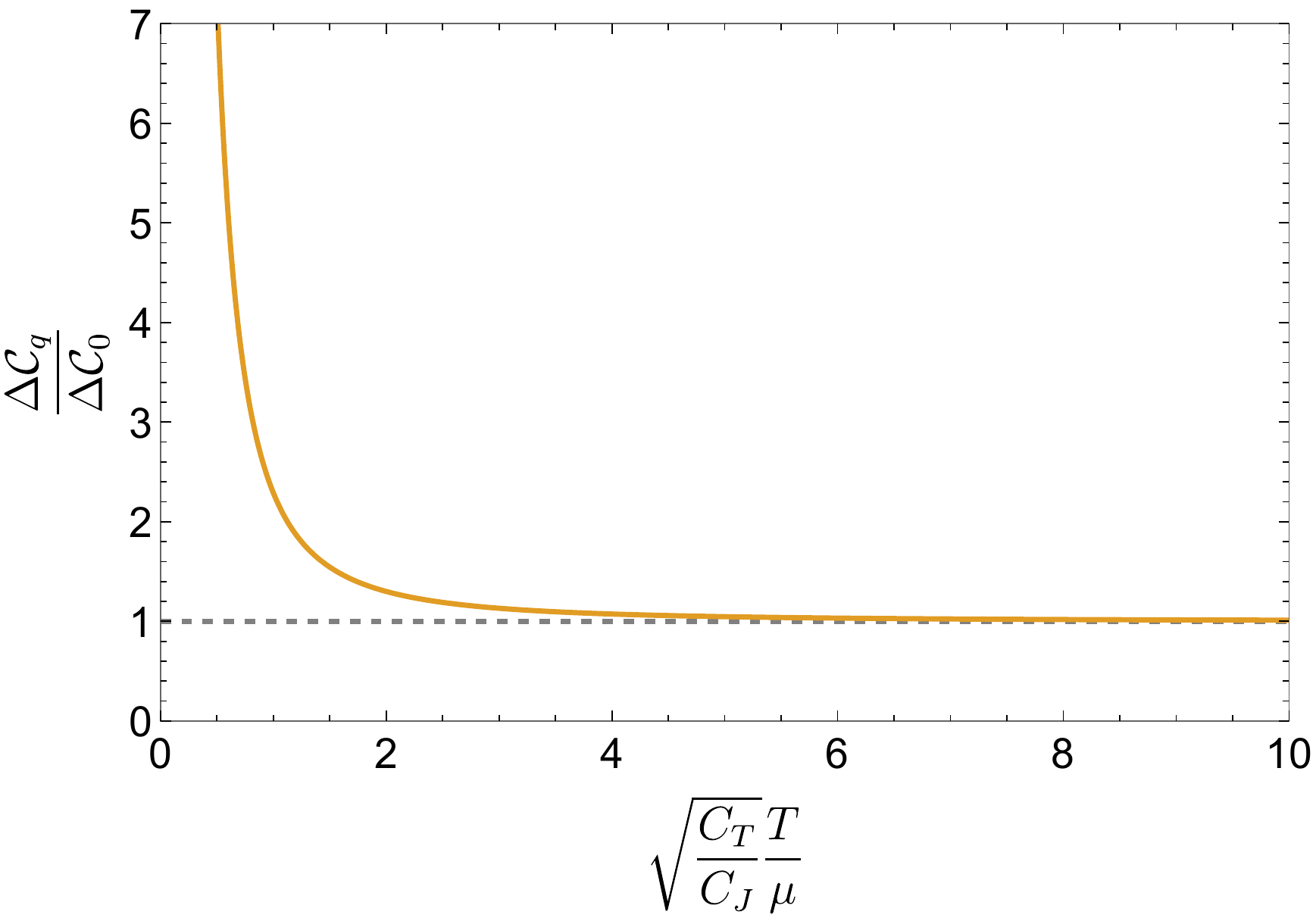}
	\caption{
		Complexity of formation from the CV conjecture normalized by $\Delta \mathcal{C}_0$ for planar ($k=0$) charged black holes in $d=4$ as a function of the dimensionless ratio of boundary quantities $\nu\equiv\sqrt{\frac{C_{J}}{C_{T}}} \frac{\mu}{T}$ and its inverse. Extremal black holes ($T=0$) have divergent complexity of formation also using the CV conjecture.}
	\label{CV:ChargedFormPlanarAdS5b}
\end{figure}

The full $\nu$ dependence of the complexity of formation is presented in figures \ref{CV:ChargedFormPlanarAdS5}
and \ref{CV:ChargedFormPlanarAdS5b} with two possible normalizations. First, we define
\begin{equation}
\Delta \mathcal{C}_S = \frac{2  \sqrt{\pi }  \Gamma
\left(
-\frac{3}{4}\right)}{\Gamma \left(-\frac{1}{4}\right)}\,S
\end{equation}
as a natural normalization to $ \Delta \mathcal{C}_V$ where $S$  in this expression denotes the entropy of the charged black hole. Another potential normalization is the corresponding complexity of formation of a neutral black hole with the same temperature.  Expressing the latter in terms of $\nu$, we have
\begin{equation}
\begin{split}
&\Delta \mathcal{C}_0=
\frac{    \Gamma  \left( -\frac{3}{4}\right)}{\Gamma \left(-\frac{1}{4}\right)}
\frac{\pi^6\sqrt{\pi }}{10}V C_T T^3
=-\frac{\sqrt{\pi }  \left(y^4+y^2-2\right)^3 \Gamma \left(-\frac{3}{4}\right)}{4 \Gamma \left(-\frac{1}{4}\right)}\,S
\\
&=\frac{27 \pi ^{7/2} \left(\sqrt{80 \nu^2+9 \pi ^2}-3 \pi \right)^3 \Gamma \left(-\frac{3}{4}\right)}{32000 \nu^6 \Gamma \left(-\frac{1}{4}\right)}\, S\ =\  \frac{27 \pi ^3 \left(\sqrt{80 \nu^2+9 \pi ^2}-3 \pi \right)^3 }{64000 \nu^6 }\, \Delta \mathcal{C}_S
\end{split}
\end{equation}
where  $V=\Omega_{k,d-1} R^3$ and again $y$ was expressed in terms of $\nu$ using eq.~\eqref{yofnu}.

The results are very similar to what we found with  the CA conjecture, see \eg figure \ref{fig:CFixedMu} and the expansions in eqs.~\eqref{eq:SeriesCoFPlan} and \eqref{eq.extd4k022}. The logarithmic divergence for near extremal black holes is present using both the CA and the CV conjectures, however the additional scale in the logarithm governing the divergence is now $\mu$ rather than $\alpha/L$ the extra scale in the boundary theory introduced there by the choice of normalization of the null normals. Just like for the CA conjecture, here as well the neutral result is recovered in the limit of vanishing chemical potential.

\subsection{Small Hyperbolic Black Holes} \label{hyperbh}
We briefly comment below on the time evolution of uncharged small hyperbolic black holes. For hyperbolic black holes with $r_h<L$, the mass parameter is negative, as can be seen from eqs.~\eqref{horiz} and \eqref{Mass}. In this case, the causal structure changes, with the appearance of an inner Cauchy horizon, and becomes similar to the one of charged black holes, as in figure  \ref{fig:charged}.
As was already pointed out in appendix [C.3] of \cite{Formation}, the CA calculation indicates that for small uncharged hyperbolic black holes the complexity does not change with time. In this subsection, we present an alternative argument for that statement using the neutral limit of charged black holes.

Consider the late time limit of the rate of change in complexity in eq.~\eqref{ChargedActionLateTimeDiff}. In general, the zero charge limit is obtained by the requirement that the chemical potential vanishes.
For small hyperbolic black holes, this limit does not coincide with the one in which the variable $y$ vanishes.
The expression for the chemical potential in general $d$ for $k=-1$ can be obtained from the multiplication of $h(y,z)$ and $\tilde h(y,z)$ in eq.~\eqref{eq:RTnuGenerald}, and it vanishes for
\begin{equation}
\mu = 0 \qquad \rightarrow \qquad z = \sqrt{\frac{1-y^d}{1-y^{d-2}}} \, .
\end{equation}
Evaluating eq.~\eqref{ChargedActionLateTimeDiff} for this value of $z$, namely, at zero chemical potential, results in a vanishing time derivative of $\mathcal{C}_{A}$ for small uncharged hyperbolic black holes.

\section{Ambiguities in the Action Calculations} \label{app:EternalAmb}

It was argued in \cite{RobLuis} that the null boundary terms in eq.~\reef{THEEACTION}, associated with
null boundary surfaces and null joints, introduce certain ambiguities in the numerical value of the gravitational action. In this appendix we consider the influence of these ambiguities on the time dependence of complexity of neutral black holes studied in this paper in section \ref{sec:EternalAction} using the CA conjecture.
The influence of the various ambiguities on the complexity of formation was studied in appendix D of \cite{Formation} and we will follow the discussion there closely. In particular it was demonstrated there that a large class of ambiguities are essentially equivalent to adding a constant to the null joint term $a$. This amounts to changing $a$ in eq.~\reef{THEEACTION} to
\begin{equation}
a_{\text{new}} = a + a_0 .
\end{equation}
This is indeed the effect of multiplying the function $\Phi(x)$, which determines the position of  the null surface according to $\Phi(x)=0$, by a constant. A similar effect is achieved by a constant rescaling of the parameter $\lambda$, which runs along the null generators.  Finally, this is also equivalent to changing the normalization constant $\alpha$, which fixes  the null normal normalization at the asymptotic boundary according to $\hat k  \cdot \hat \tau = \pm \alpha$. We reiterate here, that these ambiguities do not affect the late time rate of growth of holographic complexity. In subsection \ref{E1} we explore the influence of a constant $a_0$ on the action calculation.
In appendix B of \cite{RobLuis} it was argued that the reparametrization ambiguity can be avoided by including a certain boundary counterterm. We explore this possibility in subsection \ref{walk}.

\subsection{Influence of a Constant $a_0$}\label{E1}
\begin{figure}
\centering
		\includegraphics[scale=0.325]{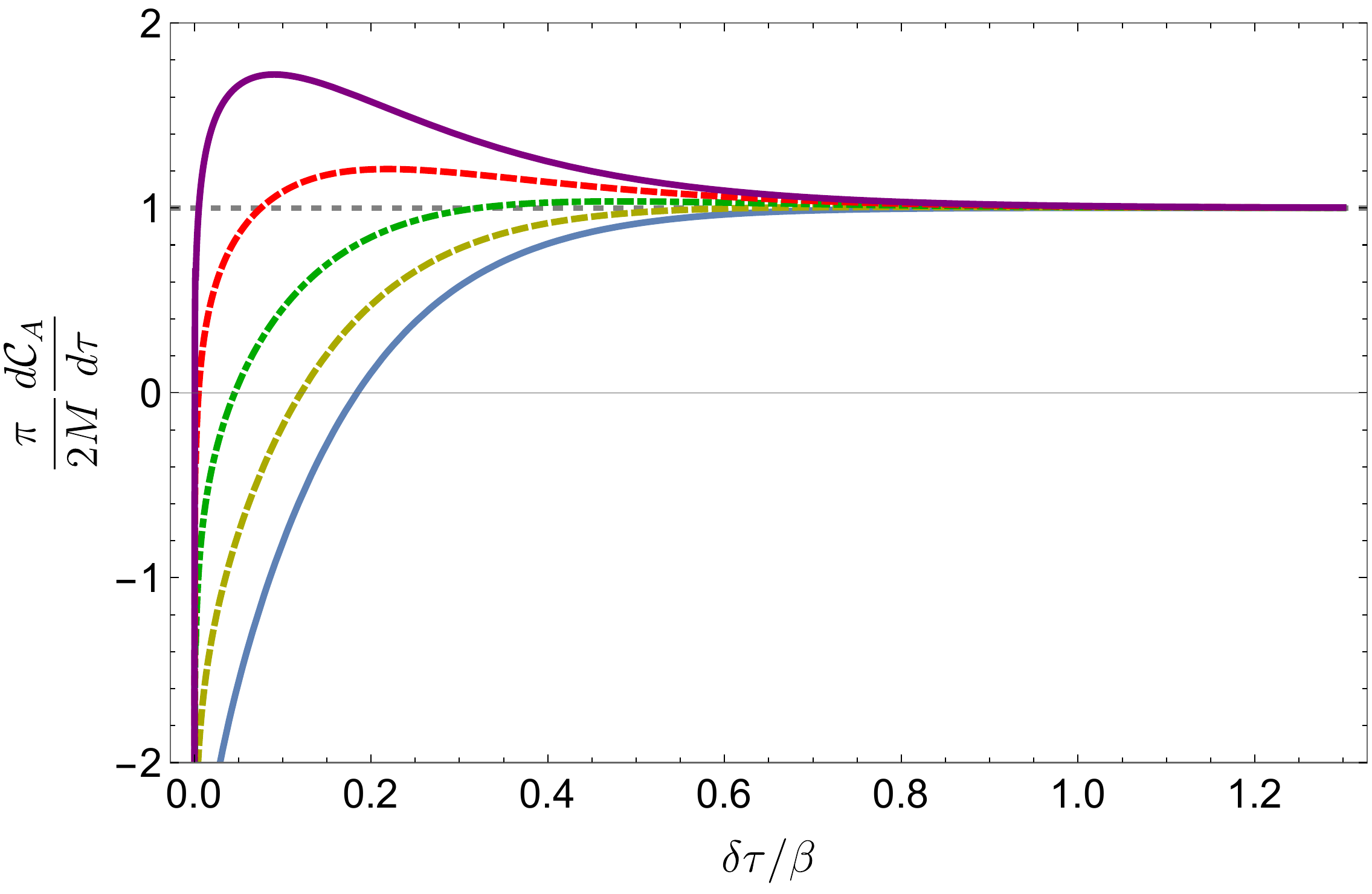}~
		\includegraphics[scale=0.36]{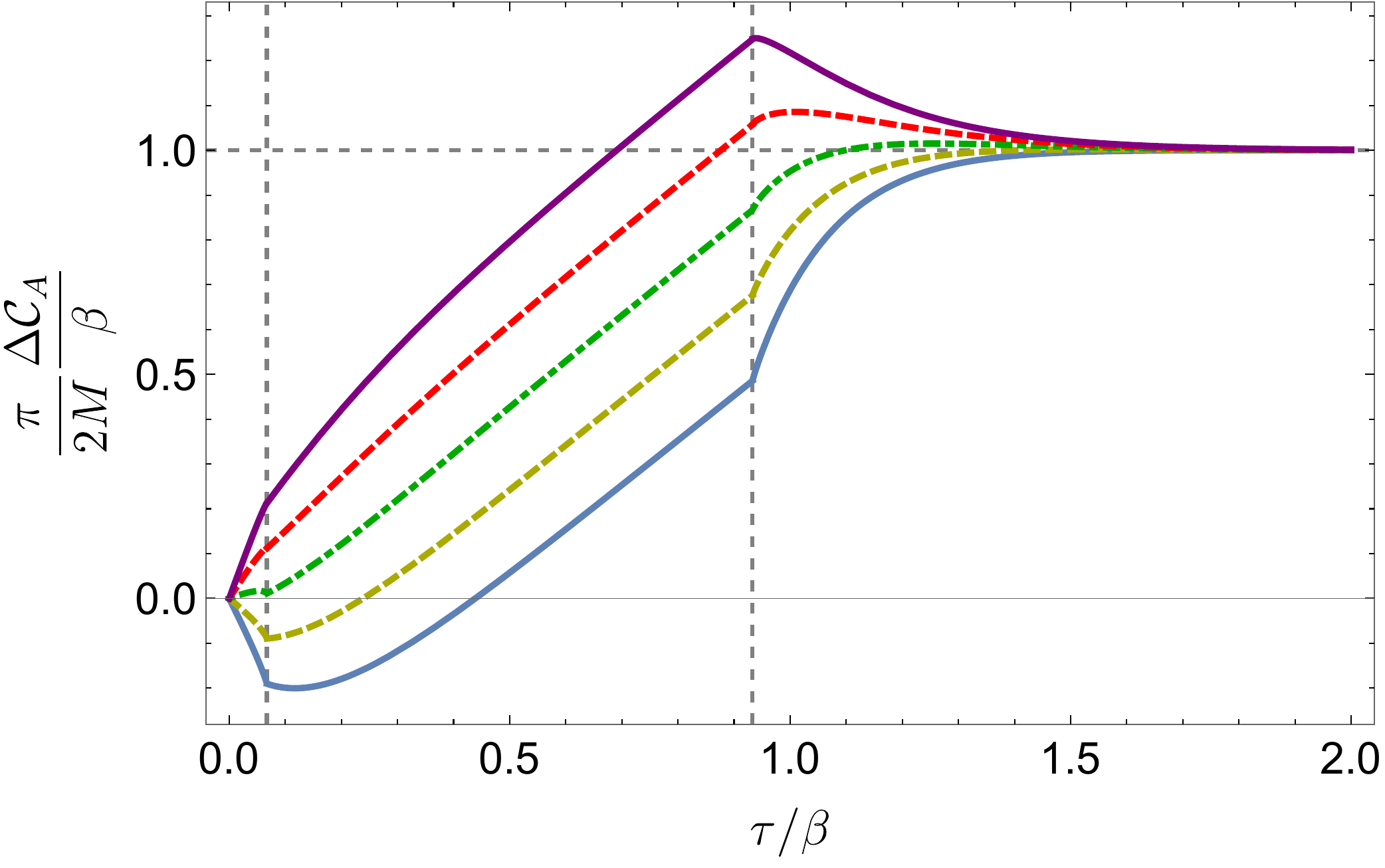}
		\caption{The rate of change of the complexity (left) and its average value (right) as a function of time for spherical black holes ($k=1$) in $d=4$ with $r_h=2 L$ for different values of the constant $a_0$ -- $a_0=-4$ (blue, solid), $a_0=-2$ (yellow, dashed), $a_0=0$ (green, dot-dashed), $a_0=2$ (red, dashed) and $a_0=4$ (purple, solid). We have set $\alpha=L/R$ for simplicity.}
\label{fig:a0}
\end{figure}

When $a_0$ is a fixed constant, the joint term at $r=r_m$ in our calculations in section \ref{sec:EternalAction} is modified by
\begin{equation}
\Delta I_{\jnt} =  a_0 \, \frac{\Omega_{k, d-1}}{8 \pi G_N}\,  r_{m}^{d-1}\, .
\end{equation}
Taking the time derivative and using eq.~\reef{forchain1} yields
\begin{equation}
\label{eq:corn34}
\Delta \Big(  \frac{d \mathcal{C}_{A}}{d\tau}  \Big)= - a_0 \, \frac{\Omega_{k, d-1}(d-1)}{16 \pi^2 G_N}\, \frac{L}{R} r_{m}^{d-2} f(r_m)\, .
\end{equation}
This shift in the corner term is equivalent to changing the normalization constant $\alpha$ in eq.~\eqref{tder44} to $\alpha_N = e^{a_0/2} \, \alpha$. Note that the term in eq.~\reef{eq:corn34} also vanishes in the late time limit since $r_m$ approaches the horizon radius $r_h$ there and so $f(r_m)$ vanishes as $\tau \to \infty$. The modification does however contribute to the rate of change of complexity at earlier times. The influence of a constant $a_0$ on the rate of change of complexity and its average for a spherical black hole in $d=4$ is studied numerically in figure \ref{fig:a0}. We note that the averaging procedure suggested in  eq.~\eqref{averageC} somewhat  reduces the effect of changing $a_0$, however the bound is still approached from above at late times.

\subsection{Boundary Counterterm} \label{walk}
In this subsection we discuss the effect of adding the boundary counterterm suggested in appendix B of \cite{RobLuis} for eternal black hole backgrounds \eqref{HigherDMetric} on the rate of change of complexity.  This counterterm makes the action invariant under the reparametrization of null surfaces.  For simplicity we set in this subsection $R=L$.
The counterterm for each null surface is given by
\begin{align}
\Delta I_{\Sigma}=\frac{1}{8\pi G_N} \int_{\Sigma} \!
d \lambda d^{d-1} \sqrt{\gamma} \Theta \log(\tilde{L}|\Theta|),
\label{nullcounterterm}
\end{align}
where $\gamma_{AB}$ is the cross-sectional metric of a bundle of null generators, $\Theta$ is the expansion parameter given by $\Theta = \del_\lambda \log{\sqrt \gamma}$ and $\tilde{L}$ is an arbitrary length scale.\footnote{The choice of the length scale corresponds to the ambiguous constant $c$ in eq.~(B4) of ref.~\cite{RobLuis}.}
We take for simplicity an affine parametrization
\begin{align}
\lambda =
\frac{r}{\alpha}.
\end{align}
However, keep in mind that  the total action with the counterterm does not depend on the parametrization of null surfaces.
In this parametrization,
the expansion takes the form
\begin{align}
\Theta = \frac{(d-1)\alpha}{r}.
\end{align}
Taking into account that there are two future null boundaries and two past ones,
the counterterm \eqref{nullcounterterm} at $t>t_{c}$ becomes
\small
\begin{align}
\Delta I_{\Sigma} &= \frac{(d-1) \Omega_{k,d-1}}{4\pi G_N} \int^{r_{max}}_{0} \!\!\!\!\!
d r \, r^{d-2}  \log \frac{(d-1)\alpha\tilde{L}}{r}
+\frac{(d-1) \Omega_{k,d-1}}{4\pi G_N} \int^{r_{max}}_{r_{m}} \!\!\!\!\!
d r \, r^{d-2}  \log \frac{(d-1)\alpha\tilde{L}}{r}
\nonumber\\
&= \frac{\Omega_{k,d-1}}{2\pi G_N} r_{max}^{d-1}
\Bigl(\log \frac{(d-1)\alpha\tilde{L}}{r_{max}}
+\frac{1}{d-1}
\Bigr)
- \frac{\Omega_{k,d-1}}{4\pi G_N} r_{m}^{d-1}
\Bigl(\log \frac{(d-1)\alpha\tilde{L}}{r_{m}}
+\frac{1}{d-1}
\Bigr).
\end{align}
\normalsize
The time derivative of the counterterm is then readily evaluated using the relation \eqref{forchain1} and found to be
\begin{equation}
\frac{d \Delta I_{\Sigma}}{d t} =
-\frac{(d-1)\Omega_{k,d-1}r_m^{d-2}}{8\pi G_N}
 f(r_m) \log{\left(\frac{r_m}{(d-1)\alpha\tilde{L}} \right)}\, .
\label{dIdt_counter}
\end{equation}
If we take another parametrization of null surfaces,
the expression \eqref{dIdt_counter} changes.
However, the total action is invariant under reparametrization. The rate of change of complexity with the counterterm is given by the following expression for any parametrization:
\begin{equation}
\frac{d \mathcal{C}_A}{d t}= \frac{1}{\pi} \left(2M+
\frac{\Omega_{k,d-1} (d-1) r_{m}^{d-2}  f(r_m)}{16 \pi G_N} \left[  \log |f(r_{m})|- 2 \log{\left(\frac{r_m}{(d-1)\tilde{L}} \right)} \right] \right).
\label{dca/dt_counter}
\end{equation}
Note that the $\alpha$-dependence which appeared in eq.~\eqref{tder1} is totally canceled when including the boundary counterterm.
We see from this expression that the counterterm does not resolve the divergence in $\frac{d \mathcal{C}_A}{d t}$ at times shortly after the critical time $t_c$ which we observed in section \ref{timeder} for $d>2$.
In fact, eq.~\eqref{dca/dt_counter} behaves shortly after $t_c$ as
\begin{equation}
\frac{d \mathcal{C}_A}{d t} \sim
\frac{\Omega_{k, d-1} d (d-1) \omega^{d-2}}{16 \pi^2 G_N} \log r_m+ \text{finite}\,,
\end{equation}
where $r_m$ is very close to $r=0$ at times right after $t_{c}$.


\begin{thebibliography}{99}

\bibitem{Ryu:2006bv}
  S.~Ryu and T.~Takayanagi,
  ``Holographic derivation of entanglement entropy from AdS/CFT,''
  Phys.\ Rev.\ Lett.\  {\bf 96}, 181602 (2006)
  \href{https://arxiv.org/abs/hep-th/0603001}{hep-th/0603001}.

\bibitem{Casini:2011kv}
  H.~Casini, M.~Huerta and R.~C.~Myers,
  ``Towards a derivation of holographic entanglement entropy,''
  JHEP {\bf 1105}, 036 (2011)
  \href{https://arxiv.org/abs/1102.0440}{hep-th/1102.0440}.


\bibitem{Lewkowycz:2013nqa}
  A.~Lewkowycz and J.~Maldacena,
  ``Generalized gravitational entropy,''
  JHEP {\bf 1308}, 090 (2013),
  \href{https://arxiv.org/abs/1304.4926}{hep-th/1304.4926}.

\bibitem{Dong:2016hjy}
  X.~Dong, A.~Lewkowycz and M.~Rangamani,
  ``Deriving covariant holographic entanglement,''
  JHEP {\bf 1611} (2016) 028
    \href{https://arxiv.org/abs/1607.07506}{hep-th/1607.07506}.

\bibitem{Susskind:2014rva}
  L.~Susskind,
  ``Computational Complexity and Black Hole Horizons,''
  Fortsch.\ Phys.\  {\bf 64}, 24 (2016)
  \href{https://arxiv.org/abs/1402.5674}{hep-th/1402.5674}.

\bibitem{Stanford:2014jda}
  D.~Stanford and L.~Susskind,
  ``Complexity and Shock Wave Geometries,''
  Phys.\ Rev.\ D {\bf 90}, no. 12, 126007 (2014),
    \href{https://arxiv.org/abs/1406.2678}{hep-th/1406.2678}.


\bibitem 
 {Brown1}  A.~R.~Brown, D.~A.~Roberts, L.~Susskind, B.~Swingle and Y.~Zhao,
  ``Holographic Complexity Equals Bulk Action?,''
  Phys.\ Rev.\ Lett.\  {\bf 116} (2016) no.19,  191301,
  \href{https://arxiv.org/abs/1509.07876}{hep-th/1509.07876}.


\bibitem 
 {Brown2}  A.~R.~Brown, D.~A.~Roberts, L.~Susskind, B.~Swingle and Y.~Zhao,
  ``Complexity, action, and black holes,''
  Phys.\ Rev.\ D {\bf 93} (2016) no.8,  086006,
  \href{https://arxiv.org/abs/1512.04993}{hep-th/1512.04993}.

   \bibitem 
 {johnw} J.~Watrous, ``Quantum Computational Complexity,'' pp 7174-7201 in {\it Encyclopedia of Complexity and Systems Science}, ed., R.~A.~Meyers  (Springer, 2009)  \href{https://arxiv.org/abs/0804.3401}{quant-ph/0804.3401}.

\bibitem 
 {AaronsonRev}   S.~Aaronson,
  ``The Complexity of Quantum States and Transformations: From Quantum Money to Black Holes,''
  \href{https://arxiv.org/abs/1607.05256}{quant-ph/1607.05256}.

  \bibitem 
 {qft1} R.~A.~Jefferson and R.~C.~Myers,
  ``Circuit complexity in quantum field theory,''
    \href{https://arxiv.org/abs/1707.08570}{hep-th/1707.08570}.

\bibitem 
 {qft2} S.~Chapman, M.~P.~Heller, H.~Marrochio and F.~Pastawski,
  ``Towards Complexity for Quantum Field Theory States,''
   \href{https://arxiv.org/abs/1707.08582}{hep-th/1707.08582}.

\bibitem 
 {koji} K.~Hashimoto, N.~Iizuka and S.~Sugishita,
  ``Time Evolution of Complexity in Abelian Gauge Theories - And Playing Quantum Othello Game -,''
  \href{https://arxiv.org/abs/1707.03840}{hep-th/1707.03840}.

   \bibitem{Chemissany:2016qqq}
  W.~Chemissany and T.~J.~Osborne,
  ``Holographic fluctuations and the principle of minimal complexity,''
  JHEP {\bf 1612}, 055 (2016)
  \href{https://arxiv.org/abs/1605.07768}{hep-th/11605.07768}.

\bibitem{EuclideanComplexity1}
  P.~Caputa, N.~Kundu, M.~Miyaji, T.~Takayanagi and K.~Watanabe,
  ``Anti-de Sitter Space from Optimization of Path Integrals in Conformal Field Theories,''
  Phys.\ Rev.\ Lett.\  {\bf 119}, no. 7, 071602 (2017)
  \href{https://arxiv.org/abs/1703.00456}{hep-th/1703.00456}.

  \bibitem{EuclideanComplexity2}
  P.~Caputa, N.~Kundu, M.~Miyaji, T.~Takayanagi and K.~Watanabe,
  ``Liouville Action as Path-Integral Complexity: From Continuous Tensor Networks to AdS/CFT,''
   \href{https://arxiv.org/abs/1706.07056}{hep-th/1706.07056}.

  \bibitem{EuclideanComplexity3}
  B.~Czech,
  ``Einstein's Equations from Varying Complexity,''
    \href{https://arxiv.org/abs/1706.00965}{hep-th/1706.00965}.


  \bibitem 
 {prep9} D.~A.~Roberts and B.~Yoshida, ``Chaos and complexity by design,''
  JHEP {\bf 1704}, 121 (2017)
  \href{https://arxiv.org/abs/1610.04903}{quant-ph/1610.04903}.

\bibitem{Yang:2017nfn}
  R.~Q.~Yang,
  ``A Complexity for Quantum Field Theory and Application in Thermofield Double States,''
  \href{https://arxiv.org/abs/1709.00921}{hep-th/1709.00921}


\bibitem{MaldacenaEternal}
  J.~M.~Maldacena,
  ``Eternal black holes in anti-de Sitter,''
  JHEP {\bf 0304}, 021 (2003)
  \href{https://arxiv.org/abs/hep-th/0106112}{hep-th/0106112}.

\bibitem{Hartman:2013qma}
T.~Hartman and J.~Maldacena,
``Time Evolution of Entanglement Entropy from Black Hole Interiors,''
JHEP {\bf 1305}, 014 (2013),
\href{https://arxiv.org/abs/1303.1080}{hep-th/1303.1080}.


\bibitem{Maldacena:2013xja}
  J.~Maldacena and L.~Susskind,
  ``Cool horizons for entangled black holes,''
  Fortsch.\ Phys.\  {\bf 61}, 781 (2013)
  \href{https://arxiv.org/abs/1306.0533}{hep-th/1306.0533}.

\bibitem 
 {2LawComp}  A.~R.~Brown and L.~Susskind,
  ``The Second Law of Quantum Complexity,''
  \href{https://arxiv.org/abs/1701.01107}{hep-th/1701.01107}.

\bibitem 
 {TheTaka}  M.~Miyaji, T.~Numasawa, N.~Shiba, T.~Takayanagi and K.~Watanabe,
  ``Distance between Quantum States and Gauge-Gravity Duality,''
  Phys.\ Rev.\ Lett.\  {\bf 115}, no. 26, 261602 (2015),
      \href{https://arxiv.org/abs/1507.07555}{hep-th/1507.07555}.

\bibitem 
{RobLuis}   L.~Lehner, R.~C.~Myers, E.~Poisson and R.~D.~Sorkin,
  ``Gravitational action with null boundaries,''
  Phys.\ Rev.\ D {\bf 94}, no. 8, 084046 (2016),
\href{https://arxiv.org/abs/1609.00207}{hep-th/1609.00207}.

\bibitem 
 {Formation} S.~Chapman, H.~Marrochio and R.~C.~Myers,
  ``Complexity of Formation in Holography,''
  JHEP {\bf 1701} (2017) 062,
  \href{https://arxiv.org/abs/1610.08063}{hep-th/1610.08063}

\bibitem 
 {diverg} D.~Carmi, R.~C.~Myers and P.~Rath,
  ``Comments on Holographic Complexity,''
  JHEP {\bf 1703} (2017) 118,
  \href{https://arxiv.org/abs/1612.00433}{hep-th/1612.00433}.

\bibitem{Reynolds:2016rvl}
  A.~Reynolds and S.~F.~Ross,
  ``Divergences in Holographic Complexity,''
  Class.\ Quant.\ Grav.\  {\bf 34}, no. 10, 105004 (2017)
  \href{https://arxiv.org/abs/1612.05439}{hep-th/1612.05439}.

\bibitem 
 {ying1}  Y.~Zhao,
  ``Complexity, boost symmetry, and firewalls,''
    \href{https://arxiv.org/abs/1702.03957}{hep-th/1702.03957}.

 \bibitem 
  {Lloyd} S.~Lloyd, ``Ultimate physical limits to computation,'' Nature 406 (2000), no. 6799 1047–1054 \href{https://arxiv.org/abs/quant-ph/9908043v3}{quant-ph/9908043}.

  \bibitem 
 {energy} R.~C.~Myers,
  ``Stress tensors and Casimir energies in the AdS/CFT correspondence,''
  Phys.\ Rev.\ D {\bf 60} (1999) 046002,
\href{https://arxiv.org/abs/hep-th/9903203}{hep-th/9903203}.

\bibitem 
 {count}   R.~Emparan, C.~V.~Johnson and R.~C.~Myers,
  ``Surface terms as counterterms in the AdS/CFT correspondence,''
  Phys.\ Rev.\ D {\bf 60} (1999) 104001,
\href{https://arxiv.org/abs/hep-th/9903238}{hep-th/9903238}.

\bibitem 
 {York} J.~W.~York, Jr.,
  ``Role of conformal three geometry in the dynamics of gravitation,''
  Phys.\ Rev.\ Lett.\  {\bf 28} (1972) 1082,
  \href{http://journals.aps.org/prl/abstract/10.1103/PhysRevLett.28.1082}{PhysRevLett.28.1082}.

\bibitem 
 {GH} G.~W.~Gibbons and S.~W.~Hawking,
  ``Action Integrals and Partition Functions in Quantum Gravity,''
  Phys.\ Rev.\ D {\bf 15} (1977) 2752,
  \href{http://journals.aps.org/prd/abstract/10.1103/PhysRevD.15.2752}{PhysRevD.15.2752}.



\bibitem 
 {Hay1} G.~Hayward,
  ``Gravitational action for space-times with nonsmooth boundaries,''
  Phys.\ Rev.\ D {\bf 47} (1993) 3275,
  \href{http://journals.aps.org/prd/abstract/10.1103/PhysRevD.47.3275}{PhysRevD.47.3275}.

\bibitem 
 {Hay2} D.~Brill and G.~Hayward,
  ``Is the gravitational action additive?,''
  Phys.\ Rev.\ D {\bf 50} (1994) 4914,
\href{https://arxiv.org/abs/gr-qc/9403018}{gr-qc/9403018}.

\bibitem{Parattu:2015gga}
  K.~Parattu, S.~Chakraborty, B.~R.~Majhi and T.~Padmanabhan,
  ``A Boundary Term for the Gravitational Action with Null Boundaries'',
  Gen.\ Rel.\ Grav.\  {\bf 48}, no. 7, 94 (2016),
  \href{https://arxiv.org/abs/1501.01053}{gr-qc/1501.01053}.
  %

\bibitem{Hopfmuller:2016scf}
  F.~Hopfmuller and L.~Freidel,
  ``Gravity Degrees of Freedom on a Null Surface,''
  Phys.\ Rev.\ D {\bf 95}, no. 10, 104006 (2017)
  \href{https://arxiv.org/abs/1611.03096}{gr-qc/1611.03096}.

\bibitem{Jubb:2016qzt}
  I.~Jubb, J.~Samuel, R.~Sorkin and S.~Surya,
  ``Boundary and Corner Terms in the Action for General Relativity,''
  Class.\ Quant.\ Grav.\  {\bf 34}, no. 6, 065006 (2017)
  \href{https://arxiv.org/abs/1612.00149}{gr-qc/1612.00149}.

\bibitem{Wieland:2017zkf}
  W.~Wieland, ``New boundary variables for classical and quantum gravity on a null surface,''
  \href{https://arxiv.org/abs/1704.07391}{gr-qc/1704.07391}

\bibitem 
 {sken1} S.~de Haro, S.~N.~Solodukhin and K.~Skenderis,
  ``Holographic reconstruction of space-time and renormalization in the AdS / CFT correspondence,''
  Commun.\ Math.\ Phys.\  {\bf 217} (2001) 595,
\href{https://arxiv.org/abs/hep-th/0002230}{hep-th/0002230}.

\bibitem 
 {sken2} K.~Skenderis,
  ``Lecture notes on holographic renormalization,''
  Class.\ Quant.\ Grav.\  {\bf 19} (2002) 5849,
\href{https://arxiv.org/abs/hep-th/0209067}{hep-th/0209067}.

  \bibitem 
{Brown:2016wib}  A.~R.~Brown, L.~Susskind and Y.~Zhao,
``Quantum Complexity and Negative Curvature,''
  Phys.\ Rev.\ D {\bf 95}, no. 4, 045010 (2017)
    \href{https://arxiv.org/abs/1608.02612}{hep-th/1608.02612}.

\bibitem 
 {couscous} O.~Coussaert and M.~Henneaux,
  ``Supersymmetry of the (2+1) black holes,''
  Phys.\ Rev.\ Lett.\  {\bf 72} (1994) 183,
\href{https://arxiv.org/abs/hep-th/9310194}{hep-th/9310194}.

\bibitem{Alishahiha:2015rta}
  M.~Alishahiha,
  ``Holographic Complexity,''
  Phys.\ Rev.\ D {\bf 92}, no. 12, 126009 (2015)
   \href{https://arxiv.org/abs/hep-th/1509.06614}{hep-th/1509.06614}.



  \bibitem{DeanSub}
  O.~Ben-Ami and D.~Carmi,
  ``On Volumes of Subregions in Holography and Complexity,''
  JHEP {\bf 1611}, 129 (2016)
   \href{https://arxiv.org/abs/hep-th/1609.02514}{hep-th/1609.02514}.



\bibitem{Chamblin:1999tk}
	  A.~Chamblin, R.~Emparan, C.~V.~Johnson and R.~C.~Myers,
  ``Charged AdS black holes and catastrophic holography'',
  Phys.\ Rev.\ D {\bf 60} (1999) 064018,
   \href{https://arxiv.org/abs/hep-th/9902170}{hep-th/9902170}.

\bibitem 
 {HartLectures} S.~A.~Hartnoll,
  ``Lectures on holographic methods for condensed matter physics,''
  C Class.\ Quant.\ Grav.\  {\bf 26}, 224002 (2009),
\href{https://arxiv.org/abs/0903.3246}{hep-th/0903.3246}.

\bibitem 
{erd} J.~Erdmenger and H.~Osborn,
  ``Conserved currents and the energy momentum tensor in conformally invariant theories for general dimensions,''
  Nucl.\ Phys.\ B {\bf 483} (1997) 431
  \href{https://arxiv.org/abs/hep-th/9605009}{hep-th/9605009}.

\bibitem 
 {pet}   H.~Osborn and A.~C.~Petkou,
  ``Implications of conformal invariance in field theories for general dimensions,''
  Annals Phys.\  {\bf 231} (1994) 311
  \href{https://arxiv.org/abs/hep-th/9307010}{hep-th/9307010}.

\bibitem 
 {RobMisha} A.~Buchel, J.~Escobedo, R.C.~Myers, M.F.~Paulos, A.~Sinha and M.~Smolkin,
``Holographic GB gravity in arbitrary dimensions,''
JHEP {\bf 1003}, 111 (2010)
\href{https://arxiv.org/abs/0911.4257}{hep-th/0911.4257}.


\bibitem{Barnes:2005bw}
E.~Barnes, E.~Gorbatov, K.~A.~Intriligator and J.~Wright,
  ``Current correlators and AdS/CFT geometry,''
  Nucl.\ Phys.\ B {\bf 732}, 89 (2006)
\href{https://arxiv.org/abs/hep-th/0507146}{hep-th/0507146}.

\bibitem{FreedmanMathur}
D.~Z.~Freedman, S.~D.~Mathur, A.~Matusis and L.~Rastelli,
``Correlation functions in the CFT(d) / AdS(d+1) correspondence,''
Nucl.\ Phys.\ B {\bf 546}, 96 (1999)
\href{https://arxiv.org/abs/hep-th/9804058}{hep-th/9804058}.

\bibitem{Cai:2016xho}
  R.~G.~Cai, S.~M.~Ruan, S.~J.~Wang, R.~Q.~Yang and R.~H.~Peng,
  ``Action growth for AdS black holes,''  JHEP {\bf 1609} (2016) 161,
     \href{https://arxiv.org/abs/1606.08307}{gr-qc/1606.08307}.

   \bibitem 
 {vaid} S.~Chapman, H.~Marrochio and R.~C.~Myers, in preparation.

 \bibitem 
 {NonComm} J.~Couch, S.~Eccles, W.~Fischler and M.~Xiao, "Holographic Complexity of Non-Commutative SYM", in preparation.

\bibitem 
 {tsun1} H.~Liu and S.~J.~Suh,
  ``Entanglement Tsunami: Universal Scaling in Holographic Thermalization,''
  Phys.\ Rev.\ Lett.\  {\bf 112} (2014) 011601
  \href{https://arxiv.org/abs/1305.7244}{hep-th/1305.7244}.


\bibitem 
 {tsun2} H.~Liu and S.~J.~Suh,
 ``Entanglement growth during thermalization in holographic systems,''
  Phys.\ Rev.\ D {\bf 89} (2014) no.6,  066012
    \href{https://arxiv.org/abs/1311.1200}{hep-th/1311.1200}.

\bibitem 
 {mark1} M.~Mezei,
  ``On entanglement spreading from holography,''
  JHEP {\bf 1705} (2017) 064
    \href{https://arxiv.org/abs/1612.00082}{hep-th/1612.00082}.


\bibitem 
 {mark2} J.~S.~Cotler, M.~P.~Hertzberg, M.~Mezei and M.~T.~Mueller,
  ``Entanglement Growth after a Global Quench in Free Scalar Field Theory,''
  JHEP {\bf 1611} (2016) 166
     \href{https://arxiv.org/abs/1609.00872}{hep-th/1609.00872}.

\bibitem
 {SimpleComplexity}
 W.~Cottrell, M.~Montero, ``Complexity Is Simple!'', in preparation.

\bibitem 
 {rot1} S.~W.~Hawking, C.~J.~Hunter and M.~Taylor,
  ``Rotation and the AdS / CFT correspondence,''
  Phys.\ Rev.\ D {\bf 59} (1999) 064005
  \href{https://arxiv.org/abs/hep-th/9811056}{hep-th/9811056}.

\bibitem
 {TimeDepQFTUs}
 S.~Chapman, J.~Eisert, M.~P.~Heller, R.~A.~Jefferson, H.~Marrochio, R.~C.~Myers, F.~Pastawski, in preparation.

\bibitem 
 {rot2} G.~W.~Gibbons, H.~Lu, D.~N.~Page and C.~N.~Pope,
  ``Rotating black holes in higher dimensions with a cosmological constant,''
  Phys.\ Rev.\ Lett.\  {\bf 93} (2004) 171102
    \href{https://arxiv.org/abs/hep-th/0409155}{hep-th/0409155}.



\end{thebibliography}
\end{document}